\documentclass{aa}

\usepackage{txfonts}
\usepackage{textcomp}
\usepackage{tensor}
\usepackage[inline]{enumitem}
\usepackage{amsmath}
\usepackage[caption=false]{subfig}
\usepackage{booktabs}
\usepackage{hyperref}
\usepackage[figuresright]{rotating}
\usepackage{float}
\usepackage{placeins}
\usepackage{verbatim}

\usepackage{tabu, multirow}

\usepackage{cmdDef}

\newcommand{\magarcsq}[1]{{mag arcsec$^{-2}$ #1}}

\usepackage{xcolor}
\usepackage[normalem]{ulem} % use normalem to protect \emph
\newcommand\hl{\bgroup\markoverwith
  {\textcolor{yellow}{\rule[-.5ex]{2pt}{2.5ex}}}\ULon}
\hypersetup{draft}
\begin{document}

    \title{The Fornax3D project: Environmental effects on the assembly of dynamically cold disks in Fornax cluster galaxies}
    \titlerunning{F3D: the environmental effects on the assembly of disks}
    \authorrunning{Y. Ding et al.}

    \author{%
        Y.~Ding,\inst{1,2}\thanks{Email:ycding@shao.ac.cn},
        L.~Zhu\inst{1}\thanks{Corr author: lzhu@shao.ac.cn},
        G.~van de Ven\inst{3},
        L.~Coccato\inst{4},
        E.~M.~Corsini\inst{5,6},
        L.~Costantin\inst{7,8},
        K.~Fahrion\inst{4,9},
        J.~Falc{\'o}n-Barroso\inst{10,11},
        D.~A.~Gadotti\inst{4,8},
        E.~Iodice\inst{4},
        M.~Lyubenova\inst{4},
        I.~Mart{\'i}n-Navarro\inst{10,11},
        R.~M.~McDermid\inst{12},
        F.~Pinna\inst{13},
        M.~Sarzi\inst{14,15}
%        P.~T.~de~Zeeuw\inst{7,8}
        %%%
    }
    \authorrunning{Y. Ding et al.}

    \institute{%
        Shanghai Astronomical Observatory, Chinese Academy of Sciences, 80 Nandan Road, Shanghai 200030, China\and
        School of Astronomy and Space Sciences, University of Chinese Academy of Sciences, No. 19A Yuquan Road, Beijing 100049, China\and
        Department of Astrophysics, University of Vienna, T{\"u}rkenschanzstra{\ss}e 17, 1180 Wien, Austria\and
        European Southern Observatory, Karl-Schwarzschild-Stra{\ss}e 2, D-85748 Garching bei M{\"u}nchen, Germany\and
        Dipartimento di Fisica e Astronomia `G. Galilei', Universit{\`a} di Padova, vicolo dell'Osservatorio 3, I-35122 Padova, Italy\and
        INAF - Osservatorio Astronomico di Padova, vicolo dell'Osservatorio 5, I-35122 Padova, Italy\and
        Centro de Astrobiolog\'ia (CSIC-INTA), Ctra de Ajalvir km 4, Torrej\'on de Ardoz, E-28850 Madrid, Spain\and
        INAF - Astronomical Observatory of Brera, via Brera 28, I-20121 Milano, Italy
        Centre for Extragalactic Astronomy, Department of Physics, Durham University, South Road, Durham DH1 3LE, UK\and
        %Sterrewacht Leiden, Leiden University, Postbus 9513, 2300 RA Leiden, The Netherlands\and
        %Max-Planck-Institut f{\"u}r Extraterrestrische Physik, Gie{\ss}enbachstra{\ss}e 1, 85748 Garching bei M{\"u}nchen, Germany\and
        European Space Agency (ESA), European Space Research and Technology Centre (ESTEC), Keplerlaan 1, 2201 AZ Noordwijk, The Netherlands\and
        Instituto de Astrof{\'i}sica de Canarias, Calle Via L{\'a}ctea s/n, 38200 La Laguna, Tenerife, Spain\and
        Depto. Astrof{\'i}sica, Universidad de La Laguna, Calle Astrof{\'i}sico Francisco S{\'a}nchez s/n, 38206 La Laguna, Tenerife, Spain\and
        Research Centre for Astronomy, Astrophysics, and Astrophotonics, Department of Physics and Astronomy, Macquarie University, NSW 2109, Australia\and
        Max-Planck-Institut f{\"u}r Astronomie, K{\"o}nigstuhl 17, 69117 Heidelberg, Germany\and
        Armagh Observatory and Planetarium, College Hill, Armagh BT61 9DG, UK\and
        Centre for Astrophysics Research, University of Hertfordshire, College Lane, Hatfield AL10 9AB, UK
    }

    \date{Received XXXXX; accepted XXXXX}

    \abstract{%
    We apply a population-orbit superposition method to 16 galaxies in the Fornax cluster observed with MUSE/VLT in the context of the Fornax3D project. By fitting the luminosity distribution, stellar kinematics, and age and metallicity maps simultaneously, we obtained the internal stellar orbit distribution, as well as the age and metallicity distribution of stars on different orbits for each galaxy. Based on the model, we decompose each galaxy into a dynamically cold disk (orbital circularity $\lambda_z\ge0.8$) and a dynamically hot non-disk component (orbital circularity $\lambda_z<0.8$), and obtain the surface-brightness, age, and metallicity radial profiles of each component. 
    The galaxy infall time into the cluster is strongly correlated with galaxy cold-disk age with older cold disks in ancient infallers. 
    We quantify the infall time $t_{\rm infall}$ of each galaxy with its cold-disk age using a correlation calibrated with TNG50 cosmological simulations.
    For galaxies in the Fornax cluster, we found that the luminosity fraction of cold disk in galaxies with $t_{\rm infall}>8$ Gyr are a factor of $\sim 4$ lower than in more recent infallers while controlling for total stellar mass.
Nine of the 16 galaxies have spatially extended cold disks, and most of them show positive or zero age gradients; stars in the inner disk are $\sim 2-5$ Gyr younger than that in the outer disk, in contrast to the expectation of inside-out growth. Our results indicate that the assembly of cold disks in galaxies is strongly affected by their infall into clusters, by either removal of gas in outer regions or even tidally stripping or heating part of the pre-existing disks. 
Star formation in outer disks can stop quickly after the galaxy falls into the cluster, while star formation in the inner disks can last for a few Gyrs more, building the positive age gradient measured in cold disks. 
    }

    \keywords{%
        galaxies: kinematics and dynamics --
        galaxies: elliptical and lenticular, cD --
        galaxies: stellar population --
        galaxies: formation --
        galaxies: structure --
        galaxies: evolution
    }

    \maketitle

%-----------------------------------------------
\section{Introduction}
\label{sec:intro}
%-----------------------------------------------

Galaxies in cluster environments have different star formation histories and display different morphology types with respect to galaxies in the field. Generally speaking, galaxies in clusters are redder, more likely to be quenched, and exhibit an elliptical morphology \citep[e.g.,][]{dressler1980,Butcher1984,Dressler1997,peng2010,Lewis2002,blanton2005,alpaslan2015}.
Galaxy evolution is mainly affected by three physical processes related to the cluster environment.
First, in a process known as  ``harassment'', tidal forces induced by
the central halo \citep{Gunn1972,toomre1972,Barnes1992,Bournaud2004} or neighbouring flyby galaxies can strip away part of the stellar and gas particles of the satellite galaxies \citep{Gunn1972}. 
Second, the interaction with the intracluster medium (ICM) known as ``ram pressure'' can strip away the hot gas and sometimes even the cold gas in the disk of the satellite galaxies \citep{Gunn1972,Abadi1999,Yun2019}. Third, the accretion rate is much lower in the cluster environment and it could cause the cut-off of gas supply to a galaxy leading to ``strangulation''
\citep{Larson1980,Balogh2000,Kawata2008}. All three processes could ultimately lead to the cessation of star formation and affect the morphology evolution of the satellite galaxies in the cluster. 

The timescale of the star-formation quenching in cluster environments and the relative contribution of the aforementioned physical processes are still under debate. The current gas content in cluster galaxies provides some direct clues. \cite{Maier2019} calculated the fraction of star-forming galaxies of seven nearby clusters through H$\alpha$ observation with Local Cluster Substructure Survey (LoCuSS). By comparing with the Millennium simulations, they suggested that star formation in cluster galaxies can last for 1-2 Gyr after their infall into the cluster, but then they quickly get quenched. This ``slow-then-rapid quenching'' is consistent with the strangulation scenario.
On the other hand, \citet{Reynolds2021} found evidence that HI gas in the outer regions of the galaxies in the Hydra I cluster is partially removed. However, these galaxies still lie on the star-forming main sequence and gas removal is not yet affecting the inner star forming disks. In the same cluster, \citet{Wang2021} modeled the ram-pressure stripping strength through the HI mass from WALLABY observation. They found that the ram pressure stripping can significant change the total HI gas mass of satellite galaxies within 600 Myr after falling into the cluster. 

 Cosmological hydrodynamical simulations allow us to investigate the details of the infalling of galaxies into the cluster as well as of the gas removal and structure formation of galaxies. \cite{Joshi2020} found that the pre-exisiting stellar disks in cluster galaxies are largely disrupted by impulsive tidal shocking and stripping at pericentres within 0.5-4 Gyr after falling into the massive clusters with $M_{200} \sim 10^{14-14.3}$ M$_{\odot}$ in the Illustris TNG100 simulation. Meanwhile, all the cluster galaxies quenched their star formation after the disk disruption. The quenching timescale could be 1-1.5 Gyr for gas-poor massive galaxies, while the star formation could last for $\sim$4.5 Gyr for gas-rich galaxies and a dynamically cold disk could regrow. Similar results apply also to Fornax-mass analogues in the TNG50 simulations \citep{GalandeAnta2022} 

 The detailed study of S0 galaxies in TNG100 simulation shows two major formation scenarios \citep{Deeley2021}: S0 galaxies in clusters formed via starvation or stripping, whereas S0s in field formed via mergers. This finding is also supported by observations \citep{coccato2020,coccato2022}. Star formation quenching is different in these two formation scenarios \citep{Deeley2021}: when a blue and gas-rich galaxy falls into a cluster, its gas in the outer regions is quickly stripped, while the star formation in the inner regions could last for a long time, whereas in case of galaxy-galaxy merger, galaxies start quenching from the inner regions, and star formation continues in the outer ring-like regions.
 
These studies suggest that cluster environment may significantly affect the internal dynamical structure of galaxies, especially the persistent growth of dynamically cold disks. The fraction of stars in cold disks and their age distributions may imprint the long-term star-formation quenching process happening in cluster galaxies. 
We expect that the disk stars are older in the inner regions and younger in the outer ones for a normal inside-out growth \citep[e.g.,][]{bird2013, Stinson2013}.
Cluster quenching may turn over the age gradient in the stellar disks, the age difference between inner and outer regions can thus tell us the star formation duration in galaxies after falling into the cluster. Although some other studies show that complicated physical processes in field galaxies might cause similar age gradient \citep{Lagos2022}.

In the past decades, integral-field unit (IFU) surveys, such as SAURON \citep{deZeeuw2002}, ATLAS$^{\text{3D}}$ \citep{cappellari2011a}, CALIFA \citep{sanchez2012}, SAMI \citep{Bryant2015}, and MaNGA \citep{Bundy2015}, have provided us the stellar kinematics, age, and metallicity maps for thousands of nearby galaxies. Early-type galaxies (ETGs) show a large variety of kinematical structures \citep{cappellari2007,emsellem2007,Emsellem2011},
some lenticular galaxies have a rapidly rotating disk similar to that of spiral galaxies. This suggests that gas removal by cluster environment may play a major role in transforming spirals into lenticulars \citep{Cappellari2016}.
Overall, ETGs are found to be old and with shallow age gradients \citep{mcDermid2015,Gonzalez2015, Zibetti2019}, while late-type galaxies (LTGs) generally have negative age gradients, which is consistent with the inside-out scenario \citep{Gonzalez2015}.
The environmental dependence of the galaxy age gradient is still controversial. A positive age gradient was found in SAURON and MaNGA samples \citep[e.g.,][]{Goddard2017, Li2018, Kuntschner2010}, with a mixture of galaxies in cluster and field environments.
On the other hand, there is no significant difference found for age gradient of ETGs in different environments \citep{Zheng2017, Ferreras2019, Santucci2020}, with the above surveys covering the inner $1R_\mathrm{e} \sim 2R_\mathrm{e}$ for most galaxies.

The Fornax3D survey \citep{sarzi2018} measured high-quality age and metallicity maps covering at least the inner $2R_\mathrm{e}$ of 23 ETGs in the Fornax cluster. These galaxies show significant positive age gradients between $R_\mathrm{e}$ and $2R_\mathrm{e}$ (Spavone et al 2022), while they have negative metallicity gradients, which are shallower than the control sample galaxies in the field. On the other hand, positive age gradients together with negative metallicity gradients are widely found in the dwarf galaxies of the Local Group \citep[e.g.,][]{Koleva2011,Kirby2011,Zhuang2021}. These positive age gradients could be the direct consequence of ram pressure stripping \citep{Genina2019, Deeley2021}, although others argue theses are the result of gas-rich mergers for dwarf galaxies \citep{Yozin2012}. 

The observation of external galaxies provides integrated properties along the line-of-sight. The stellar kinematics and populations of a galaxy are a combination of different dynamical components, displaying a range of possible physical origins. 
As learned from the simulations aforementioned, the luminosity fraction and age distribution of stellar disk might be a good probe of environmental affects. However, there are relatively small disk fractions in ETGs \citep{Zhu2018c}. The age and metallicity maps of the whole galaxy may not reveal the properties of the disk component.

Dynamical models are powerful tools to uncover the galaxies' underlying mass profiles, as well as the internal 3D structures which could lead to physical-motivated decomposition of different components. There are a few methods widely used for modelling the IFU kinematic data of external galaxies, including the Jeans-Anisotripic-MGE (JAM) models \citep[e.g.,][]{Cappellari2008,Li2017}
%, the action-angle distribution function method \citep[e.g.,][]{Binney2010}
, the particle-based Made-to-Measure (M2M) models \citep[e.g.,][]{Syer1996,deLorenzi2007,Long2010,Long2012,Zhu2014}, and the orbit-superposition Schwarzschild method used in this work \citep{schwarzschild1979}. 
The Schwarzschild's orbit superposition method can be used in different geometries such as the spherical systems \citep{Richstone1985, Breddels2013, Kowalczyk2017}, axisymmetric systems \citep{cretton1999,Gebhardt2000,valluri2004,Cappellari2006,Thomas2007,Saglia2016}, and triaxial systems \citep{vandenbosch2008, Neureiter2021}.
The orbit-superposition model reconstructs the backbone of galaxies without ad-hoc assumptions of the underlying distribution functions, it has been widely used to uncover galaxies' underlying dark matter distributions \citep[e.g.,][]{cappellari2006b,Yang2020}
, central black hole mass \citep[e.g.,][]{vandermarel1998,cretton1999,verolme2002,gebhardt2003,valluri2004,krajnovic2009,Ahn2018,Thater2022b}, and internal stellar orbit distributions \citep[e.g.,][]{zhu2018a, zhu2018, jin2020}.
Recently it has been modified to include the barred structures explicitly, thus also for uncovering the bar pattern speed \citep{vasiliev2020,Tahmasebzadeh2022}.

Based on the Schwarzschild method, a population-orbit superposition method \citep{poci2019,zhu2020} was recently developed by tagging age and metallicity to the orbits, we can thus simultaneously fit the stellar kinematic, age, and metallicity maps from IFU survey, and obtain the internal stellar orbit distribution of galaxies as well as the age and metallicity distributions. This allows for a physically motivated chemo-dynamical decomposition of galaxies.
In this work, we apply the population-orbit superposition method to galaxies in Fornax cluster with the IFU data obtained with MUSE/VLT in the context of the Fornax3D survey \citep{iodice2019}. Using this modeling technique, age-dispersion profiles of dynamically decomposed cold disks in a few edge-on galaxies have been obtained \citep{poci2019,Poci2021} and a dynamically decomposed ``hot inner stellar halo'' has been used to weigh and time the ancient massive mergers the galaxies experienced in NGC~1380 and NGC~1427 \citep{zhu2022b}.

In this paper, we set our focus on dynamically cold disks. By studying the luminosity fraction, age, and metallicity of these stars, we aim to quantify how the cluster environment has affected the formation of cold disks. This results will also lead to a direct and in-depth comparison with galaxies in cosmological simulations and, thus, to a better understanding of the physical processes driving galaxy evolution in cluster environments. This work is undertaken in the context of $\Lambda$CDM cosmology, with density parameters of $\Omega_m=0.3089$, $\Omega_{\Lambda}=0.6911$, $\Omega_b=0.0486$, normalization $\sigma_8=0.8159$ and spectral index $n_s=0.9667$ \citep{Collaboration2014}.

The paper is organized as follows. We introduce the data set in \cref{sec:data}. We describe the population-orbit superposition model and its relevance to the orbital decomposition in \cref{sec:methods}. 
We show the orbital decomposition for all galaxies and present the surface-brightness, age, and metallicity radial profiles of each component in \cref{sec:Sb18}. 
We present an explanation of how the cold disk age can be used as a novel proxy for galaxy infall time into a cluster in \cref{sec:infall_time}. We further show the dependence of cold disk properties on galaxy stellar mass and cluster environment in \cref{sec:mass_dependence}. We discuss the results in \cref{sec:discussion} and present our summary in \cref{sec:conclusion}.

%-----------------------------------------------
%\section{Deep photometric and spectroscopic data}\label{sec:data}
\section{Sample and Data}\label{sec:data}
%-----------------------------------------------

We set our focus on galaxies in the Fornax cluster, which has a virial radius of $R_{\mathrm{vir}} \sim 0.7$ Mpc , virial mass of $M_{\mathrm{vir}} \sim 7\times10^{13}$ M$_{\odot}$, and a distance of $D \sim 20$ Mpc \citep{Diaferio1999,drinkwater2001}.

We used the deep photometric data of the Fornax Deep Survey (FDS, \citet{venhola2018}). FDS observed all the galaxies in the area of 9 square degrees around the core of the cluster, with the VLT Survey Telescope down to a surface-brightness level of 27 \magarcsq in the $r$ band \citep{iodice2016}. 

We used the spectroscopic data of the Fornax3D survey \citep{sarzi2018}. Using the MUSE instrument on the VLT, Fornax3D observed all the 23 ETGs and 10 LTGs within the virial radius of the cluster down to 25 \magarcsq in $B$ band. 
MUSE has a field-of-view of $1\times 1$ $\mathrm{arcmin}^2$ and spatial scale of 0.2 arcsec $\mathrm{pixel}^{-1}$. The wavelength covers the range of 4650 $\AA$ to 9300 $\AA$, with a resolution of FWHM$_{7000\AA}=2.5\ \AA$. Twenty-eight galaxies were observed by more than one MUSE pointing. The data extend to at least $2R_\mathrm{e}$ for 20 galaxies.

\begin{table*}\label{tab:galaxy_prop}
   \caption{Properties of the sample galaxies.}
\renewcommand\arraystretch{1.5}
   \centering
   \begin{tabular}{cccccccccccccc}
   \hline\hline
   Object & Type & Distance & $R_{\mathrm{e}}$ & $M_\star$ & $i$ & PA & S/N\\
       & &[Mpc] &[kpc] & [10$^{10}$M$_{\odot}$] & [$^{\circ}$] & [$^{\circ}$] & &\\
     (1)  &  (2) & (3) & (4) & (5) & (6) & (7) & (8)\\
   \hline
   FCC\,083 & E5  & 19.2 & 3.46 & 2.27 & 68.2$\pm$13.3 & 139.7 & 100\\
   FCC\,119 & S0  & 20.9 & 1.17  & 0.05 & 64.9$\pm$5.4 & 45.4 & 40\\ 
   FCC\,143 & E3  & 19.3 & 1.07 & 0.28 & 64.6$\pm$7.1 & 124.6 & 60\\
   FCC\,147 & E0  & 19.6 & 2.40 & 2.40 & 33.3$\pm$6.0 & 120.9 & 100\\
   FCC\,148 & S0  & 19.9 & 2.74  & 0.58 & 75.8$\pm$3.1 & 89.2 & 60\\
   FCC\,153 & S0  & 20.8 & 1.92 & 0.76 & 87.5$\pm$0.1 & 82.6 & 100\\
   FCC\,161 & E0  & 19.9 & 2.77 & 2.63 & 33.1$\pm$3.3 & 0.0 & 100\\
   FCC\,167 & S0/a & 21.2 & 5.47 & 9.85 & 78.0$\pm$2.1 & 5.0 & 100\\
   FCC\,170 & S0 & 21.9 & 1.54 & 2.25 & 84.9$\pm$1.5 & -42.0 & 100\\
   FCC\,177 & S0 & 20.0 & 3.48 & 0.85 & 85.8$\pm$1.3 & 177.8 & 100\\
   FCC\,179 & Sa  & 20.9 & 2.91 & 1.58 & 69.9$\pm$0.7 & 64.34 & 100\\
   FCC\,182 & SB0 & 19.6 & 0.96 & 0.15 & 63.2$\pm$4.4 & 169.3 & 60\\
%   FCC\,184 & SB0 & 19.3 & 3.44 & 4.70 & 52.1±\pm0.5 & 60.0\\
   FCC\,249 & E0  & 20.9 & 0.93 & 0.98 & 37.8$\pm$4.1 & 150.6 & 60\\
   FCC\,255 & S0  & 20.9 & 1.34 & 0.31 & 78.1$\pm$5.8 & 172.9 & 60\\
   FCC\,263 & SBcdIII  & 20.9 & 2.64 & 0.04 & 66.4$\pm$1.9 & 87.9 & 60\\ 
   FCC\,276 & E4  & 19.6 & 4.33 & 1.81 & 66.6$\pm$9.0 & 75.4 & 100\\
   FCC\,290 & ScII  & 20.9 & 4.70 & 0.64 & 52.8$\pm$1.1 & 122.5 & 60\\ 
   FCC\,301 & E4  & 19.7 & 1.13 & 0.20 & 82.0$\pm$5.7 & 155.8 & 60\\
   FCC\,308 & Sd  & 20.9 & 3.60 & 0.04 & 73.5$\pm$1.4 & 88.5 & 60\\ 
%   FCC\,310 & SB0 & 20.9 & 3.45 & 0.54 & 82.8±\pm2.7 & 67.6\\
   FCC\,312 & Scd  & 20.9  & 10.62 & 1.48 & 81.0$\pm$0.9 & 136.2 & 100\\ 
   \hline
   \end{tabular}
   \tablefoot{Name (1); Hubble type (2) taken from \citep{ferguson1989}; distance (3); Effective radius (4); and stellar mass (5) is from \citet{iodice2019b} and \citet{Martin-Navarro2021}. The inclination angle (6) obtained from our orbit-superposition model. The positional angle (7) determined by photometric isophotes and adopted for the multi-Gaussian expansion fitting and target S/N (8) used in binning the spectra to obtain kinematic maps.}
   \end{table*}

 \begin{sidewaystable*}\label{tab:galaxy_model_prop}
   \caption{Properties of the sample galaxies extracted by the models.}
\renewcommand\arraystretch{1.5}
   \centering
   \begin{tabular}{m{3.0em}m{4.0em}m{4.0em}m{4.0em}m{4.0em}m{3.5em}m{3.5em}m{3.5em}m{3.5em}m{4.5em}m{4.5em}m{4.5em}m{4.5em}m{3.9em}}
   \hline\hline
   Object & $M_{\mathrm{tot,e}}$ & $f_{\mathrm{DM,e}}$ & $f_\mathrm{cold,e}$ & $f_\mathrm{cold,2e}$ & $\langle t_{\mathrm{cold,e}}\rangle$ & $\langle t_{\mathrm{hot,e}}\rangle$ & $\langle Z_{\mathrm{cold,e}}\rangle$ & $\langle Z_{\mathrm{hot,e}}\rangle$ &$\nabla t_\mathrm{{cold,e}}$ &$\nabla t_\mathrm{{hot,e}}$ & $\nabla Z_\mathrm{{cold,e}}$ & $\nabla Z_\mathrm{{hot,e}}$ & $t_{\mathrm{infall}}$ \\
       & [10$^{10}$M$_{\odot}$] &  &  &  & [Gyr] & [Gyr] & [$Z_{\odot}$] & [$Z_{\odot}$] & [Gyr/R$_\mathrm{e}$] & [Gyr/R$_\mathrm{e}$] & $\mathrm{Z_{\odot}/R_e}$ & $\mathrm{Z_{\odot}/R_e}$ & [Gyr ago]\\
     (1)  &  (2) & (3) & (4) & (5) & (6) & (7) & (8) & (9) & (10) & (11) & (12) & (13) & (14)\\
   \hline
   FCC\,083 & 2.67$\pm$0.19 & 0.40$\pm$0.05 & 0.18$\pm$0.02 & 0.19$\pm$0.02 & 11.8$\pm$0.6 & 11.4$\pm$0.5 & 0.72$\pm$0.24 & 0.38$\pm$0.20 & -2.52$\pm$1.73 & -2.92$\pm$0.48 & -0.99$\pm$0.67 & -0.91$\pm$0.15 & 8.7$\pm$3.1\\

   FCC\,119 & 0.04$\pm$0.02 & 0.66$\pm$0.05 & 0.17$\pm$0.06 & 0.12$\pm$0.03 & 5.9$\pm$1.1 & 4.2$\pm$0.5 & 0.36$\pm$0.05 & 0.43$\pm$0.05 & 18.24$\pm$5.75 & 10.02$\pm$1.14 & -1.22$\pm$0.28 & -0.7$\pm$0.07 &  2.4$\pm$2.3\\ 
   FCC\,143 & 0.15$\pm$0.03 & 0.24$\pm$0.06 & 0.06$\pm$0.01 & 0.05$\pm$0.01 & 11.9$\pm$1.3 & 13.0$\pm$0.5 & 0.98$\pm$0.71 & 0.36$\pm$0.19 & ``` & -2.05$\pm$0.44 & ``` & -0.49$\pm$0.25 & 11.4$\pm$1.2\\
   FCC\,147 & 2.88$\pm$0.30 & 0.19$\pm$0.03 & 0.11$\pm$0.02 & 0.12$\pm$0.02 & 12.7$\pm$0.8 & 13.5$\pm$0.3 & 0.91$\pm$0.33 & 0.46$\pm$0.13 & -2.07$\pm$1.87 & -1.01$\pm$0.26 & -1.19$\pm$0.84 & -1.32$\pm$0.04 & 9.6$\pm$3.0\\
   FCC\,148 & 0.56$\pm$0.03 & 0.47$\pm$0.03 & 0.27$\pm$0.02 & 0.33$\pm$0.02 & 4.9$\pm$0.6 & 5.7$\pm$0.6 & 0.90$\pm$0.11 & 0.95$\pm$0.18 & 0.97$\pm$2.05 & 4.13$\pm$0.87 & -1.37$\pm$0.6 & -1.47$\pm$0.1 & 3.8$\pm$2.4\\
   FCC\,153 & 1.03$\pm$0.14 & 0.43$\pm$0.01 & 0.32$\pm$0.01 & 0.49$\pm$0.01 & 9.4$\pm$0.2 & 9.9$\pm$0.6 & 1.52$\pm$0.09 & 0.51$\pm$0.19 & 3.69$\pm$2.11 & 2.85$\pm$0.6 & -0.81$\pm$0.50 & -1.24$\pm$0.34 & 8.9$\pm$1.7\\
   FCC\,161 & 1.33$\pm$0.07 & 0.25$\pm$0.04 & 0.10$\pm$0.02 & 0.09$\pm$0.02 & 11.6$\pm$1.1 & 12.9$\pm$0.4 & 0.49$\pm$0.14 & 0.47$\pm$0.04 & ``` & -1.41$\pm$0.23 & ``` & -0.37$\pm$0.03 & 9.0$\pm$3.0\\
   FCC\,167 & 12.16$\pm$0.87 & 0.25$\pm$0.03 & 0.12$\pm$0.01 & 0.20$\pm$0.01 & 11.6$\pm$0.1 & 13.6$\pm$0.1 & 1.95$\pm$0.01 & 1.61$\pm$0.01 & -4.08$\pm$1.88 & -0.32$\pm$0.28 & -0.3$\pm$0.83 & -0.73$\pm$0.13 & 8.8$\pm$3.1\\
   FCC\,170 & 1.31$\pm$0.3 & 0.14$\pm$0.02 & 0.03$\pm$0.01 & 0.15$\pm$0.01 & 13.2$\pm$0.6 & 13.2$\pm$0.1 & 1.05$\pm$0.44 & 0.47$\pm$0.03 & -0.1$\pm$0.20 & -0.5$\pm$0.08 & -0.02$\pm$0.30 & -0.48$\pm$0.08 & 9.8$\pm$3.0\\
   FCC\,177 & 0.74$\pm$0.05 & 0.62$\pm$0.03 & 0.32$\pm$0.01 & 0.30$\pm$0.01 & 6.7$\pm$0.4 & 8.1$\pm$0.8 & 1.25$\pm$0.20 & 0.81$\pm$0.23 & -0.12$\pm$2.80 & 5.71$\pm$0.40 & 0.65$\pm$0.85 & -1.30$\pm$0.13 & 6.3$\pm$2.2\\
   FCC\,179 & 2.2$\pm$0.09 & 0.19$\pm$0.02 & 0.31$\pm$0.02 & 0.47$\pm$0.03 & 5.0$\pm$0.5 & 7.2$\pm$0.7 & 1.12$\pm$0.15 & 0.77$\pm$0.05 & 1.59$\pm$1.59 & -2.45$\pm$1.03 & -1.11$\pm$0.67 & -0.22$\pm$0.10 & 3.7$\pm$2.4\\
   FCC\,182 & 0.05$\pm$0.01 & 0.34$\pm$0.08 & 0.05$\pm$0.01 & 0.03$\pm$0.01 & 11.6$\pm$1.3 & 10.3$\pm$0.8 & 0.81$\pm$0.46 & 0.44$\pm$0.09 & ``` & -0.84$\pm$0.24 & ``` & -0.3$\pm$0.05 & 10.4$\pm$0.9\\
   %FCC\,184 & 4.43±\pm0.68 & 0.23±\pm0.02 & 0.05±\pm0.01 & 13.6±\pm0.4 & 12.8±\pm0.3 & 2.08±\pm0.64 & 1.22±\pm0.12 & 9.9±\pm3.0\\
   FCC\,249 & 0.60$\pm$0.02 & 0.17$\pm$0.03 & 0.07$\pm$0.03 & 0.05$\pm$0.02 & 13.3$\pm$0.7 & 12.0$\pm$0.5 & 1.26$\pm$1.18 & 0.37$\pm$0.13 & ``` & -1.97$\pm$0.25 & ``` & -0.41$\pm$0.21 & 12.3$\pm$1.7\\
   FCC\,255 & 0.11$\pm$0.01 & 0.44$\pm$0.03 & 0.12$\pm$0.02 & 0.16$\pm$0.02 & 3.7$\pm$2.5 & 8.2$\pm$1.5 & 0.98$\pm$0.27 & 0.37$\pm$0.05 & 6.42$\pm$3.51 & 3.5$\pm$0.93 & -0.64$\pm$0.32 & -0.45$\pm$0.06 & 1.0$\pm$1.0\\
   FCC\,263 & 0.21$\pm$0.02 & 0.88$\pm$0.01 & 0.08$\pm$0.01 & 0.10$\pm$0.02 & ``` & ``` & ``` & ``` & ``` & ``` & ``` & ``` & ``` \\ 
   FCC\,276 & 4.23$\pm$0.46 & 0.33$\pm$0.04 & 0.04$\pm$0.01 & 0.03$\pm$0.01 & 13.1$\pm$0.7 & 12.7$\pm$0.3 & 0.79$\pm$0.52 & 0.35$\pm$0.05 & ``` & -2.01$\pm$0.37 & ``` & 0.74$\pm$0.06 & 9.4$\pm$3.0\\
   FCC\,290 & 2.08$\pm$0.13 & 0.68$\pm$0.05 & 0.25$\pm$0.03 & 0.37$\pm$0.04 & ``` & ``` & ``` & ``` & ``` & ``` & ``` & ``` & ``` \\ 
   FCC\,301 & 0.15$\pm$0.01 & 0.28$\pm$0.05 & 0.07$\pm$0.01 & 0.11$\pm$0.01 & 6.4$\pm$0.9 & 7.8$\pm$0.8 & 0.87$\pm$0.01 & 0.38$\pm$0.05 & 1.66$\pm$3.80 & 0.3$\pm$0.65 & -1.16$\pm$0.36 & 0.35$\pm$0.04 & 6.5$\pm$2.1\\
   FCC\,308 & 0.33$\pm$0.03 & 0.51$\pm$0.03 & 0.09$\pm$0.01 & 0.09$\pm$0.01 & ``` & ``` & ``` & ``` & ``` & ``` & ``` & ``` & ``` \\ 
   %FCC\,310 & 0.59±\pm0.08 & 0.51±\pm0.03 & 0.06±\pm0.02 & 10.0±\pm1.0 & 10.8±\pm0.2 & 1.07±\pm0.30 & 0.29±\pm0.02 & 9.8±\pm1.3\\
   FCC\,312 & 6.16$\pm$0.72 & 0.63$\pm$0.06 & 0.32$\pm$0.04 & 0.32$\pm$0.05 & ``` & ``` & ``` & ``` & ``` & ``` & ``` & ``` & ``` \\ 
   \hline
   \end{tabular}
   \tablefoot{Name (1); total dynamical mass (2); dark matter fraction (3); $1R_{\mathrm{e}}$ luminosity fraction of the cold disk (4); $2R_{\mathrm{e}}$ luminosity fraction of the cold disk (5); mean age of the cold disk (6); mean age of the non-disk component (7); mean metallicity of the cold disk (8); mean metallicity of the non-disk component (9); age gradient of the cold disk (10); age gradient of the non-disk component (11); metallicity gradient of the cold disk (12); metallicity gradient of non-disk component (13). All values in columns (2)-(4) and (6)-(13) are calculated within the effective radius $R_{\mathrm{e}}$, except for (5) which is extracted within $2R_{\mathrm{e}}$ . Columns (10) and (12) of FCC~170 and FCC~255 are extracted within the whole data coverage as their disks have very low surface-brightness at $r<R_{\rm e}$. For five ETGs FCC~143, FCC~161, FCC~182, FCC~249, and FCC~276, we only show the age and metallicity gradient of non-disk component because of the low surface-brightness of cold disk at $r<R_{\rm e}$. Column (13) gives the infall time into the cluster based on the cold-disk age using a correlation calibrated with TNG50 simulations. For four LTGs FCC~263, FCC~290, FCC~308, and FCC~312, we only have the stellar kinematics and thus only orbital models without coloring.}
   \end{sidewaystable*}
   
%-----------------------------------------------
\subsection{Stellar kinematics maps}
%-----------------------------------------------
To obtain reliable stellar kinematics, we first bin the spectra in nearby pixels to reach a target signal-to-noise ratio (S/N) taken a Voronoi binning scheme \citep{cappellari2003}. The stellar kinematics of all the galaxies in the Fornax3D survey was published in \citet{iodice2019} with target S/N=40. The only difference in the kinematic maps we use here is that different S/N for different galaxies were chosen as indicated in Table~1.

We took a target S/N = 100 or 200 for the bright galaxies with more than one MUSE pointing and S/N =60 or 40 for the remaining galaxies. This results in a number of bins from $\sim 100$ to $\sim 1000$ for each galaxy with a spatial resolution from $\sim 100$ pc in the inner region to $\sim 1$ kpc in the outer faint regions. This choice is convenient for dynamical modeling without too many bins, while the spatial resolution is still good enough to meet our science goals for resolving sub-kpc scale structures.

The stellar kinematics was then extracted by applying the pPXF full-spectral fitting \citep{cappellari2004} to the wavelength range 4750-5500 \AA. This yielded high-quality maps of the stellar mean velocity, $V$, velocity dispersion, $\sigma$, and higher order velocity moments parameterized through the Gauss-Hermite (GH) coefficients $h_3$ and $h_4$ \citep{Gerhard1993, Rix1997}.

Such stellar kinematic maps were derived in a consistent way for the 15 ETGs and 5 LTGs included in this paper.
We did not consider the 8 ETGs -- FCC~213 (i.e., the BCG), FCC~090, FCC~184, FCC~190, FCC~193, FCC~219, FCC~277, and FCC~310 -- due to existence of bright foreground stars, limited data coverage or the presence of strong bar, and the 5 LTGs FCC~113, FCC~176, FCC~267, FCC~285, and FCC~306 due to the presence of a strong bar or irregular morphology. We excluded the strongly barred galaxies because our current dynamical models cannot fit their kinematic maps well with bars that are not included explicitly. However, we did keep the three weakly barred galaxies (FCC~179, FCC~182, and FCC~263), since their stellar kinematic maps are not strongly affected by the bar (see figures in Appendix C) and our current kinematic models can still provide reasonably good fits.
Some basic properties of the final sample of galaxies are listed in Table~1.

\subsection{Age and metallicity maps}
We used the age and metallicity maps derived in \citet{Martin-Navarro2021}.  Here, we briefly describe how they were derived.

To obtain high-quality maps of the age and metallicity, the spectra were spatially rebinned to reach a minimum of $\mathrm{S/N}=100$ for each galaxy. We note that the spaxels with S/N$<5$ were not included in this process in order to retain the highest quality of data. Thus, for most of the galaxies, only the inner region map was available.
The single stellar population synthesis models (SSP) by \citet{vazdekis2016} based on the MILES stellar library by \citet{Sanchez-Blazquez2006} at a constant spectral resolution of 2.51 $\AA$ (FWHM) 
\citep{falcon-barroso2011} were used for the analysis.

The luminosity-weighted age and metallicity maps were derived in a two-step spectral fitting of the wavelength range between 4800 $\AA$ and 6400 $\AA$, considering the initial mass function (IMF) variation. First pPXF was run with an un-regularized combination of MILES SSPs in order to measure $V$ and $\sigma$. A second pPXF was run again with stellar kinematics fixed to that measured in the first step, and this time regularizing the age-metallicity-IMF slope plane.
The mean luminosity-weighted age and metallicity of the spectra were measured using this regularized second pPXF run.  The metallicity, $Z$, is added up linearly rather than in the commonly used logarithmic way. 
Templates with a variable IMF were intentionally included to account for the possible effect of the IMF in the age determination. The abundance ratio [$\alpha$/Fe] was not fitted, but we used the so-called base MILES models which inherit the [$\alpha$/Fe]-[M/H] relation of the solar neighborhood \citep{Vazdekis2015}. This choice avoids non-local equilibrium uncertainties introduced by the theoretical response functions needed to compute stellar population models with variable abundance ratios, uncertainties that can be particularly problematic for Balmer lines and thus for the age determination. We note that in \cite{Martin-Navarro2021}, the metallicity maps were also derived in a third step with index fitting when determining the IMF.
However, we used the luminosity-weighted metallicity maps derived from the pPXF fitting, in a consistent way as the luminosity-weighted age maps. 

Alternatively, we have another version of the age and metallicity maps for FCC~153, FCC~167, FCC~170, and FCC~177, which was obtained by pPXF fitting regularizing age-metallicity-[$\alpha$/Fe] assuming a Kroupa IMF \citep{Pinna2019a,Pinna2019}.

We compared the two versions of age and metallicity maps for these four galaxies, they generally show the same age and metallicity gradients but with a systematic offset of 1-2 Gyr in age. This offset will not affect our results.
The version of the age and metallicity maps by \cite{Martin-Navarro2021} is used consistently for all galaxies in this paper. 

Finally, we have the age and metallicity maps for 16 galaxies, including all the 15 ETGs and the LTG FCC~179 only. The remaining 4 LTGs of our sample appear to be dominated by young stars at all radii, for which the uncertainties of age and metallicity derived from the spectra fitting are large \citep{Ge2018},
we will build their dynamical models without associating age and metallicity to the stellar orbits.

%-----------------------------------------------
\section{Population-orbital superposition method}\label{sec:methods}
%-----------------------------------------------
\begin{figure*}
    \centerline{
        \includegraphics[width=2.2\columnwidth, clip=true, trim=35 0 60 0]{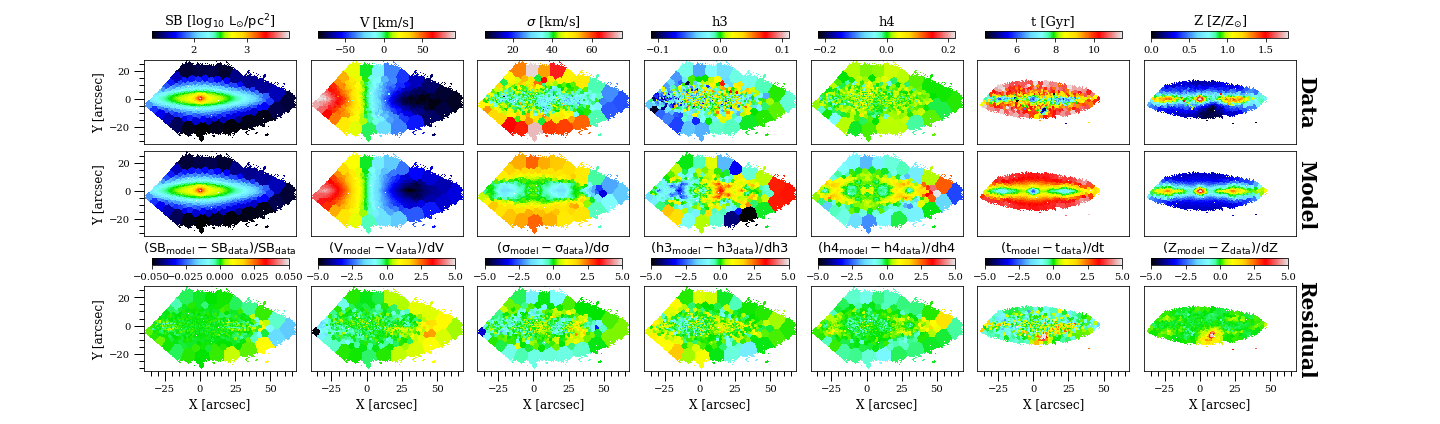}}
        \caption{
        Best-fit population-orbit superposition model of
        FCC\,177. From top to bottom: Maps of the data, model, and
        residuals (model$-$data). From left to right: Maps of the
        surface-brightness, SB, mean velocity, $V$, velocity dispersion,
        $\sigma$, Gauss-Hermite coefficients $h_3$ and $h_4$, light-weighted
        age and metallicity of the stars. Similar plots for the other sample
        galaxies are shown in Figs.~\ref{img:fitting083}-\ref{img:fitting312}.} 
        \label{img:fitting177}
\end{figure*}

\begin{figure*}
    \centerline{
        \includegraphics[width=1.6\columnwidth, clip=true, trim=20 0 20 0]{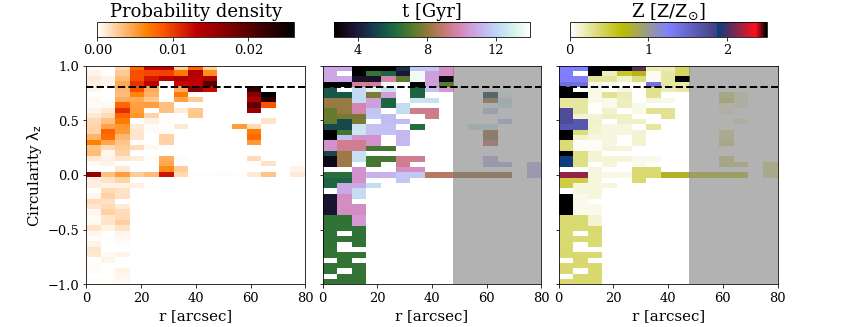}}
    \caption{
    Orbital decomposition of FCC\,177. Probability density
    distribution $p(r, \lambda_z)$ in the left panel, age distribution
    $p(r, t)$ in the central panel, and metallicity distribution $p(r,
    Z)$ in the right panel of the stellar orbits in the phase space of
    time-averaged radius, $r,$ versus circularity, $\lambda_z$. The
    probability densities are normalized to unity within the data
    coverage. All the distributions are averages of multiple best-fit
    models that fall within the $1\sigma$ confidence level of the model
    hyper-parameters. The dashed line marks our orbit-based division into
    two components: a dynamically cold disk component ($\lambda_z \geq
    0.8$) and a dynamically hot non-disk component ($\lambda_z <
    0.8$). 
    The shadow regions are beyond the data coverage of age and metallicity maps.
    Similar plots for the other galaxies are shown in
    Figs.~\ref{img:fitting083}-\ref{img:fitting312}.}
    \label{phasespace177}
\end{figure*}

We follow \citet{zhu2020}\footnote{We have fixed the bug in the orbit mirroring as reported in \citet{Quenneville2022}
although our results are unaffected as confirmed in \citet{Thater2022}.} in constructing population-orbit superposition models that simultaneously fit the luminosity density, kinematic, age, and metallicity maps of each galaxy. From the best-fit models, we obtain the internal stellar orbit distribution associated with the age and metallicity distributions.
Based on the model, we then decomposed each galaxy into a dynamically cold disk and a dynamically hot non-disk component, based on circularity distribution of the orbits, and extracted the face-on surface-brightness, age, and metallicity radial profiles for each component separately. We used FCC~177 as an example to illustrate the whole process as shown in Figs.~\ref{img:fitting177} and \ref{phasespace177}.

The construction of a population-orbit superposition model consists of four steps. First, constructing the model of gravitational potential. Second, calculating the orbital library under the gravitational potential. Third, fitting the luminosity density and kinematic maps by weighting the orbits. Fourth, fitting the age and metallicity maps by ``coloring'' the orbits. The method has been presented in great detail and carefully validated in \citet{zhu2020}. Here, we just briefly describe some of the key steps.

%-----------------------------------------------
\subsection{Constructing the gravitational potential}\label{schwarz_model}
%-----------------------------------------------
The gravitational potential combines the contribution of the stars, dark matter halo, and a central black hole. 

To construct the stellar mass distribution, we use the multi-Gaussian expansion (MGE, \citet{cappellari2002}) to fit the $r$-band image from FDS.
The galaxies in our sample have a variety of morphological and kinematic properties, which can be classified in three categories: galaxies which are generally flat and with strong rotation indicating an extended disk, galaxies which are generally round 
but flatter, and with stronger rotation in the inner regions indicating an inner disk, galaxies with no clear evidence of a disk. 
For the first two categories, 
we consider an axisymmetric solution by adopting a constant positional angle (PA) for the Gaussians to match that of the photometric and kinematic signature of the inner or outer disk component. Photometric PA from the disk region is taken, which is generally consistent with the kinematic PA in that region. For the third category, we still consider axisymmetric solution, but we did not put any extra constrain on the PA; thus, the photometric PA directly derived from the whole galaxy image was taken. The best-fit Gaussian parameters are listed in the Appendix for all the galaxies.

Then we de-project the 2D Gaussian components to 3D by assuming a set of viewing angles ($\theta$, $\psi$, $\phi$), where $\theta$ and $\phi$ define the orientation of the line of sight with respect to the principal axes of the galaxy and $\psi$ is chosen to specify the rotation of the galaxy around the line-of-sight in the projected sky-plane. By combining the 3D Gaussian components, we obtained the 3D luminosity density distribution of the galaxy and we multiply a stellar mass-to-light ratio $M_{*}/L$ to get the 3D mass density distribution. The gravitational potential could then be calculated by the classical Chandrasekhar formula \citep{vandenbosch2008}. 

In practice, we do not directly use the three viewing angles as free parameters. In \cite{vandenbosch2008}, three intrinsic shape parameters $(p,q,u)$ are used instead of the viewing angles ($\theta$, $\psi$, $\phi$), where $q=Z/X$, $p=Y/X$, and $u=X'/X$, where $X$, $Y$, $Z$ are the intrinsic long, intermediate, and short axis of the galaxy and $X'$ is the projected major axis. The conversion of the two sets of parameters follows Eq. 10 in \cite{vandenbosch2008}. Exploring the space of intrinsic shape parameters is more efficient than that of the three viewing angles. For instance, the deprojection of an axisymmetric system would have no constrain on the parameter $\phi$ but have a finite axis-ratio between $Y$ and $X$. In this work, we follow the approach by \cite{vandenbosch2008} and allow for some degree of triaxiality of the galaxy by setting a non-unity $u$.

We use a spherical Navarro-Frenk-White (NFW, \citet{navarro1996})
dark matter halo, with one free parameter dark matter virial mass $M_{200}$, while concentration $C$ is fixed by the correlation between $M_{\rm 200}$ and $C$ \citep{Dutton:2014xda}. Although there is a large scatter in the $M_{\rm 200}-C$ plane for real galaxies, the choice of a fixed $C$ will not significantly affect our results because the two parameters are degenerated and will not be constrained separately with our kinematic data covering out to $2-3 R_\mathrm{e}$. The potential also includes a central black hole characterized
by a Plummer potential \citep{vandenbosch2008}, with the back hole mass, $M_\bullet$, fixed according the $M_\bullet$-$\sigma$ relation \citep{kormendy2013}. This choice does not affect our results, because the black hole sphere of influence is mostly not resolved by our kinematic data and we cannot directly constrain the black hole mass. 

In summary, we have five free "hyper-parameters," for the gravitational potential: the mass-to-light ratio, $M_{*}/L$, three parameters on the intrinsic shape of stellar distribution, $p$, $q$, and $u$, and the dark matter virial mass, $M_{200}$.

%-----------------------------------------------
\subsection{Computing the orbit library}
%-----------------------------------------------

For each model, with a set of hyper-parameters, we calculated an orbit library with tens of thousands orbits. The orbits sampling follows the way described in \cite{vandenbosch2008}.

We first sampled regular orbits according to a separable triaxial potential. The orbits are sampled from the three integrals of motion: energy, $E$, second integral of motion, $I_2$, and third integral of motion, $I_3$ \citep{Binney2008}.
We sample two sets of $55 \times 11 \times 11 $ combinations of $(E,I_2,I_3)$ as initial conditions, which include co-rotating and counter-rotating orbits. 

Box orbits are crucial for supporting the triaxial structures, we sampled another set of box orbits by constructing the initial conditions on equipotential surfaces with the energy, $E$, two spherical angles, $\theta$ and $\phi$, which gives another set of $55 \times 11 \times 11 $ orbits. 

The number of orbits, especially across $E$, we sample here is larger than that used in previous works \citep[e.g.,][]{vandenbosch2008, zhu2018, jin2019}. For instance, $21 \times 7 \times 7$ orbits were used in fitting the CALIFA data \citep{zhu2018}.
Because the kinematic data used in this work have larger spatial coverage and higher spatial resolution than those from the previous IFU surveys, more freedom is required by the model to fit the data. 
To reduce the Poisson noise of the model, we dither every orbit by slightly perturbing the initial conditions to give $5^3$ orbits.

%-----------------------------------------------
\subsection{Fitting stellar luminosity density and kinematics}
%-----------------------------------------------
We fit the luminosity density and stellar kinematics of the galaxy by weighting the orbits. The 2D surface-brightness, 3D luminosity density deprojected from the MGE, and kinematic maps are used as model constrains. The kinematic maps include the line-of-sight mean velocity, $V$,
velocity dispersion, $\sigma$, Gauss-Hermite coefficients, $h_3$ and $h_4$. It is worth noting that we did not fit $V$ and $\sigma$ maps directly, but the Gauss-Hermite coefficients $h_1$, $h_2$, $h_3$, and $h_4$. We extracted similar luminosity and kinematic maps from the model superposed by orbits, then obtained a solution of the orbit weights by minimizing the $\chi^2$ between the data and model using a non-negative least squares (NNLS) method \citep{Lawson1974,lawson1995}.

We used an optimized grid searching process \citep{zhu2018a} to adjust the free hyper-parameters of the gravitational potential. 
We started with a model with an initial guess of the hyper-parameters, then we performed an iterative searching process with intervals of 0.05, 0.02, 0.01, 0.01, and 0.05 for the hyper-parameters, $M_{*}/L$, $p$, $q$, $u$, and $\log{M_{200}}$. We note that the concentration parameter, $C,$ in the NFW dark matter halo is fixed by the $M_{\rm 200}-C$ relation. After the previous sampled models were completed, we selected the best-fit models with $\chi^2-\chi^2_{\mathrm{min}}<200$ and sampled the new models around the selected ones. We continued the iterative process until an area of $\chi^2$ minimum was found and all models within $3-\sigma$ confidence level around the minimum $\chi^2$ were calculated. 
At the end, we calculated a few hundred models for each galaxy, the resulting parameter grid for FCC 177 is shown in Fig.~\ref{img:grid177} in the appendix.
The models with least-$\chi^2$ are selected as the best-fit model. In Fig.~\ref{img:fitting177}, we show the best-fit model of FCC~177, where the model matches the observed kinematic data in great detail. Considering the model uncertainties, we defined the $1\sigma$ confidence level in a similar way as \citet{Zhu2018c}: 
\begin{align}
\delta\chi^2 = \chi^2-\chi^{2}_{\mathrm{min}}<f \sqrt{2\times n_{\mathrm{GH}} \times N_{\mathrm{obs}}},
\label{eqn:chi2}
\end{align}
where $ n_{\mathrm{GH}}=4$ is the number of stellar kinematic moments and $N_{\mathrm{obs}}$ is the number of bins in the kinematic data. We adopt $f=1$ for the galaxies with $N_{\mathrm{obs}}<200$, similar to the previous models for the CALIFA data; for galaxies with a much larger value for $N_{\mathrm{obs}}$, we found that the $\chi^2$ fluctuation caused by numerical noise is much higher and thus we adopted $f=4,$ following the results of a bootstrapping analysis. \footnote{For galaxies with $N_{\mathrm{obs}} \sim 1000$, the $\delta \chi^2$ is obtained by a bootstrapping process in the following: in a single model with fixed potential and orbit library, we perturb the kinematic data with its errors and fit the model to the perturbed data for many times. The standard deviation of $\chi^2$ obtained from these fittings are taken as the $\chi^2$ fluctuation caused by numerical noise of the model. Unlike the classic statistical analysis for analytic models, numerical noise is dominating the $\chi^2$ in our models, and it could be different for data with different spatial resolution. The confidence level we adopt is not motivated by robust statistical consideration (more discussion on it could see \citet{Lipka2021}), but practically it works well in covering the true values in our model test.}

Once we fit the stellar kinematics, we obtained the intrinsic stellar orbit distribution of the galaxy. We use the circularity, $\lambda_z$, and time-averaged radius, $r,$ to characterize the orbits, where $\lambda_z$ is defined as the orbital angular momentum around the $z$ direction, normalized by the maximum that is allowed by a circular orbit with the same binding energy. Here, $\lambda_z \sim 1$ represents highly rotating short-axis tube orbits and the $\lambda_z \sim 0$ represents mostly long-axis tube or box orbits. The stellar orbit distribution from the best-fit model of FCC~177 is shown in the left panel of Fig. \ref{phasespace177}.

%-----------------------------------------------
\subsection{Tagging orbits with age and metallicity}
%-----------------------------------------------
Next, we fit the age and metallicity maps by tagging age and metallicities to the orbits. 
We took the models within the $1\sigma$ confidence level selected by the kinematic fitting as given by Eq.~(\ref{eqn:chi2}). 
For each model, we have the stellar orbit distribution. We applied a Voronoi binning scheme to the orbits in the phase-space of $\lambda_z$ vs. $r$ and decompose them into $\sim 100$ orbital bundles.
Orbits with similar $r$ and $\lambda_z$ are included in the same orbital bundle and each bundle has a minimum orbital weight of 0.005.

After the Voronoi binning, we assume that each orbital bundle $k$ has a single value of age, $t_k$, and metallicity, $Z_k$.
Then the age, $t^i_{\text{obs}}$, and metallicity, $Z^i_{\text{obs}}$, of each observational aperture, $i,$ can be expressed as:
\begin{align}
t^i_{\text{obs}} &= \Sigma^{N_{\text{bundle}}}_{k=1} t_k f^i_k/\Sigma_k f^i_k, \\
Z^i_{\text{obs}} &= \Sigma^{N_{\text{bundle}}}_{k=1} Z_k f^i_k/\Sigma_k f^i_k,\quad k=1,...,N_{\text{bundle}},
\end{align}
where $N_{\text{bundle}}$ is the total number of orbital bundles and $f^i_k$ is the luminosity contribution of orbital bundle, $k,$ at an aperture, $i$. 

We applied a Bayesian statistical analysis (Python package pymc3)\citep{Salvatier2016}
to first solve $t_k$ by fitting the observed age map and then to solve $Z_k$ by fitting the observed metallicity map.
To use the Bayesian theorem to compute the posterior probability of a model, it is necessary to include the prior probability and data likelihood. For the prior probability of $t_k$, we adopted a bounded normal distribution, where the mean value $\mu(t_k)$ is randomly sampled around the average of the observed $t^i_{\rm obs}$.
We adopted a student's t-distribution for the data likelihood \citep{Salvatier2016}, which allows for some outliers in the data and results in a robust fitting. For each $t_k$,
we ran a chain with 2000 steps and take the last 500 steps to calculate the mean and $1\sigma$ values of $t_k$.

For a few galaxies, we did not get a good match of the age maps with the above fitting process. Therefore, we tried a second fitting round with $\mu(t_k)$ in the bounded normal distribution chosen as the $t_k$ value obtained from the first fitting round \citep{zhu2020}. We obtained a good match of age maps for all the galaxies after the second round of fitting.

Then we fit the metallicity map. For the prior probability of $Z_k$, we adopt a log-normal distribution, where the mean value $\mu(\log(Z_k))$ is set by a build-in age-metallicity relation using $t_k$ obtained from the fitting of age map and following \citet{zhu2020}. The student’s t-distribution is adopted again for the data likelihood, and we run a chain with 2000 steps and take the last 500 steps to calculate the mean and $1\sigma$ values of $Z_k$.

The best-fit model of FCC 177 is shown in Fig.\ref{img:fitting177}.
At the end, we obtained models that simultaneously fit well the surface-brightness, stellar kinematics, age, and metallicity maps of all the galaxies. We note that the age and metallicity maps we used have different binning schemes and have less data coverage than the kinematic data. For this reason, the age and metallicity of the orbits beyond the data coverage are not well constrained.

%-----------------------------------------------
\subsection{Orbital decomposition}
%-----------------------------------------------

\begin{figure}
    \centerline{
        \includegraphics[width=1.05\columnwidth, clip=true, trim=30 20 60 20]{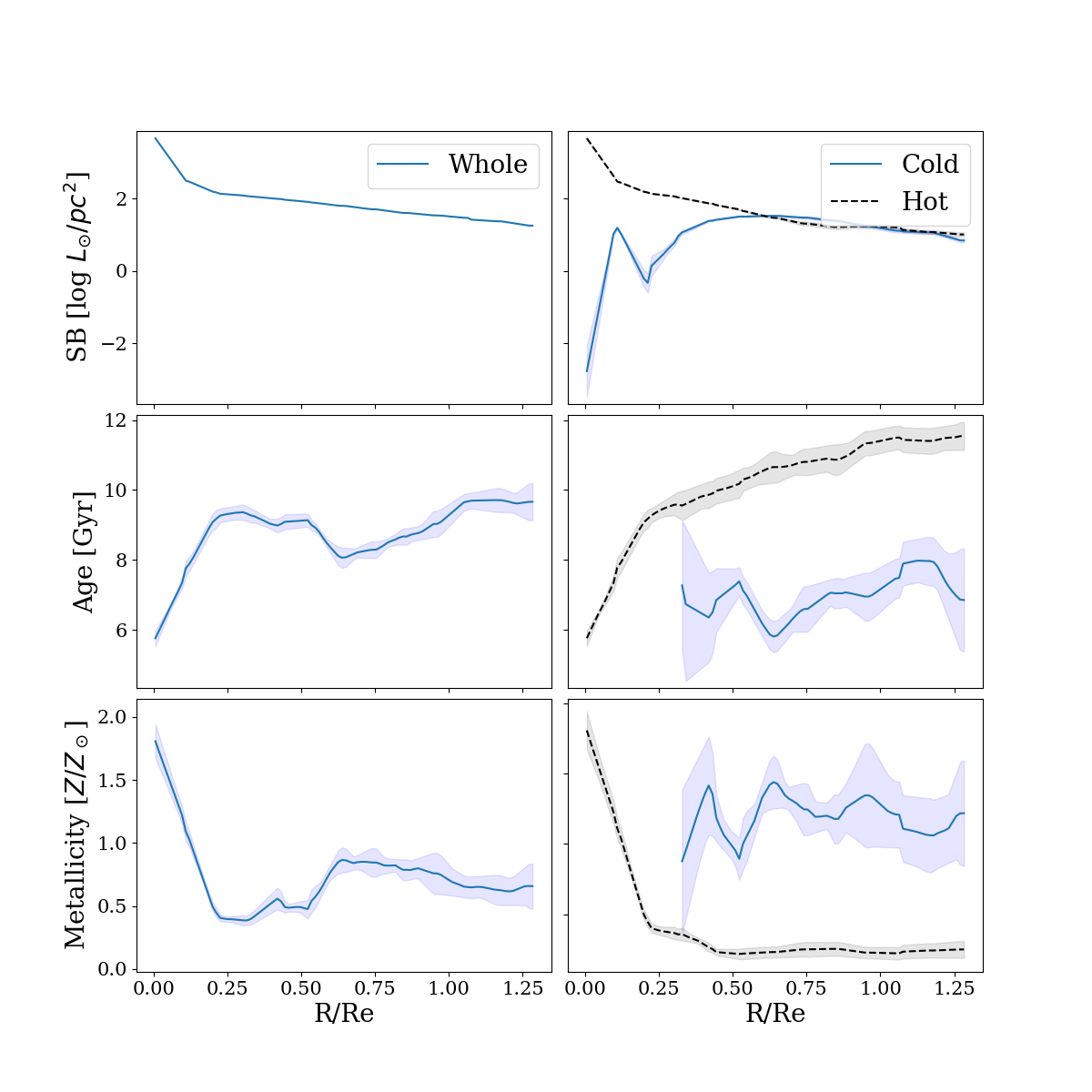}}
    \caption{
     Surface-brightness, age, and metallicity radial profiles of the whole galaxy, the dynamically cold disk, and the dynamically hot non-disk component for FCC~177. We show the profiles of the whole galaxy (left panels) with the blue solid curve, and the profiles of dynamically cold disk and dynamically hot non-disk component (right panels) with the blue solid curve and black dashed curve. The shadowed areas indicate the scatter of the profiles of models that fall within the 1$\sigma$ confidence level. The age and metallicity profiles of the dynamically cold disks are considered reliable and shown in the regions where the dynamically cold disk contributes at least 10\% of the total surface brightness.}
    \label{profile177}
\end{figure}

In Fig.~\ref{phasespace177}, we show the probability density, age, and metallicity distributions of the orbits of the best-fit model of FCC~177 in the phase-space of $\lambda_z$ versus $r$. We consider only the age and metallicity distributions within the data coverage ($r\lesssim 45$ arcsec) to be reliable. The models are usually more noisy (than smooth) in the phase-space of $\lambda_z$ vs. $r$. As we tested with mock data in \citet{zhu2020}, not all the substructures in the model are real, but we can trust the model considering the distributions of orbits and of these age and metallicity in a statistical way.

Based on the stellar orbit distribution, we coarsely decompose the galaxy into two components: the orbits with $\lambda_z\ge0.8$ are taken as a dynamically-cold component which is usually a disk. All the remaining orbits with $\lambda_z<0.8$ are considered as a dynamically-hot non-disk component. We note that here the dynamically cold component is defined as in \citet{zhu2018}, whereas the dynamically hot component actually includes their dynamically warm, hot, and counter-rotating orbits.
In Fig.~\ref{profile177}, we show the radial profiles of surface-brightness, age and metallicity of the whole galaxy for FCC~177, as well as those profiles for the dynamically cold and hot components, separately.
At $r\sim 0.3 R_{\rm e}$, the cold disk is about $2$ Gyr younger and significantly more metal-rich than the hot component and the age difference has risen to $\sim 4$ Gyr at $r\sim 0.7 R_{\rm e}$.  However, there is a young and metal-rich nuclear star cluster in the galactic center ($r\lesssim 0.2 R_{\rm e}$) which contributes to the hot component. The luminosity-weighted mean age and metallicity of the hot component are dominated by this young and metal-rich central regions, as a result, the difference of mean age (metallicity) of the cold disk and the hot component becomes smaller than that revealed in the profiles at $r\gtrsim 0.3 R_{\rm e}$. Within $R_{\rm e}$, the stars on the cold disk orbits have an luminosity-weighted average age of 6.7 Gyr and an luminosity-weighted metallicity of 1.25 $Z_{\odot}$ while the stars on dynamically warmer orbits have an average age of 8.1 Gyr and an average metallicity of 0.81 $Z_{\odot}$.

%-----------------------------------------------
\section{Results of the orbital decomposition for the sample galaxies}
\label{sec:Sb18}
%-----------------------------------------------
We have 20 galaxies with orbit-superposition models.
For 16 of those (15 ETGs and one LTG, FCC 179), we also have age and metallicity maps and subsequently build a population-orbit model as well.
In this section, we show the surface-brightness, age, and metallicity radial profiles of the dynamically cold disk component and dynamically hot non-disk component for all 16 galaxies with their orbits tagged with age and metallicity.
For the four LTGs without orbits tagged with age and metallicity, we are limited to the surface-brightness profiles of each component.

\begin{figure*}
    \centering{
        \hspace{-10pt}
        \includegraphics[width=2.\columnwidth, clip=true, trim=6 0 5 0]{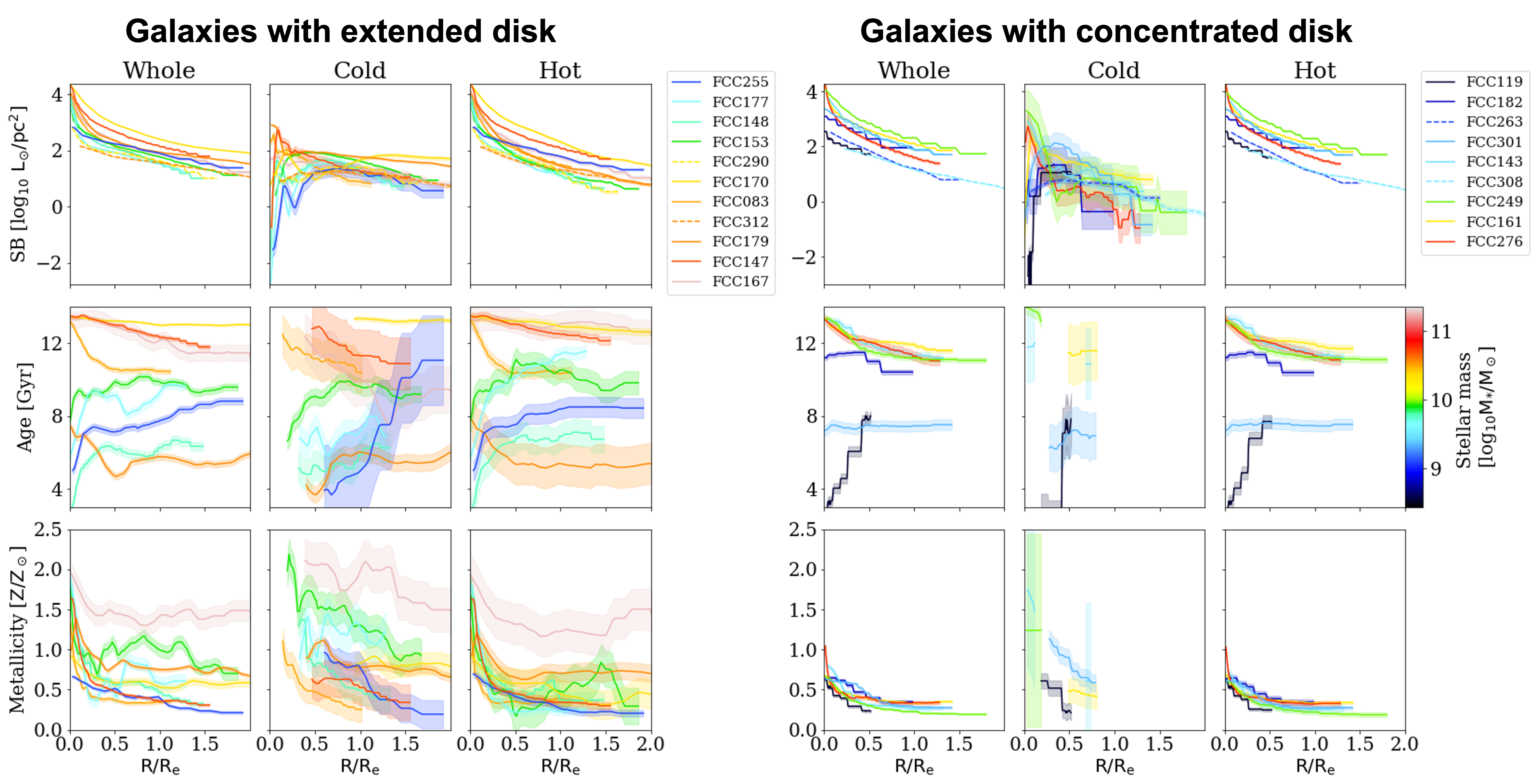}}
        \caption{Surface-brightness, age, and metallicity radial profiles of
        the galaxy, dynamically cold disk, and dynamically hot non-disk
        component for the sample galaxies. We divided the galaxies into two
        groups: 11 galaxies with a dynamically cold disk extended out to large
        radii (left panels) and 9 galaxies with a dynamically cold
        disk concentrated in the inner regions (right panels). The
        radial profiles are color-coded by the galaxy stellar mass. The
        shadowed areas indicates the scatter of the profiles of models that
        fall within the $1\sigma$ confidence level of the model
        hyper-parameters. Dashed lines refer to the four LTGs without orbits 
        tagged with age and metallicity. The age and metallicity
        radial profiles of the dynamically cold disks are considered reliable
        and shown in the regions where the dynamically cold disk contributes
        at least $10\%$ of galaxy surface-brightness. Therefore, these
        profiles are available only for a limited radial range for the
        galaxies with a concentrated cold disk.}
    \label{img:posi_grd}
\end{figure*}

%-----------------------------------------------
\subsection{Surface-brightness, age, and metallicity radial profiles}
%-----------------------------------------------
As illustrated by the best-fit model of FCC 177, we decomposed each model into a dynamically cold disk (with $\lambda_z\ge0.8$) and a dynamically hot non-disk component (including all the remaining orbits with $\lambda_z<0.8$). 
Then, we reconstruct the 3D luminosity, age and metallicity distributions of each orbital component. 
We project each orbital component to be face-on and then extract the radial profiles of surface-brightness, age, and metallicity from the projected maps. 

We note that the model already takes into the account the information for the 2D maps of the observational data. When we extract the radial profiles from the face-on projection, the radial profile along different directions are the same for the disk which is axisymmetric, while very similar for the hot non-disk component which could be moderately triaxial.
For each galaxy, we have tens to hundreds of models within the $1\sigma$ confidence level.
The aforementioned radial profiles are extracted for each model. We take the average of these profiles from models within the $1\sigma$ confidence level as the mean profile, and their scatter as the $1\sigma$ uncertainty. 

There are significant variations of internal structure from galaxy to galaxy.
We define the luminosity fraction of cold disk as:
\begin{align}
f_{\text{cold}} =\sum_{k}^{\lambda_z\ge0.8} f_k,
\end{align}
 where $f_k$ is the luminosity fraction of orbital bundle $k$. The cold-disk fraction of a galaxy varies with radius, we calculated $f_{\rm cold}(r<R_{\mathrm{e}})$ and $f_{\rm cold}(r<2R_{\mathrm{e}})$ with orbits within $R_{\rm e}$ and $2R_{\rm e}$, respectively.

We classify the galaxies into two categories: 11 of them have a relatively large cold-disk fraction and their cold disks have an extended surface-brightness, whereas the other 9 galaxies have a relatively small cold-disk fraction and their cold disks are concentrated in the very inner regions. 
Although the two categories are classified by eye, 
this is fairly consistent with a separation based on the cold-disk fraction $f_{\rm cold}(r<2R_{\mathrm{e}})$:
the galaxies in the first category generally have $f_{\rm cold}(r<2R_{\mathrm{e}}) > 0.1$,
most galaxies in the second category have $f_{\rm cold}(r<2R_{\mathrm{e}}) < 0.1$, but a few $>0.1$. We take this visual classification for practical reasons that we can obtain reliable age and metallicity gradients for the extended cold disks in the following analysis, but not for the concentrated ones.

The age and metallicity observed at any radius is a combination of different components. Our model tests in \citet{zhu2020} show that we cannot recover the age and metallicity of a component at radii where it contributes $\leq10\%$ to the total luminosity.
For the galaxies with extended disks, the surface-brightness of the cold disk component contributes at least $10\%$ of the total surface-brightness over a radial region extending to at least one $R_\mathrm{e}$.
Reliable age and metallicity profiles of the disk are thus obtained over extended radial regions. 
On the contrary, for the galaxies with small cold-disk fractions and concentrated cold components, the age and metallicity radial profiles of the cold component are only derived in inner radial regions. 

In Fig.~\ref{img:posi_grd}, we show the radial profiles of surface-brightness of the whole galaxy and of the dynamically cold and hot components, for all the 20 galaxies, as well as the age and metallicity radial profiles for the 16 galaxies with their orbits tagged with age and metallicity.
The galaxies with extended cold disks are shown in the left panels, while those with concentrated cold components are shown in the right panels. The dynamically cold and hot components are well separated with different surface-brightness, age, and metallicity radial profiles, which will be discussed in the following sections.

We note that the surface-brightness profiles of the cold disks do not follow the traditional exponential law: the surface-brightness in the central regions is much lower than the inward extrapolation of the exponential profile.
This finding is consistent with the results of orbital decomposition of the CALIFA galaxies \citep{Zhu2018c} and with those from direct spectrum fitting, where bulge and disk are separated through their stellar populations \citep{Breda2020}. 

%-----------------------------------------------
\subsection{Quantitative description of the radial profiles}
%-----------------------------------------------
We quantify the luminosity fraction $f_{\mathrm{cold}}$ of the cold disk component within $R_{\rm e}$ (or $2R_{\rm e}$), with respect to the cumulative luminosity of the galaxy within that radius. The $1\sigma$ uncertainty of $f_{\rm cold}$ is calculated by the scatter of the models within the $1\sigma$ confidence level.
We calculate $f_{\mathrm{cold}}$ for all the 20 sample galaxies.

For the 16 galaxies with their orbits tagged with age and metallicity,
we obtain the mean value and gradient of the radial profiles of age and metallicity of each component.
We calculated the mean age and mean metallicity of the cold disk by
\begin{align}
\left\langle t_{\text{cold}} \right\rangle =& \frac{1}{f_{\text{cold}}} \sum_{k}^{\lambda_z\ge0.8} t_k f_k,\\
\left\langle Z_{\text{cold}} \right\rangle =&\frac{1}{f_{\text{cold}}} \sum_{k}^{\lambda_z\ge0.8} Z_k f_k,
\end{align}
where $t_k$ is the age of the orbital bundle, $k$.
The lumonisity-weighted mean age $\left\langle t_{\mathrm{hot}} \right\rangle$ and metallicity $\left\langle Z_{\mathrm{hot}} \right\rangle$ of the hot component with the orbits $\lambda_z<0.8$ are calculated in a similar manner.
Note that the mean age is luminosity-weighted. Due to the different surface-brightness profiles of the cold and hot component, the difference in mean age of these two components sometimes appears differently from that seen in their age profiles, similar to the case of FCC 177 (as we explained in Section 3.5).

The gradients of age and metallicity are calculated as follows. We first uniformly interpolated 100 data points along the radial profile within the data coverage to smooth the profile, then we calculated the linear slope (gradient) between adjacent data points. In the end, we took the average of the gradients from the data points. We obtain the uncertainty of the gradient by bootstrapping within the shadowed regions of the profiles as shown in Fig.~\ref{img:posi_grd}. All these parameters, together with other basic parameters obtained from the population-orbit superposition models, are included in Table~\ref{tab:galaxy_prop}.

%-----------------------------------------------
\section{Cold-disk age as a proxy of galaxy infall time into a cluster}\label{sec:infall_time}
%-----------------------------------------------
\subsection{Calibration from simulations}
\begin{figure*}
    \centering{
        \includegraphics[width=1.8\columnwidth]{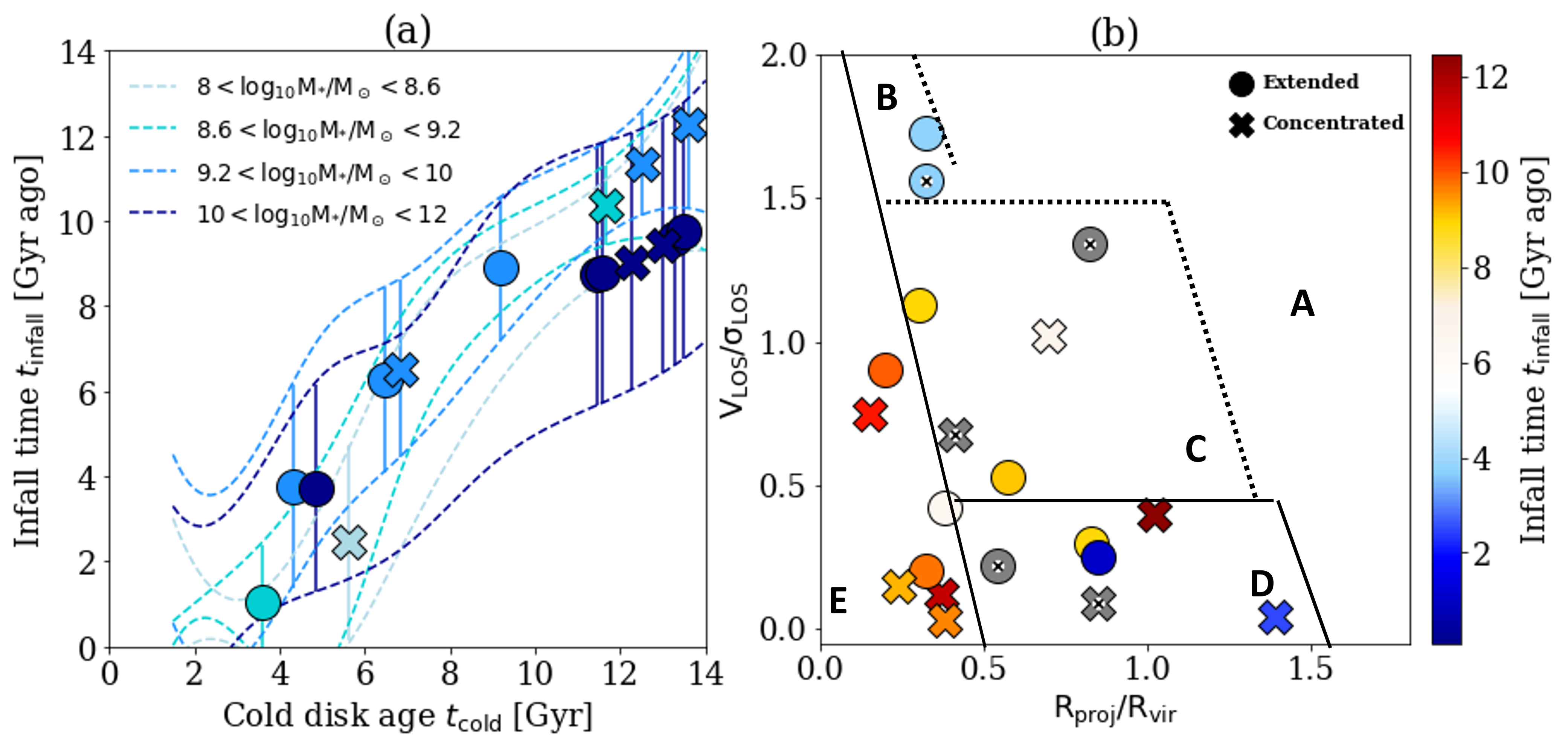}} 
    \caption{
    Infall time into the Fornax cluster of the sample galaxies for which we obtained the age of the dynamically cold disk. Left panel: Correlation between the galaxy infall time $t_{\rm infall}$ and age of the dynamically cold disk $t_{\rm cold}$ for four different mass bins: $8 < \log{M_\ast/{\rm M}_\odot}<8.6$ (light cyan), $8.6 <\log{M_\ast/{\rm M}_\odot}<9.2$ (cyan), $9.2 < \log{M_\ast/{\rm M}_\odot}<10$ (blue), and $10 < \log{M_\ast/{\rm M}_\odot}<12$ (dark blue). Dashed lines mark the $1\sigma$ confidence limits of each correlation. Circles and crosses correspond to galaxies with extended and concentrated dynamically cold disks, respectively. Each galaxy is plotted by the age of its dynamically cold disk and infall time given by the median of the correlation corresponding to its stellar mass. The error bars corresponds to the $1\sigma$ uncertainty of the infall time inferred from the correlation. Right panel: Distribution of the left-panel galaxies in the phase-space of the projected line-of-sight velocity of the galaxy normalized by line-of-sight
    velocity dispersion of the cluster $V_{\rm LOS}/\sigma_{\rm LOS}$ versus the projected clustercentric radius of the galaxy normalized by the cluster virial radius $R_{\rm proj}/R_{\rm vir}$. Each galaxy is color-coded according to its infall time except for the 4 LTGs without age and metallicity information shown in gray. The symbols are the same as in the left panel and the LTGs are marked by gray symbols. The boundaries of the regions A, B, C, D, and E are defined as in \citet{Rhee2017}.}
    \label{img:infall_time}
\end{figure*}

It is hard to accurately estimate the infall time into the cluster for each galaxy from observations, although we can statistically infer the likelihood of being ancient or recent infallers from their projected position and line-of-sight velocity in the cluster \citep{iodice2019}.
By analyzing the cluster galaxies in the cosmological simulation Illustris TNG50 \citep{pillepich2019, nelson2019}, we find that the cold-disk age is tightly correlated with the infall time of the galaxy into the cluster, 
as a result of star formation quenching in disks associated with galaxy infall into the cluster (Ding et al., in prep.).

To make direct comparison between the Fornax and TNG50 galaxies. We defined cold disks of TNG50 galaxies in exactly the same way as here for the Fornax cluster galaxies, with the probability density distribution of stars in the phase-space of circularity $\lambda_z$ versus $r$ calculated from the 6D position-velocity information of particles in the simulation. We note that we use $\lambda_z$ and $r$ of the particles' orbits, not directly the instantaneously values of each particle, following previous work in comparing the stellar orbit distribution of observed and simulated galaxies \citep{zhu2018}. The cold-disk fraction and cold-disk age are thus defined exactly the same way as described in Section~\ref{sec:Sb18}. 
All galaxies in the 14 clusters with virial mass $M_{\rm 200} >10^{13.3}\,\mathrm{M}_{\odot}$ in the TNG50 simulations are chosen. We define the galaxies' infall time into the cluster as the time when it first reaches the virial radius $r_{\rm 200}$ of the cluster at that time. For the pre-processed galaxies, we define their infall time as their first time of falling into the pre-cluster. 

We found that the correlation between the cold-disk age and the infall time of the galaxy into the cluster does not strongly depend on the cluster mass.
We divided all the galaxies in the 14 clusters into four mass bins with  $\log_{10}M_{*}/\mathrm{M_\odot}\in(8,8.6),\ (8.6, 9.2), \ (9.2,10), \mathrm{and} \ (10,11)$, respectively. 
In the left panel of Fig.~\ref{img:infall_time}, we show the running median and $\pm1\sigma$ scatter
of the correlation in the four mass bins color-coded by the stellar mass.
The $1\sigma$ scatter is 0.63, 0.63, 0.94, and 1.34 Gyr from the low to high mass bins. 
The correlation is tighter for the least-massive galaxies. The scatter becomes significantly larger in the most massive galaxies because a significant fraction of them are already quenched or had little gas left before they fell into the cluster.  

For galaxies with the cold-disk age obtained from the population-orbit superposition model, we can thus use this correlation to infer their infall time into the Fornax cluster. As shown in Fig.~\ref{img:infall_time}, we obtained the infall time of each galaxy by means of the correlation for the mass bin including its stellar mass. The values inferred from the median and $1\sigma$ scatter of the correlation are taken as the median and $1\sigma$ uncertainty of the galaxy infall time $t_{\rm infall}$. They are listed in the last column of Table~\ref{tab:galaxy_model_prop}.

\subsection{Comparison with galaxy locations in the projected phase-space}

\begin{table}
\label{tab:pps_comparison}
    \caption{Fractions of ancient, intermediate, and recent infallers in the regions of E, D, and B+C we find in our sample and a comparison with that predicated from \citet{Rhee2017} in brackets.}
   \centering
    \begin{tabular}{|c|c|c|c|}
        \hline
        Region & Ancient & Intermediate & Recent\\
        \hline
        E & 85.7\% (50\%) &  14.3\% (20\%) & 0\% (25\%) \\
        \hline
        D & 50\% (20\%) & 0\% (30\%) & 50\% (50\%) \\
        \hline
        B+C & 40\% (30\%) & 20\% (20\%) & 40\% (50\%) \\
        \hline
    \end{tabular}
\end{table}

We cross-checked the infall time inferred from the cold-disk age with the galaxy location in the phase-space $V_{\mathrm{LOS}}/\sigma_{\mathrm{LOS}}$ vs. $R_{\mathrm{proj}}/R_{\mathrm{vir}}$, as shown in the right panel of Fig.~\ref{img:infall_time}. Here, $V_{\mathrm{LOS}}/\sigma_{\mathrm{LOS}}$ is the projected line-of-sight velocity of the galaxy normalized by line-of-sight velocity dispersion of the cluster and $R_{\mathrm{proj}}/R_{\mathrm{vir}}$ is the projected clustercentric radius of the galaxy normalized by the cluster virial radius. The galaxies are color-coded by the infall time inferred from the cold-disk age in the left panel of Fig.~\ref{img:infall_time}.

Following \citet{iodice2019b}, we define ancient infallers as galaxies that fell into the cluster at 8-12 Gyr ago, intermediate infallers at 4-8 Gyr ago, and recent infallers at 0-4 Gyr ago.
We divided the phase-space into five regions like in \citet{iodice2019b}, with the fraction of ancient infallers decreaseing from regions E and D to B+C (based on \citet{Rhee2017}).
About $50\%$ of the galaxies in region E are assumed to be ancient infallers in our sample, while six out of seven galaxies in region E are ancient infallers and one is intermediate infaller. Galaxies in region D are supposed to be dominated by recent and intermediate infallers, while we have two ancient infallers and two recent infallers. Galaxies in region B+C are suggested to be dominated by recent infallers, while we have two ancient, one intermediate, and two recent infallers. The fractions of ancient, intermediate and ancient infallers in different regions we obtained and in comparison to that predicted from \citet{Rhee2017} are shown in Table~3.

 Although we have a low amount of statistics at hand, we remain consistent with \citet{Rhee2017} in that the fraction of ancient infallers is highest in region E and then lower in D and B+C. However, the fraction of ancient infallers in our sample is generally higher, as we only included ETGs -- and not the whole sample of galaxies in the Fornax cluster. 
The infall time inferred from cold-disk age seems reasonable for these Fornax cluster galaxies. Importantly, the correlation $t_{\rm infall}$ versus $t_{\rm cold}$ goes beyond a mere statistical estimation, providing a proxy of the cluster infall time per individual galaxy. 

%-----------------------------------------------
\section{Dependence of cold disk assembly on galaxy mass and cluster environment}\label{sec:mass_dependence}
%-----------------------------------------------
In this section, we show the dependence of the cold disk properties, including the luminosity fraction, age, and metallicity on galaxy stellar mass and cluster environment.

%-----------------------------------------------
\subsection{Cold-disk luminosity fraction}
%-----------------------------------------------

\begin{figure*}
    \centering{
        \includegraphics[width=1.8\columnwidth]{{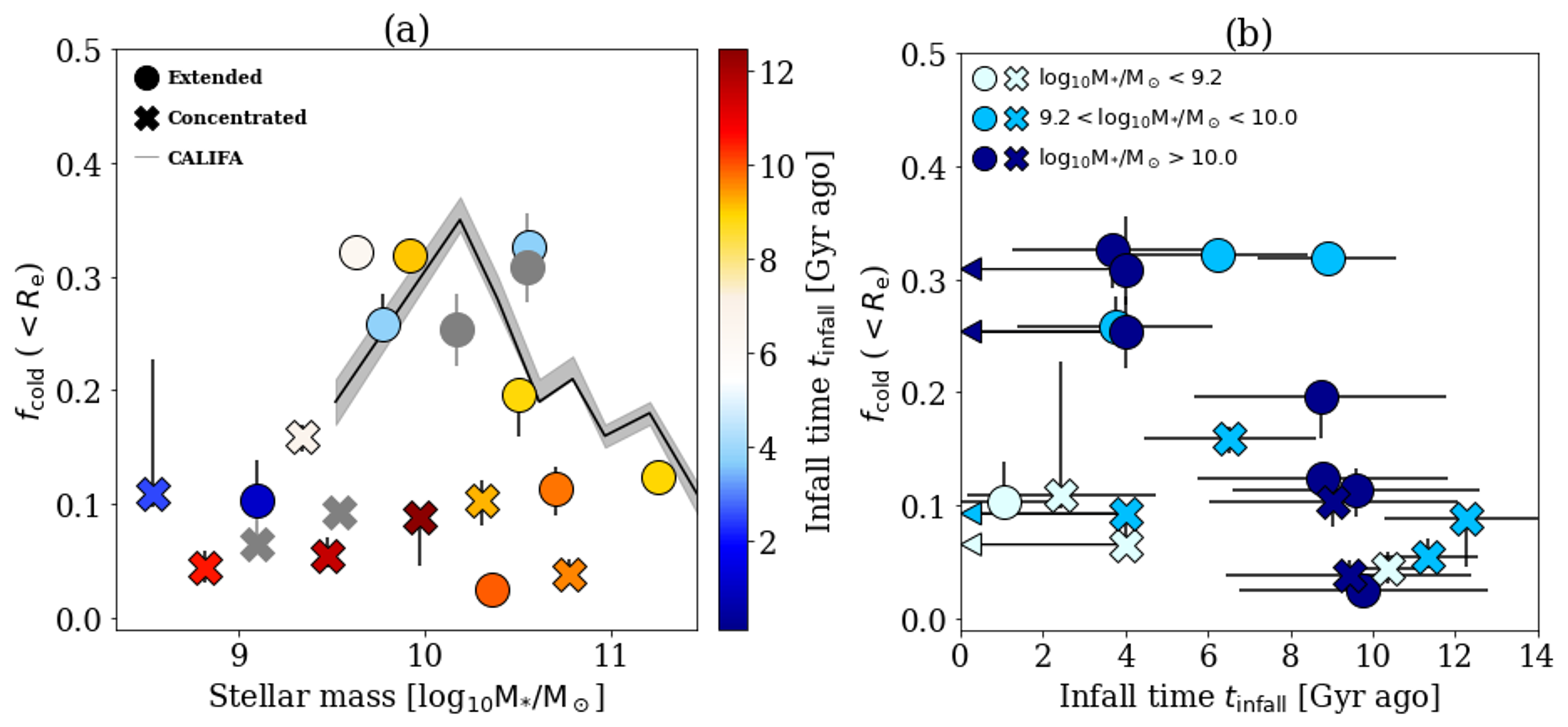}}}
    \caption{
    Dependence of cold-disk fraction (within $1R_{\rm e}$) on galaxy stellar mass and infall time. Left panel:
    Cold-disk fraction of the sample galaxies as a function of stellar mass, color-coded by infall time into the Fornax Cluster to highlight the ancient
    (red), intermediate (white), and recent (blue) infallers. Gray symbols correspond to the four LTGs
    without colored model which are considered as recent infallers. The gray curve represents the cold-disk fraction with $1R_{\rm e}$ of CALIFA galaxies \citep{zhu2018a}.
    Circles and crosses correspond to galaxies with extended and
    concentrated dynamically cold disks, respectively. 
    Right panel:
    Cold-disk fraction within $1R_{\rm e}$ as a function of infall time color-coded by stellar mass, divided into three mass bins: $\log{M_\ast/{\rm M}_\odot}<9.2$ (light blue), $9.2 <
    \log{M_\ast/{\rm M}_\odot}<10$ (blue), and $\log{M_\ast/{\rm M}_\odot}>10$ (dark blue). The four LTGs without colored model are plotted with an infall time upper limit of 4 Gyr ago.
    }
    \label{img:fcold}
\end{figure*}

\begin{figure*}
    \centering{
        \includegraphics[width=1.8\columnwidth]{{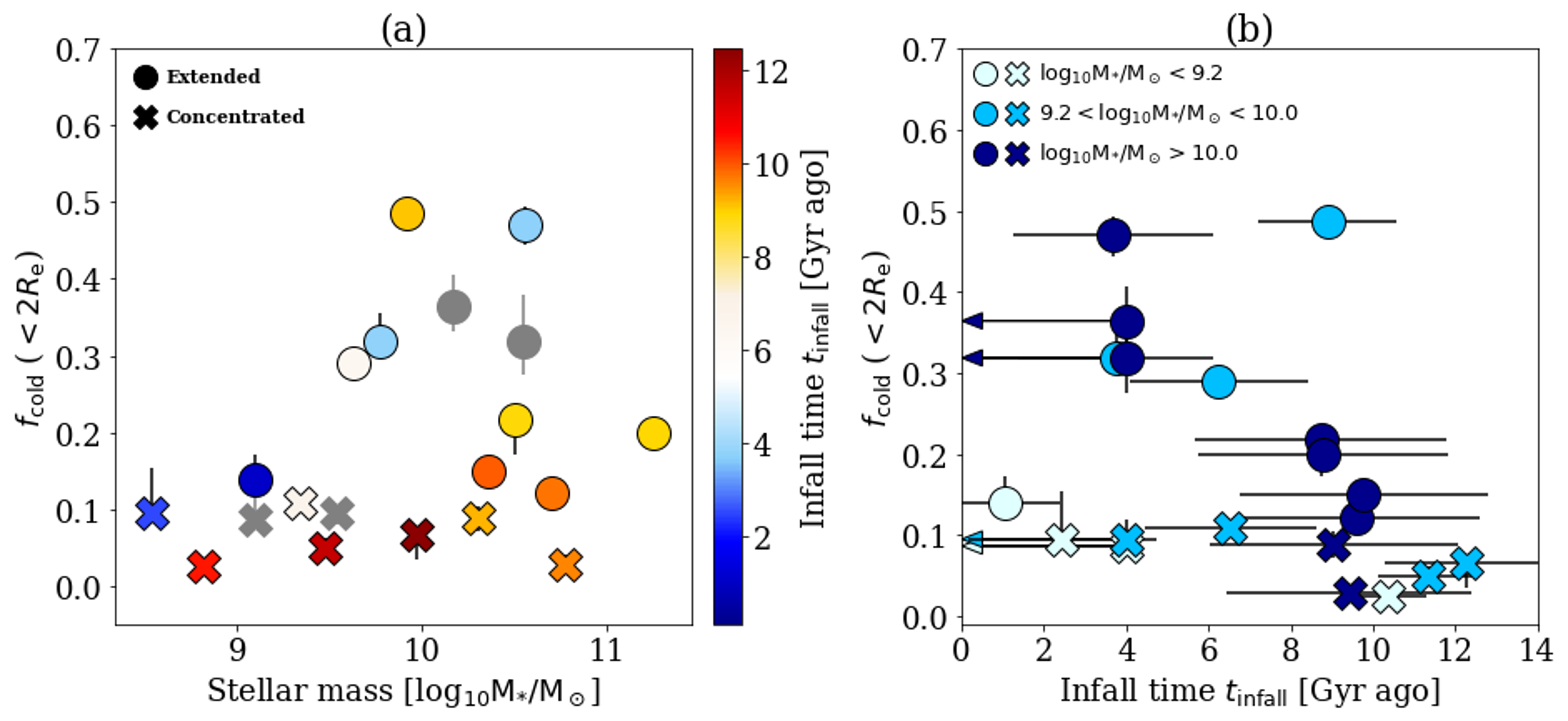}}}
    \caption{Dependence of cold-disk fraction (within $2R_{\rm e}$) on galaxy stellar mass and infall time. Details are similar to Fig.~\ref{img:fcold}, but for cold-disk fractions within $2\,R_\mathrm{e}$, which are about 0.1 higher than that within $1\,R_\mathrm{e}$ for galaxies with extended disks.
    }
    \label{img:fcold_2Re}
\end{figure*}

\begin{figure*}
    \centering{
        \includegraphics[width=1.8\columnwidth]{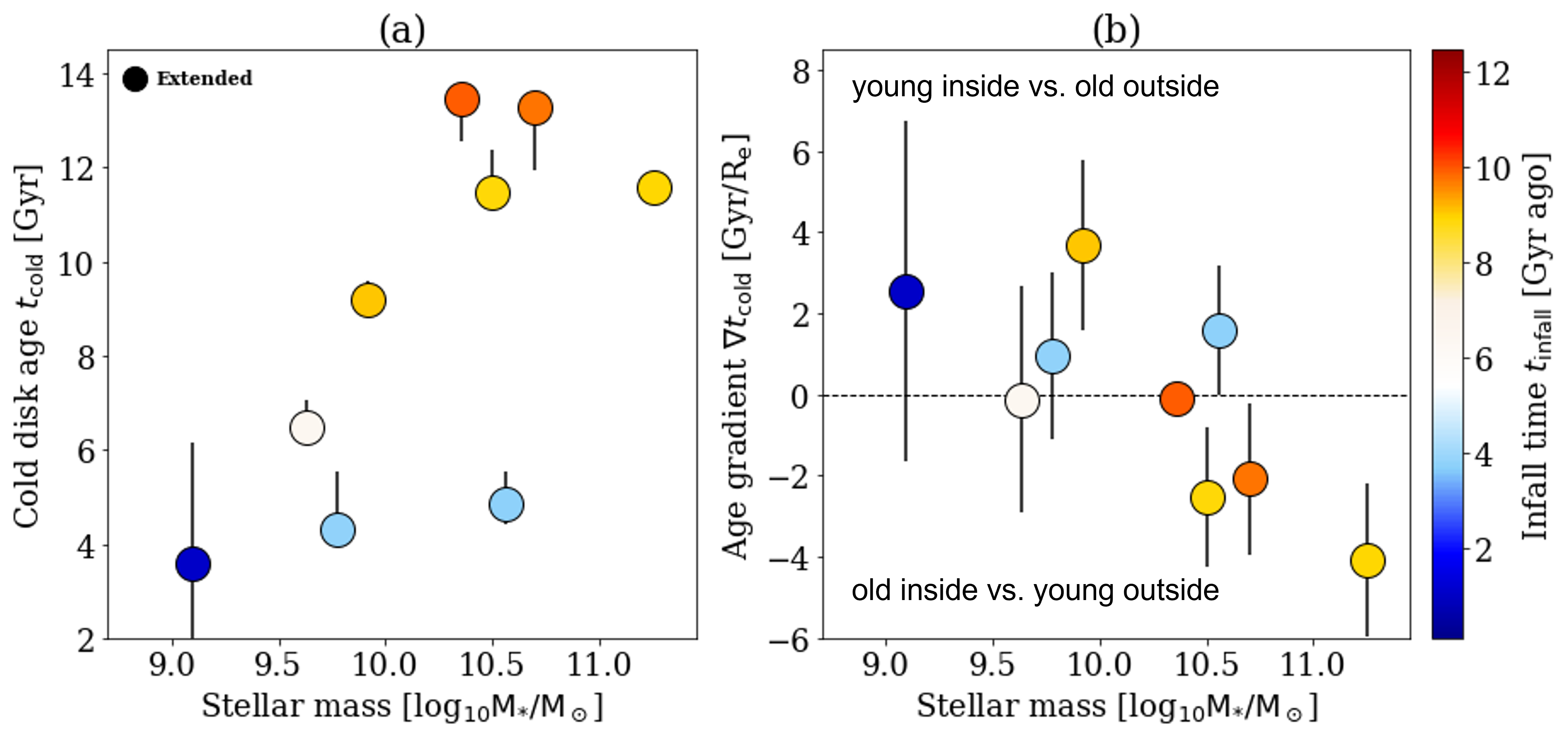}}
    \caption{
     Mean age (left panel) and age gradient
    (right panel) of the extended cold disks in the sample galaxies
    as a function of stellar mass and infall time into the Fornax Cluster.
    Dashed line in the right panel marks the zero age gradient.
    }
    \label{img:age_gradient}
\end{figure*}

We studied the dependence of cold-disk fraction on galaxy stellar mass and infall time into the cluster. The cold-disk fraction is calculated as the fraction of cold orbits ($\lambda_z\ge0.8$) within $1\,R_\mathrm{e}$ and $2\,R_\mathrm{e}$, respectively, of each galaxy.

We show the cold-disk fraction within $1\,R_\mathrm{e}$ in Fig.~\ref{img:fcold}. In the left panel, we show the cold-disk fraction as a function of stellar mass. The 16 galaxies with colored models are color-coded by infall time inferred from their cold-disk age, while the four LTGs without age information in the models are considered as recent infallers, as they are dominated by younger stars at all radii. 

Generally, the recent and intermediate infallers have higher cold-disk fractions, the galaxies with $t_{\rm infall} < 9$ Gyr have cold-disk fractions, which vary as a function of stellar mass, comparable to the cold-disk fractions of field galaxies in the CALIFA survey \citep{zhu2018a}. 
The most ancient infallers (red and orange points with $t_{\rm infall} > 9$ Gyr) have low cold-disk fractions of $f_{\mathrm{cold}}(r<R_\mathrm{e}) \lesssim 0.1$ at all mass regions. 

In the right panel, we show cold-disk fraction as a function of galaxy infall time by dividing the galaxies into three mass bins. For galaxies with similar stellar mass, the cold-disk fraction decreases from recent to ancient infallers. Although there is some scatter, especially in the case of FCC 153, with $t_{\rm infall} \sim 9$ Gyr but a high cold-disk fraction of $f_{\mathrm{cold}} (<1R_\mathrm{e}) > 0.3$.

Similar figures but for cold-disk fraction within $2\,R_\mathrm{e}$ are shown in Fig.~\ref{img:fcold_2Re}. The cold-disk fraction within $2\,R_\mathrm{e}$ is about 0.1 higher than that within $1\,R_\mathrm{e}$ for the galaxies with extended disks. 
The cold-disk fractions of $f_{\mathrm{cold}} (<1R_\mathrm{e})\sim 0.3$ and $f_{\mathrm{cold}} (<2R_\mathrm{e})\sim 0.4$ at $M_* \sim 10^{10}\,\mathrm{M}_{\odot}$ for the recent infallers are a factor of $\sim 3\times$ or $\sim 4\times$ higher than the ancient infallers mostly with both $f_{\mathrm{cold}} (<1R_\mathrm{e}) < 0.1$ and $f_{\mathrm{cold}} (<2R_\mathrm{e}) < 0.1$.

%------------------------------------------
\subsection{Cold-disk age and age gradient}
%------------------------------------------
We further studied the age and age gradient of cold disks in these galaxies. 
We calculated the age gradients of cold disks $\nabla t_{\rm cold}$ for the nine galaxies with extended disks and colored models. As we can see from Fig.~\ref{img:posi_grd}, the age profiles of disks are diverse in the inner regions and they become flat at the outer regions for most galaxies. We calculated the age gradients with profiles at $r<R_\text{e}$ consistently for all galaxies, except for FCC~170 and FCC~255, the age gradient is calculated within the whole data coverage as their disks have very low surface-brightness at $r<R_{\rm e}$.

In the left panel of Fig.~\ref{img:age_gradient}, we show the mean age of the cold disk versus stellar mass for the nine ETGs with extended disks and with colored model. The disks are generally older in the more massive galaxies.
In the right panel of Fig.~\ref{img:age_gradient}, we show the age gradient of the cold disk versus stellar mass colored by the galaxy infall time into the cluster.
We get a positive or zero age gradient of the cold disk for most galaxies. 
The least-massive galaxies show significantly positive $\nabla t_{\rm cold}$ for $r<R_\mathrm{e}$ and the age profiles become shallower at the outer regions. Three of the most massive galaxies FCC~083, FCC~147, and FCC~167
still have negative $\nabla t_{\rm cold}$ at $r<R_\mathrm{e}$, while their age profiles become flat or even turn to be positive (for FCC 167) at the outer regions.

\begin{figure*}
    \centering{
        \includegraphics[width=1.8\columnwidth]{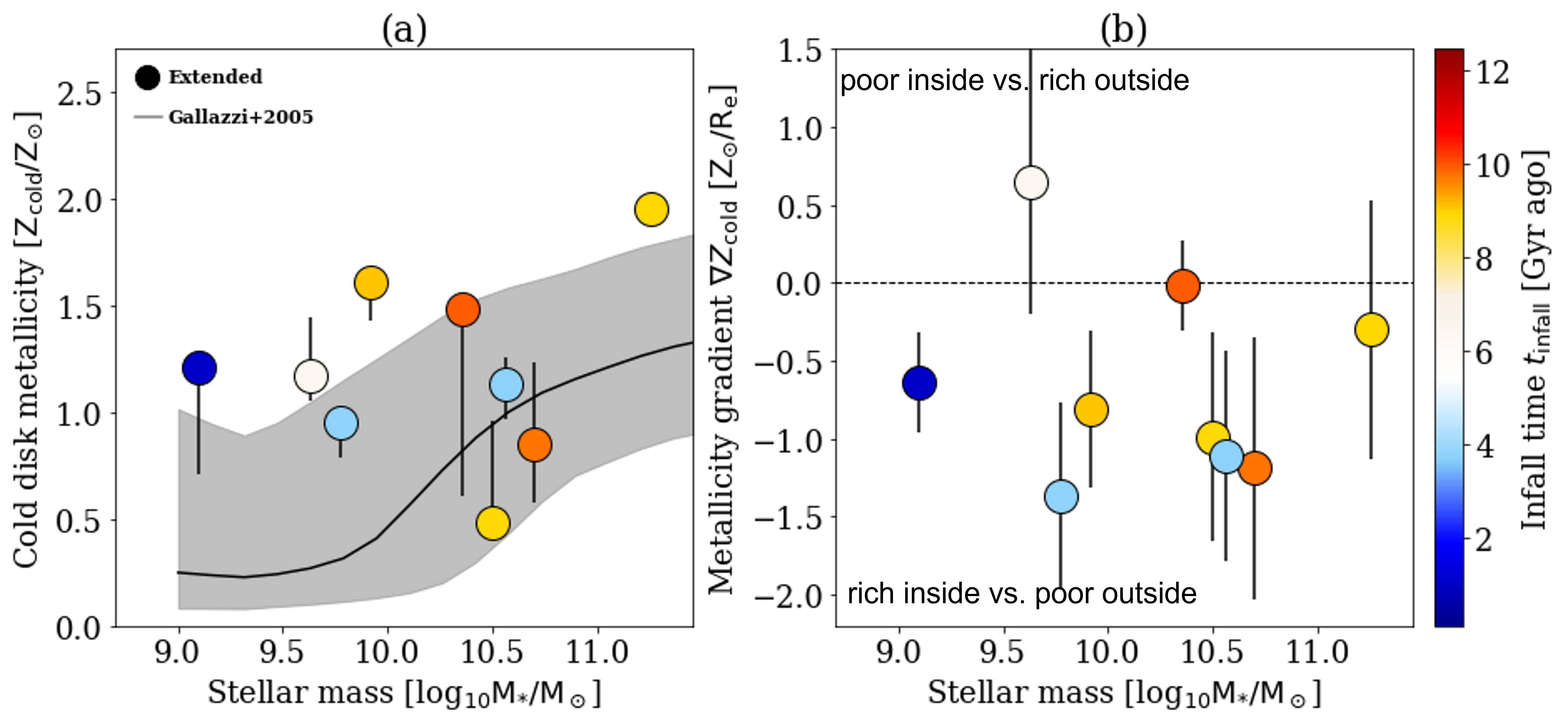}}
    \caption{
     Mean metallicity (left panel) and metallicity gradient
    (right panel) of the extended cold disks in the sample galaxies
    as a function of stellar mass and infall time into the Fornax Cluster. The black curve and gray shadow represent the metallicity-stellar mass relation and the $1\sigma$ uncertainty for the galaxies in \citet{gallazzi2005}. Symbols are same as Fig.~\ref{img:age_gradient}.
    }
    \label{img:mean_met_vs_mass}
\end{figure*}

%-----------------------------------------------
\subsection{Cold-disk metallicity and metallicity gradient}
%-----------------------------------------------
In the left panel of Fig.~\ref{img:mean_met_vs_mass}, we show the mean metallicity of the cold disks versus stellar mass $M_*$ colored by the galaxy infall time into the cluster.  We only show the nine galaxies with extended cold disks and colored models. The cold disks appear to be similarly metal rich from low mass to high mass galaxies. 
The average metallicity of the cold disks is $\langle Z_{\text{cold}}\rangle = 0.91\pm 0.6\,Z_{\odot}$, with no significant correlation with galaxy stellar mass or infall time into the cluster. 
Comparing to the general metallicity-mass relation of galaxies \citep{gallazzi2005}, the cold disks in galaxies with $M_* \lesssim 10^{10}\,\Msun$ in our sample are systematically more metal-rich than the general population of galaxies, while in more massive galaxies, they are more similar in composition.

In the right panel of Fig.~\ref{img:mean_met_vs_mass}, we show the metallicity gradient of the cold disks versus stellar mass.
Metallicity gradients are calculated in a way similar to age gradients and only for the nine galaxies with extended disks and colored models. Almost all these galaxies show negative metallicity gradients for the cold disks, which are more metal-rich in the inner regions than in the outer one. There is no significant correlation seen between the metallicity gradient of cold disk and stellar mass -- other than the change of metallicity gradients of the whole galaxy from positive to negative for low to high mass galaxies \citep{Goddard2017, Zhuang2019}, however, with low statistics of only nine galaxies included.

%-----------------------------------------------
\subsection{Cold versus hot galaxy components}
\begin{figure*}
    \centerline{
        \includegraphics[width=1.8\columnwidth]{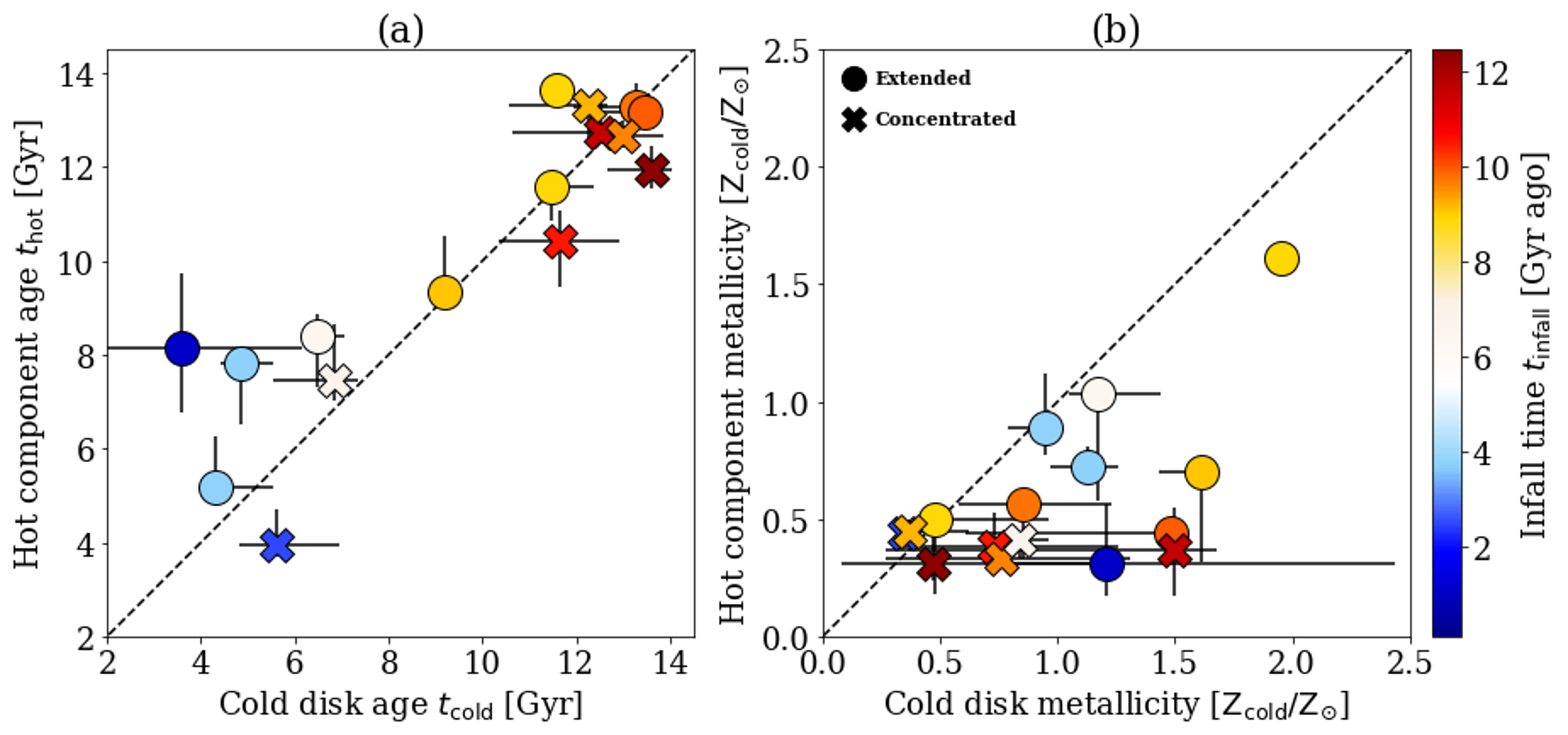}}
    \caption{
    Comparison of the mean ages (left panel) and mean
    metallicities (right panel) of the dynamically cold disks and
    dynamically hot non-disk components. Circles and crosses correspond to galaxies with extended and
    concentrated cold disks, respectively. The mean ages and metallicities
    are calculated for $r < R_{\rm e}$ (filled circles). Black dashed lines represent $y=x$.
    }
    \label{img:cold_vs_hot}
\end{figure*}

%-----------------------------------------------
We compare the age and metallicity of the dynamically cold disk and dynamically hot non-disk component in Fig.~\ref{img:cold_vs_hot}. The stars in the cold disks are as old as those of the hot components for the ancient infallers, whereas they are significantly younger than those of the hot components in recent infallers. The cold disks are in general more metal-rich that the hot components, which could explain the difference between our results on cold disk and the general mass-metallicity relation \citep{gallazzi2005}, as shown in Fig.~\ref{img:mean_met_vs_mass}.

\begin{figure*}
    \centerline{
        \includegraphics[width=1.8\columnwidth]{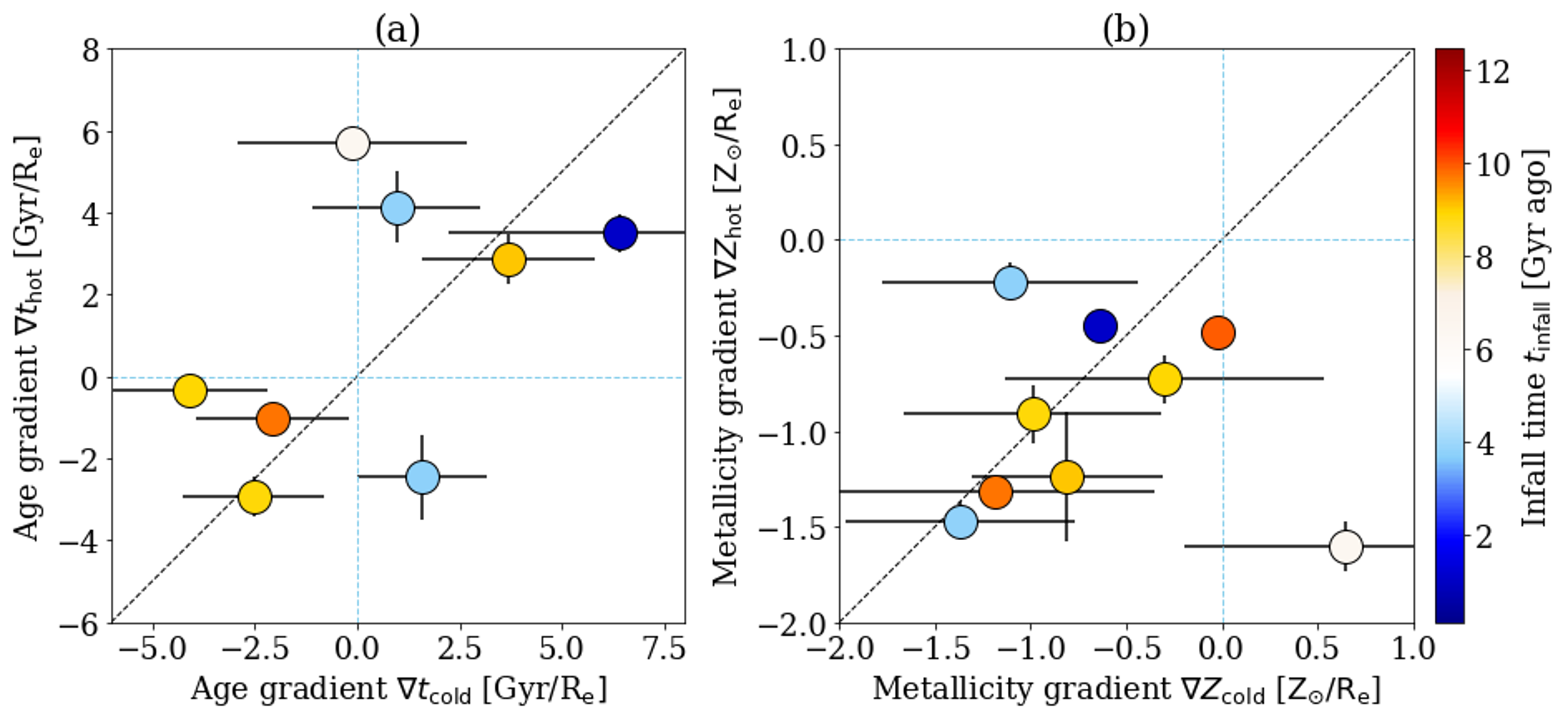}}
    \caption{
    Comparison of the age gradient (left panel) and metallicity gradient (right panel) of the dynamically cold disks and dynamically hot non-disk components, for the galaxies with extended disks. Black dashed lines represent $y=x$. Blue dashed lines represent zero gradients.
    }
    \label{img:cold_vs_hot_grad}
\end{figure*}

We compared the age and metallicity gradients of the two components in Fig.~\ref{img:cold_vs_hot_grad}. For the galaxies with negative age gradients in cold disks, they also have negative age gradients in the hot component, while for the galaxies with positive age gradient in cold disks, the age gradients in their hot component could be very different; whereas both the cold disk and hot component have negative metallicity gradients in most galaxies and are generally consistent with each other. 

\begin{figure*}
    \centerline{
        \includegraphics[width=1.8\columnwidth]{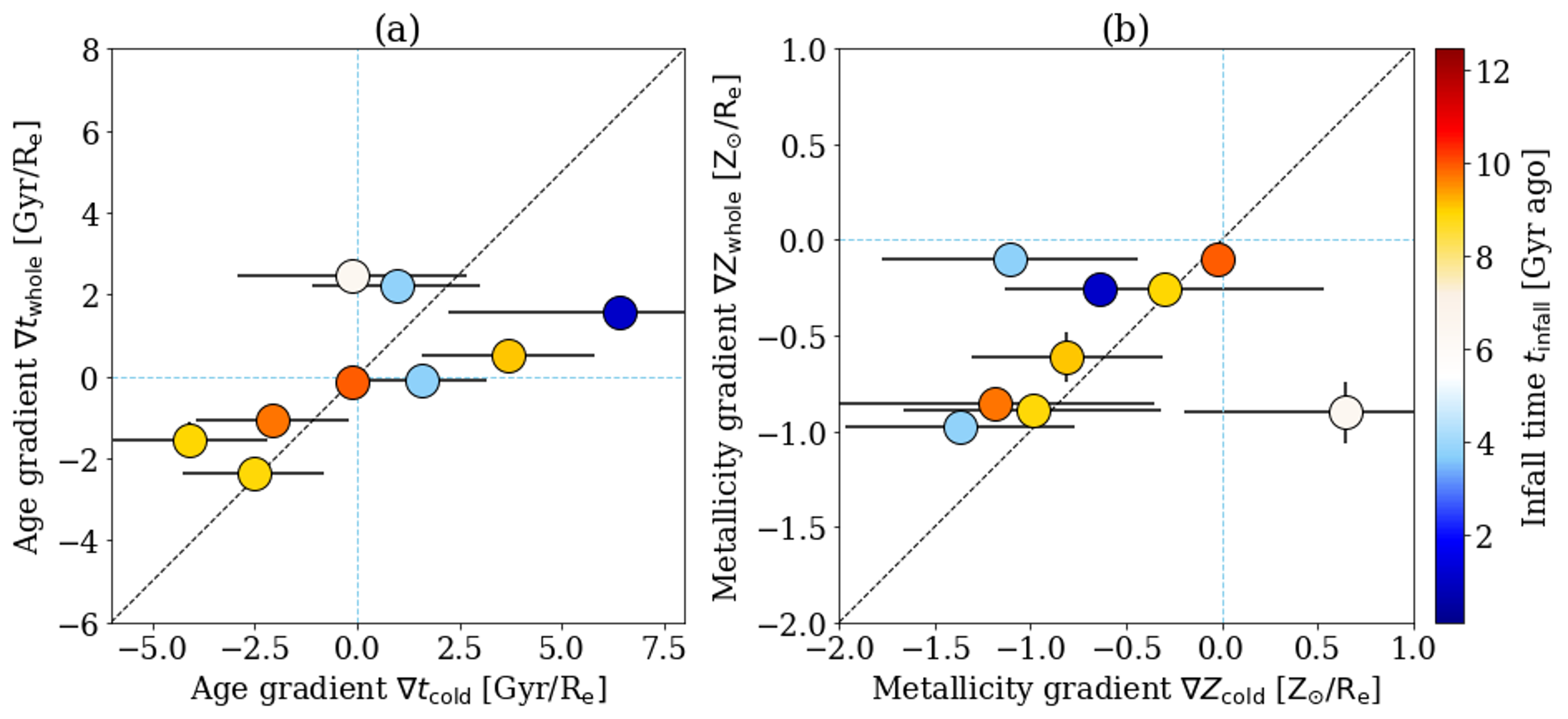}}
    \caption{
    Comparison of the age gradient (left panel) and metallicity gradient (right panel) of the dynamically cold disks and the whole galaxy, for the galaxies with extended disks. Symbols are the same as in Fig.~\ref{img:cold_vs_hot_grad}.
    }
    \label{img:cold_vs_whole_grad}
\end{figure*}

 The age and metallicity gradients of the cold disks are similar to that of the whole galaxy, as shown in Fig.~\ref{img:cold_vs_whole_grad}. The signs of the gradients in cold disks are the same as what is seen for the whole galaxy in most cases. In this sense, the gradients directly measured for the whole galaxy could statistically reflect that in the cold disks, for galaxies with extended cold disks.

\section{Discussion}\label{sec:discussion}

We find that the most ancient infallers in the Fornax cluster have low cold-disk fractions of $f_{\rm cold} \lesssim 0.1$, which is lower by about a factor of 4 than those of the recent and intermediate infallers in the Fornax cluster and of the field galaxies \citep{zhu2018a},
with similar stellar mass. Although there is some scatter, especially with regard to FCC 153, which is potentially an ancient infaller, but with a high cold-disk fraction.

Our results indicate that a galaxy's infall into a cluster has a strong effect on the assembly of the cold disk, which might be caused by the cut-off of the gas accretion from the cluster environment or removal of the gas by ram pressure and tidal stripping. 
Both processes explain the above results: galaxies that have earlier fallen into the cluster have their gas earlier removed or cut off, thus their disks are older and contribute a smaller fraction to total luminosity; galaxies that have recently fallen into the cluster have more time to grow their disk and the disk stars could be younger. 
However, it should be noted that we have the luminosity and not the mass fraction of the disk. The higher luminosity fraction of the cold disks in recent infallers could be partially caused by the lower stellar mass-to-light ratio in younger stellar disks.

It has been argued that ancient infallers could have a distinct evolutionary history
before entering the cluster environment, which determined their morphology, kinematics, and gas content ``at infall`` \citep[e.g.,][]{Han2018,Su2021}.
This could be part of the reason for the difference we see between the ancient and recent infallers at present time. But this preprocessing cannot explain all the differences. Numerical simulations show that dynamically cold disks are often
partially removed or heated by tidal disruption in cluster \citep{Joshi2020}. 
With the control of galaxy properties at the time of infall,
the galaxies that fall into a cluster result in significantly smaller disk fraction at redshift zero than those not. In this sense, the ancient infallers now in the cluster center
would suffer most from the tidal heating or disruption, thus their cold-disk fraction could be further reduced.

We find positive or flat age gradients in the extended disks, which is opposite to what expected from a normal inside-out disk growth. The turnover of disk-age gradients could be the result of the continuous star formation in the galaxy inner regions, while the star formation in the outer regions is quickly reduced after falling into the cluster. This is generally consistent with the scenario found in the EAGLE simulations \citep{Pfeffer2022}. In five of our nine galaxies with extended disks, the stars in the inner disk region are about 2-5 Gyr younger than those in the outer disk region, which indicates that the star formation in galaxy inner regions could last for a few Gyrs more after the galaxy falls into the cluster. None of the 15 ETGs of our sample exhibit any detectable ongoing star formation: stars in the inner disks are younger, but their star formation stopped a few Gyrs ago.
On the other hand, the inner regions are usually blended with an old bulge component, the age gradient of the whole galaxy is thus not so obviously negative as the disk only. Lasting star formation in galaxy inner regions is usually difficult to be directly observed in galaxies at present time, while hopefully might be directly observable at high redshift \citep{Maier2019b}.

Strong positive age gradients have been reported for the SAURON ETGs in the mass range of $10^{10-12}\,\mathrm{M}_{\odot}$ using the method based on line indices  \citep{Kuntschner2010}. However, in larger sample of MaNGA galaxies, the age gradients of ETGs are found to be generally shallow \citep{Goddard2017}. Positive age gradients are mostly found in low mass ETGs with $M_*\lesssim 10^{10.5}$ \citep{Li2018} by using the technique of full spectra fitting to derive the stellar populations. As shown in \citet{Li2018}, the strong positive age gradients reported in \citet{Kuntschner2010} could disappear when uniformly using the same technique. 
As we are also using the full spectra fitting, a comparison with these recent results from MaNGA galaxies might offer a more fair outcome. We note that the age gradients found in MaNGA galaxies are different in \citet{Goddard2017} and \citet{Li2018}, while the samples these authors studied are also different.

We found positive age gradients in the cold disks of least-massive galaxies, namely, $M_*\lesssim 10^{10.5}$, whereas the disks that have negative age gradients in the inner regions and tend to be flat in the outer regions of most massive galaxies. 
This is consistent with the positive age gradient in low-mass MaNGA ETGs  \citep{Li2018}, considering age gradients of the cold disks are statistically consistent with that of the whole galaxies.
The difference of age gradients in low mass and high mass ETGs could be explained by the difference of gas content in low and high mass galaxies at infall. The low mass galaxies could have large gas fractions when they fall into the cluster. The gas in the outer regions is stripped, but the gas kept in inner regions form a significant amount of new stars that finally turn-over the age gradient.
On the other hand, massive galaxies could have already formed most of their stars before entering the cluster. They have small gas fractions when falling into the cluster, thus the age gradient is less likely to be turned over by the formation of new stars after entering the cluster. 
Although with limited statistics, there is no galaxy with significant negative age gradient in our sample with $M_*\lesssim 10^{10.5}$.
Approximately flat gradients for such galaxies are found in the MaNGA sample studied by \citet{Goddard2017}, while the MaNGA sample studied in \citet{Li2018} includes galaxies in different environment, and shows large scatter of age gradients of ETGs in the same mass range, which may indicate the effects of different environments.
Although other physical processes, such as mergers, could also cause new star formation in the center of these relatively low mass ETGs in the field \citep{Yozin2012,Lagos2022}.

Similar ages of the disk and hot components in ancient infallers indicate that stars also formed on dynamically warmer orbits after falling. Star formation onto dynamically warmer orbits is possible if the gas is compacted into the inner regions of the galaxy,
resulting in a high gas density in the galaxy center after falling into the cluster. As shown in Fig.~\ref{img:posi_grd}, we also find younger stars in the very center of the hot components for a few galaxies, which is consistent with the aforementioned scenario.

\section{Conclusions}\label{sec:conclusion}
We built orbit-superposition models for 20 galaxies (15 ETGs + 5 LTGs) in the Fornax cluster observed with MUSE/VLT in the context of the Fornax3D survey. We built a population-orbit superposition model for 16 sample galaxies by coloring the orbits with age and metallicity. By simultaneously fitting the surface-brightness, stellar kinematic, age, and metallicity maps, we obtain the internal stellar orbit distribution, as well as age and metallicity distribution of each galaxy. Based on the best-fit population-orbit superposition model, we decompose each galaxy into a dynamically cold disk ($\lambda_z \ge 0.8$) and a dynamically hot non-disk component ($\lambda_z<0.8$). Then, we obtain the surface-brightness, age, and metallicity radial profiles for each component by projecting them face-on. 
We analyze the dependence of the cold disk properties on galaxy stellar mass and infall time into the Fornax cluster. Our main results are as follows:

1. We estimate the galaxy infall time into the cluster based on a tight correlation with cold-disk age calibrated with the cosmological simulation TNG 50. The infall time inferred in this way is statistically consistent with the galaxy's location in the cluster's observational phase-space diagram.

2. The ancient infallers have significantly lower luminosity fractions of the cold disk component, regardless of stellar mass. The recent and intermediate infallers have cold-disk fraction increasing as a function of stellar mass. They reach $f_{\rm cold} (<R_\mathrm{e}) \sim 0.3$ and $f_{\rm cold} (<2R_\mathrm{e}) \sim 0.4$ at $M_* \sim 10^{10}\,\mathrm{M}_{\odot}$, consistent with the CALIFA galaxies in the field.
Most ancient infallers have $f_{\rm cold} \lesssim 0.1$, which is a factor of $\sim 4 $ lower than that of recent infallers of the same stellar mass.

3. Nine of the 16 galaxies with population-orbit superposition have extended cold disks. The five least-massive galaxies have positive or flat age gradients in their cold disks. Indeed, the stars in the inner disk regions are about 2-5 Gyr younger than those in the outer disk regions. In contrast, the four more massive galaxies, which might have a lower gas content at infall, have negative age gradients in the inner disk regions and flat age profiles in the outer disk regions. 

Our results indicate that the star formation rate in disks can be significantly reduced after the galaxies fall into a cluster in general. While the star formation in the outer regions of a galaxy could stop shortly after it falls into the cluster, star formation in the inner regions should last for longer time causing the stars in the inner disk regions to be 2-5 Gyr younger than those in the outer disk regions. This is generally consistent with the scenario found for the Virgo cluster \citep{Crowl2008}, while we have been able for the first time to quantify the age difference between the inner and outer regions of cold disks.

By taking advantage of the high-quality MUSE datacubes and structure decomposition based on the population-orbit superposition models, we have been able to isolate and accurately study the dynamically cold disk component. 
This allows for an in-depth and direct comparison of the galactic structures with galaxies formed in cosmological simulations, which will help us to further understand the physical processes driving galaxy evolution.

\begin{acknowledgements}
The models and corresponding results presented in this paper would not have been possible without the VLT-MUSE data granted through the award of DDT telescope time (ESO programme 296.B-5054(A)).
LZ acknowledges the support from the National Key
R$\&$D Program of China under grant No. 2018YFA0404501, National Natural Science Foundation of China under grant No. Y945271001, and CAS Project for Young Scientists in Basic Research under grant No. YSBR-062.
GvdV acknowledges funding from the European Research Council (ERC) under the European Union's Horizon 2020 research and innovation programme under grant agreement No 724857 (Consolidator Grant ArcheoDyn).
L.C. acknowledges financial support from Comunidad de Madrid under Atracción de Talento grant 2018-T2/TIC-11612 and the Spanish Ministerio de Ciencia, Innovación y Universidades through grant PGC2018- 093499-B-I00.
J.~F-B, IMN and FP acknowledge support through the RAVET project by the grant PID2019-107427GB-C32 from the Spanish Ministry of Science, Innovation and Universities (MCIU), and through the IAC project TRACES which is partially supported through the state budget and the regional budget of the Consejer\'ia de Econom\'ia, Industria, Comercio y Conocimiento of the Canary Islands Autonomous Community.
The F3D data is based on observations collected at the European Southern Observatory under ESO programme 296.B-5054(A), and available in the ESO Science Archive Facility.
EMC is supported by MIUR grant PRIN 2017 20173ML3WW-001 and Padua University grants DOR2019-2021.
\end{acknowledgements}

% ------------------------------------------------------------------------------

\bibliographystyle{aa}
\typeout{} %IMPORTANT! Do not remove this line!
\bibliography{main}

% ------------------------------------------------------------------------------
\begin{appendix}
%\appendix

\section{Mass density MGE of all galaxies}\label{app:MGE}

\FloatBarrier
\begin{table}[H]
 \centering
    %%%%%%%%%%%%%%%%%%%%%%%%%%%%%%%%%%%%%%%%%%%%%%%%%%%%%%%%%%%%%%%%%%%%%%%%%%%%%%%%
	\begin{tabular}{c|c|c}
	 $\Sigma$ & $\sigma$ & $q$\\
	 $[\si{\Msun\per\parsec\squared}]$ & $[{\rm arcsec}]$ & \\
	\hline
    26077.2  &   0.3800  &   0.670\\
    6911.64  &    1.330  &   0.673\\
    2183.15  &    3.290  &   0.667\\
    643.410  &    8.000  &   0.576\\
    199.660  &    15.24  &   0.666\\
    60.8800  &    30.28  &   0.612\\
    24.9400  &    57.44  &   0.554\\
    6.48000  &    91.07  &   0.867\\
	\hline
	\end{tabular}
    \caption{\protect MGE parametrization of the stellar mass distribution
of FCC\,083. The central mass surface density (1), rms (2),
and axial ratio (3) of all the model Gaussians are given.}
    \label{tab:083massMGE}
\end{table}

\begin{table}[H]
 \centering
    %%%%%%%%%%%%%%%%%%%%%%%%%%%%%%%%%%%%%%%%%%%%%%%%%%%%%%%%%%%%%%%%%%%%%%%%%%%%%%%%
	\begin{tabular}{c|c|c}
	 $\Sigma$ & $\sigma$ & $q$\\
	 $[\si{\Msun\per\parsec\squared}]$ & $[{\rm arcsec}]$ & \\
	\hline
      505.05 & 0.570 &  0.894\\
      129.23 &  4.600 &  0.846\\
      39.090 &  10.40 &  0.941\\
      7.2100 &  10.48 &  0.503\\
      9.4600 &  25.18 &  0.783\\
	\hline
	\end{tabular}
    \caption{\protect MGE parametrization of the stellar mass distribution
of FCC\,119. Details are same as Table~\ref{tab:083massMGE}.}
    \label{tab:119massMGE}
\end{table}

\begin{table}[H]
 \centering
    %%%%%%%%%%%%%%%%%%%%%%%%%%%%%%%%%%%%%%%%%%%%%%%%%%%%%%%%%%%%%%%%%%%%%%%%%%%%%%%%
	\begin{tabular}{c|c|c}
	 $\Sigma$ & $\sigma$ & $q$\\
	 $[\si{\Msun\per\parsec\squared}]$ & $[{\rm arcsec}]$ & \\
	\hline
      17884.1 & 0.2900 & 0.787\\
      3400.83 &  1.020 & 0.738\\
      1153.89 &  2.030 & 0.837\\
      426.680 &  4.330 & 0.697\\
      104.830 &  7.060 & 0.990\\
      59.5300 &  9.470 & 0.462\\
      50.9900 &  13.76 & 0.977\\
      7.08000 &  30.06 & 0.943\\
	\hline
	\end{tabular}
    \caption{\protect MGE parametrization of the stellar mass distribution
of FCC\,143. Details are same as Table~\ref{tab:083massMGE}.}
    \label{tab:143massMGE}
\end{table}

\begin{table}[H]
 \centering
    %%%%%%%%%%%%%%%%%%%%%%%%%%%%%%%%%%%%%%%%%%%%%%%%%%%%%%%%%%%%%%%%%%%%%%%%%%%%%%%%
	\begin{tabular}{c|c|c}
	 $\Sigma$ & $\sigma$ & $q$\\
	 $[\si{\Msun\per\parsec\squared}]$ & $[{\rm arcsec}]$ & \\
	\hline
      54287.4  &   0.2800 & 0.871\\
      11087.4  &    1.080 & 0.845\\
      3380.12  &    2.730 & 0.896\\
      944.320  &    7.150 & 0.889\\
      288.790  &    14.53 & 0.903\\
      90.3000  &    24.10 & 0.891\\
      43.3400  &    42.76 & 0.947\\
      6.38000  &    98.90 & 0.910\\
	\hline
	\end{tabular}
    \caption{\protect MGE parametrization of the stellar mass distribution
of FCC\,147. Details are same as Table~\ref{tab:083massMGE}.}
    \label{tab:147massMGE}
\end{table}

\begin{table}[H]
 \centering
    %%%%%%%%%%%%%%%%%%%%%%%%%%%%%%%%%%%%%%%%%%%%%%%%%%%%%%%%%%%%%%%%%%%%%%%%%%%%%%%%
	\begin{tabular}{c|c|c}
	 $\Sigma$ & $\sigma$ & $q$\\
	 $[\si{\Msun\per\parsec\squared}]$ & $[{\rm arcsec}]$ & \\
	\hline
      35949.7 & 0.2500 & 0.673\\
      3819.35 & 0.8500 & 0.717\\
      1382.12 &  2.010 & 0.629\\
      704.410 &  4.320 & 0.552\\
      385.580 &  15.15 & 0.385\\
      59.4800 &  31.93 & 0.436\\
      4.55000 &  67.21 & 0.664\\
	\hline
	\end{tabular}
    \caption{\protect MGE parametrization of the stellar mass distribution
of FCC\,148. Details are same as Table~\ref{tab:083massMGE}.}
    \label{tab:148massMGE}
\end{table}

\begin{table}[H]
 \centering
    %%%%%%%%%%%%%%%%%%%%%%%%%%%%%%%%%%%%%%%%%%%%%%%%%%%%%%%%%%%%%%%%%%%%%%%%%%%%%%%%
	\begin{tabular}{c|c|c}
	 $\Sigma$ & $\sigma$ & $q$\\
	 $[\si{\Msun\per\parsec\squared}]$ & $[{\rm arcsec}]$ & \\
	\hline
      22889.8  & 0.2800 &  0.534\\
      1889.26  &  1.300 &  0.582\\
      732.830  &  4.810 &  0.487\\
      165.960  &  8.680 &  0.672\\
      1454.73  &  17.72 & 0.0660\\
      269.200  &  23.37 &  0.152\\
      77.8500  &  34.30 &  0.264\\
      10.3700  &  48.75 &  0.452\\
	\hline
	\end{tabular}
    \caption{\protect MGE parametrization of the stellar mass distribution
of FCC\,153. Details are same as Table~\ref{tab:083massMGE}.}
    \label{tab:153massMGE}
\end{table}

\begin{table}[H]
 \centering
    %%%%%%%%%%%%%%%%%%%%%%%%%%%%%%%%%%%%%%%%%%%%%%%%%%%%%%%%%%%%%%%%%%%%%%%%%%%%%%%%
	\begin{tabular}{c|c|c}
	 $\Sigma$ & $\sigma$ & $q$\\
	 $[\si{\Msun\per\parsec\squared}]$ & $[{\rm arcsec}]$ & \\
	\hline
      41054.9  &   0.2700 &  0.990\\
      6726.13  &   0.8900 &  0.990\\
      1976.01  &    2.240 &  0.990\\
      878.060  &    5.350 &  0.990\\
      456.600  &    10.56 &  0.990\\
      154.630  &    19.45 &  0.990\\
      80.1700  &    35.79 &  0.990\\
      7.38000  &    84.77 &  0.862\\
	\hline
	\end{tabular}
    \caption{\protect MGE parametrization of the stellar mass distribution
of FCC\,161. Details are same as Table~\ref{tab:083massMGE}.}
    \label{tab:161massMGE}
\end{table}

\begin{table}[H]
 \centering
    %%%%%%%%%%%%%%%%%%%%%%%%%%%%%%%%%%%%%%%%%%%%%%%%%%%%%%%%%%%%%%%%%%%%%%%%%%%%%%%%
	\begin{tabular}{c|c|c}
	 $\Sigma$ & $\sigma$ & $q$\\
	 $[\si{\Msun\per\parsec\squared}]$ & $[{\rm arcsec}]$ & \\
	\hline
      16508.58  &   0.584  & 0.864\\
       2537.55  &   1.917  & 0.999\\
       2872.85  &   5.143  & 0.727\\
        463.50  &   13.820  & 0.807\\
        485.79  &   14.448  & 0.541\\
        120.08  &   43.797  & 0.269\\
        150.33  &   47.212  & 0.524\\
         27.77  &   58.192  & 0.972\\
         10.88  &  120.764  & 0.542\\
	\hline
	\end{tabular}
    \caption{\protect MGE parametrization of the stellar mass distribution
of FCC\,167. Details are same as Table~\ref{tab:083massMGE}.}
    \label{tab:167massMGE}
\end{table}

\begin{table}[H]
 \centering
    %%%%%%%%%%%%%%%%%%%%%%%%%%%%%%%%%%%%%%%%%%%%%%%%%%%%%%%%%%%%%%%%%%%%%%%%%%%%%%%%
	\begin{tabular}{c|c|c}
	 $\Sigma$ & $\sigma$ & $q$\\
	 $[\si{\Msun\per\parsec\squared}]$ & $[{\rm arcsec}]$ & \\
	\hline
    36302.10 & 0.15000  & 0.72549\\
    11913.97 & 0.37207  & 0.35867\\
    12273.34 & 0.37328  & 0.92307\\
    8826.98  & 0.68704  & 0.32285\\
    9179.47  & 0.97482  & 0.79328\\
    4462.54  & 1.90931  & 0.74662\\
    3495.96  & 2.83135  & 0.74571\\
    988.46   & 5.90353  & 0.59377\\
    590.84   & 6.86747  & 0.75000\\
    360.95   & 7.31068  & 0.97943\\
    26.06    & 16.27466 & 0.99990\\
    434.80   & 16.30921 & 0.18070\\
    162.51   & 26.67680 & 0.18070\\
    162.51   & 26.67696 & 0.18070\\
    162.51   & 26.67736 & 0.18070\\
    4.03     & 32.84468 & 0.99990\\
    21.70    & 36.36404 & 0.25361\\
    59.66    & 36.80145 & 0.25867\\
    13.12    & 46.28263 & 0.39076\\
	\hline
	\end{tabular}
    \caption{\protect MGE parametrization of the stellar mass distribution
of FCC\,170. Details are same as Table~\ref{tab:083massMGE}.}
    \label{tab:170massMGE}
\end{table}

\begin{table}
 \centering
    %%%%%%%%%%%%%%%%%%%%%%%%%%%%%%%%%%%%%%%%%%%%%%%%%%%%%%%%%%%%%%%%%%%%%%%%%%%%%%%%
	\begin{tabular}{c|c|c}
	 $\Sigma$ & $\sigma$ & $q$\\
	 $[\si{\Msun\per\parsec\squared}]$ & $[{\rm arcsec}]$ & \\
	\hline
      8967.82  &   0.4600 &  0.877\\
      1007.78  &    2.500 &  0.645\\
      158.910  &    11.75 &  0.493\\
      316.810  &    23.26 &  0.145\\
      78.0600  &    30.28 &  0.315\\
      15.6900  &    47.55 &  0.510\\
	\hline
	\end{tabular}
    \caption{\protect MGE parametrization of the stellar mass distribution
of FCC\,177. Details are same as Table~\ref{tab:083massMGE}.}
    \label{tab:177massMGE}
\end{table}

\begin{table}
 \centering
    %%%%%%%%%%%%%%%%%%%%%%%%%%%%%%%%%%%%%%%%%%%%%%%%%%%%%%%%%%%%%%%%%%%%%%%%%%%%%%%%
	\begin{tabular}{c|c|c}
	 $\Sigma$ & $\sigma$ & $q$\\
	 $[\si{\Msun\per\parsec\squared}]$ & $[{\rm arcsec}]$ & \\
	\hline
      14717.035 &  1.318  & 0.582\\
       2904.596 &  3.360  & 0.851\\
        512.611 & 10.647  & 0.658\\
        281.527 & 37.423  & 0.336\\
         28.881 & 47.429  & 0.493\\
	\hline
	\end{tabular}
    \caption{\protect MGE parametrization of the stellar mass distribution
of FCC\,179. Details are same as Table~\ref{tab:083massMGE}.}
    \label{tab:179massMGE}
\end{table}

\begin{table}
 \centering
    %%%%%%%%%%%%%%%%%%%%%%%%%%%%%%%%%%%%%%%%%%%%%%%%%%%%%%%%%%%%%%%%%%%%%%%%%%%%%%%%
	\begin{tabular}{c|c|c}
	 $\Sigma$ & $\sigma$ & $q$\\
	 $[\si{\Msun\per\parsec\squared}]$ & $[{\rm arcsec}]$ & \\
	\hline
      2579.35  &   0.2400 & 0.600\\
      641.810  &   0.8400 & 0.684\\
      549.780  &    2.370 & 0.516\\
      400.120  &    3.290 & 0.862\\
      68.7500  &    6.680 & 0.990\\
      42.9800  &    12.07 & 0.996\\
      5.95000  &    24.20 & 0.990\\
	\hline
	\end{tabular}
    \caption{\protect MGE parametrization of the stellar mass distribution
of FCC\,182. Details are same as Table~\ref{tab:083massMGE}.}
    \label{tab:182massMGE}
\end{table}

% \begin{table}
%  \centering
%     %%%%%%%%%%%%%%%%%%%%%%%%%%%%%%%%%%%%%%%%%%%%%%%%%%%%%%%%%%%%%%%%%%%%%%%%%%%%%%%%
% 	\begin{tabular}{c|c|c}
% 	 $\Sigma$ & $\sigma$ & $q$\\
% 	 $[\si{\Msun\per\parsec\squared}]$ & $[{\rm arcsec}]$ & \\
% 	\hline
%       34290.0 & 0.4400 & 0.837\\
%       9576.25 &  1.810 & 0.995\\
%       2793.17 &  4.620 & 0.990\\
%       842.060 &  11.05 & 0.636\\
%       142.500 &  22.39 & 0.945\\
%       78.9400 &  44.42 & 0.990\\
%       8.39000 &  91.46 & 0.990\\
% 	\hline
% 	\end{tabular}
%     \caption{\protect Same as in Table~\ref{tab:083massMGE}, but for FCC~184.}
%     \label{tab:184massMGE}
% \end{table}

\begin{table}
 \centering
    %%%%%%%%%%%%%%%%%%%%%%%%%%%%%%%%%%%%%%%%%%%%%%%%%%%%%%%%%%%%%%%%%%%%%%%%%%%%%%%%
	\begin{tabular}{c|c|c}
	 $\Sigma$ & $\sigma$ & $q$\\
	 $[\si{\Msun\per\parsec\squared}]$ & $[{\rm arcsec}]$ & \\
	\hline
      25539.2 & 0.2600 & 0.841\\
      5431.22 &  1.010 & 0.775\\
      1957.88 &  1.830 & 0.907\\
      876.430 &  3.080 & 0.990\\
      307.980 &  5.340 & 0.990\\
      186.070 &  8.180 & 0.990\\
      51.7400 &  15.45 & 0.990\\
      9.10000 &  45.49 & 0.990\\
	\hline
	\end{tabular}
    \caption{\protect MGE parametrization of the stellar mass distribution
of FCC\,249. Details are same as Table~\ref{tab:083massMGE}.}
    \label{tab:249massMGE}
\end{table}

\begin{table}
 \centering
    %%%%%%%%%%%%%%%%%%%%%%%%%%%%%%%%%%%%%%%%%%%%%%%%%%%%%%%%%%%%%%%%%%%%%%%%%%%%%%%%
	\begin{tabular}{c|c|c}
	 $\Sigma$ & $\sigma$ & $q$\\
	 $[\si{\Msun\per\parsec\squared}]$ & $[{\rm arcsec}]$ & \\
	\hline
      3301.56 & 0.3100 & 0.510\\
      309.560 &  1.910 & 0.699\\
      472.760 &  6.970 & 0.322\\
      180.810 &  12.80 & 0.405\\
      48.2500 &  21.67 & 0.591\\
      7.15000 &  43.10 & 0.855\\
	\hline
	\end{tabular}
    \caption{\protect MGE parametrization of the stellar mass distribution
of FCC\,255. Details are same as Table~\ref{tab:083massMGE}.}
    \label{tab:255massMGE}
\end{table}

\begin{table}
 \centering
    %%%%%%%%%%%%%%%%%%%%%%%%%%%%%%%%%%%%%%%%%%%%%%%%%%%%%%%%%%%%%%%%%%%%%%%%%%%%%%%%
	\begin{tabular}{c|c|c}
	 $\Sigma$ & $\sigma$ & $q$\\
	 $[\si{\Msun\per\parsec\squared}]$ & $[{\rm arcsec}]$ & \\
	\hline
        199.583 &  2.438 & 0.999\\
        244.042 &  6.932 & 0.471\\
        127.639 & 12.327 & 0.501\\
         42.153 & 24.796 & 0.522\\
          1.213 & 71.410 & 0.571\\
	\hline
	\end{tabular}
    \caption{\protect MGE parametrization of the stellar mass distribution
of FCC\,263. Details are same as Table~\ref{tab:083massMGE}.}
    \label{tab:263massMGE}
\end{table}

\begin{table}
 \centering
    %%%%%%%%%%%%%%%%%%%%%%%%%%%%%%%%%%%%%%%%%%%%%%%%%%%%%%%%%%%%%%%%%%%%%%%%%%%%%%%%
	\begin{tabular}{c|c|c}
	 $\Sigma$ & $\sigma$ & $q$\\
	 $[\si{\Msun\per\parsec\squared}]$ & $[{\rm arcsec}]$ & \\
	\hline
      66551.4 & 0.26000 & 0.690\\
      15406.9 &  1.0000 & 0.660\\
      4284.05 &  2.2900 & 0.720\\
      1621.34 &  4.8800 & 0.728\\
      576.060 &  11.500 & 0.695\\
      222.270 &  24.030 & 0.696\\
      49.4000 &  50.390 & 0.688\\
      12.6200 &  107.64 & 0.769\\
	\hline
	\end{tabular}
    \caption{\protect MGE parametrization of the stellar mass distribution
of FCC\,276. Details are same as Table~\ref{tab:083massMGE}.}
    \label{tab:276massMGE}
\end{table}

\begin{table}
 \centering
    %%%%%%%%%%%%%%%%%%%%%%%%%%%%%%%%%%%%%%%%%%%%%%%%%%%%%%%%%%%%%%%%%%%%%%%%%%%%%%%%
	\begin{tabular}{c|c|c}
	 $\Sigma$ & $\sigma$ & $q$\\
	 $[\si{\Msun\per\parsec\squared}]$ & $[{\rm arcsec}]$ & \\
	\hline
       2494.579 &  0.355 & 0.794\\
        558.645 &  2.237 & 0.836\\
        225.838 &  6.090 & 0.752\\
        102.083 & 32.165 & 0.614\\
         60.018 & 46.855 & 0.711\\
	\hline
	\end{tabular}
    \caption{\protect MGE parametrization of the stellar mass distribution
of FCC\,290. Details are same as Table~\ref{tab:083massMGE}.}
    \label{tab:290massMGE}
\end{table}

\begin{table}
 \centering
    %%%%%%%%%%%%%%%%%%%%%%%%%%%%%%%%%%%%%%%%%%%%%%%%%%%%%%%%%%%%%%%%%%%%%%%%%%%%%%%%
	\begin{tabular}{c|c|c}
	 $\Sigma$ & $\sigma$ & $q$\\
	 $[\si{\Msun\per\parsec\squared}]$ & $[{\rm arcsec}]$ & \\
	\hline
      8660.98 & 0.2500 & 0.482\\
      1943.09 &  2.390 & 0.414\\
      954.810 &  3.660 & 0.551\\
      249.790 &  7.690 & 0.561\\
      68.4400 &  13.88 & 0.850\\
      10.9000 &  30.71 & 0.870\\
	\hline
	\end{tabular}
    \caption{\protect MGE parametrization of the stellar mass distribution
of FCC\,301. Details are same as Table~\ref{tab:083massMGE}.}
    \label{tab:301massMGE}
\end{table}

\begin{table}
 \centering
    %%%%%%%%%%%%%%%%%%%%%%%%%%%%%%%%%%%%%%%%%%%%%%%%%%%%%%%%%%%%%%%%%%%%%%%%%%%%%%%%
	\begin{tabular}{c|c|c}
	 $\Sigma$ & $\sigma$ & $q$\\
	 $[\si{\Msun\per\parsec\squared}]$ & $[{\rm arcsec}]$ & \\
	\hline
         91.394 &  0.920 & 0.749\\
        184.543 & 11.217 & 0.382\\
         55.719 & 32.469 & 0.345\\
          6.282 & 91.447 & 0.333\\
	\hline
	\end{tabular}
    \caption{\protect MGE parametrization of the stellar mass distribution
of FCC\,308. Details are same as Table~\ref{tab:083massMGE}.}
    \label{tab:308massMGE}
\end{table}

% \begin{table}
%  \centering
%     %%%%%%%%%%%%%%%%%%%%%%%%%%%%%%%%%%%%%%%%%%%%%%%%%%%%%%%%%%%%%%%%%%%%%%%%%%%%%%%%
% 	\begin{tabular}{c|c|c}
% 	 Σ\Sigma & σ\sigma & qq\\
% 	 [\si\Msun\per\parsec\squared][\si{\Msun\per\parsec\squared}] & [arcsec][{\rm arcsec}] & \\
% 	\hline
%       5335.52 & 0.3800 & 0.925\\
%       858.590 &  1.760 & 0.792\\
%       349.760 &  4.900 & 0.810\\
%       38.7900 &  14.54 & 0.175\\
%       162.480 &  15.86 & 0.383\\
%       70.8500 &  19.94 & 0.940\\
%       11.2700 &  54.12 & 0.990\\
% 	\end{tabular}
%     \caption{\protect Same as in Table~???\ref{tab:083massMGE}, but for FCC~310.}
%     \label{tab:310massMGE}
% \end{table}

\begin{table}
 \centering
    %%%%%%%%%%%%%%%%%%%%%%%%%%%%%%%%%%%%%%%%%%%%%%%%%%%%%%%%%%%%%%%%%%%%%%%%%%%%%%%%
	\begin{tabular}{c|c|c}
	 $\Sigma$ & $\sigma$ & $q$\\
	 $[\si{\Msun\per\parsec\squared}]$ & $[{\rm arcsec}]$ & \\
	\hline
        374.086   &   0.917 & 0.665\\
        284.956   &   7.942 & 0.342\\
        243.135   &  31.433 & 0.184\\
         67.028   &  62.940 & 0.222\\
          9.274   &  151.033 & 0.260\\
	\hline
	\end{tabular}
    \caption{\protect MGE parametrization of the stellar mass distribution
of FCC\,312. Details are same as Table~\ref{tab:083massMGE}.}
    \label{tab:312massMGE}
\end{table}
\FloatBarrier

\section{Best-fit models of all galaxies}\label{app:profiles}
\begin{figure*}
    \centering{
        \includegraphics[width=1.5\columnwidth, clip=true, trim=80 0 60 0]{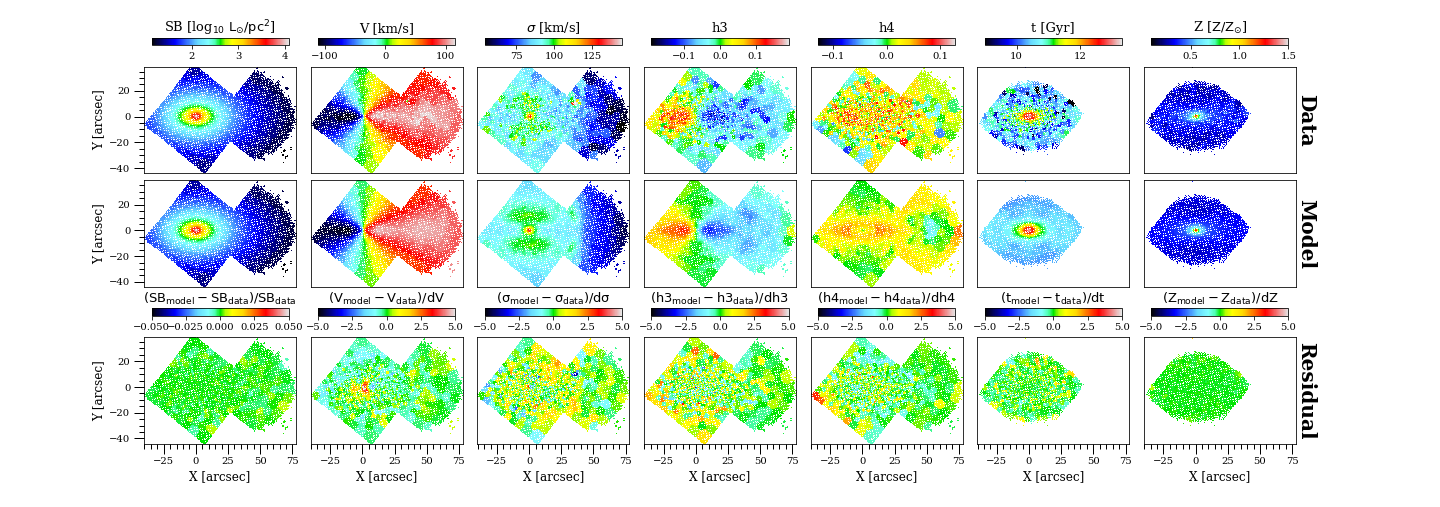}}
        \includegraphics[width=1.05\columnwidth, clip=true, trim=20 0 20 0]{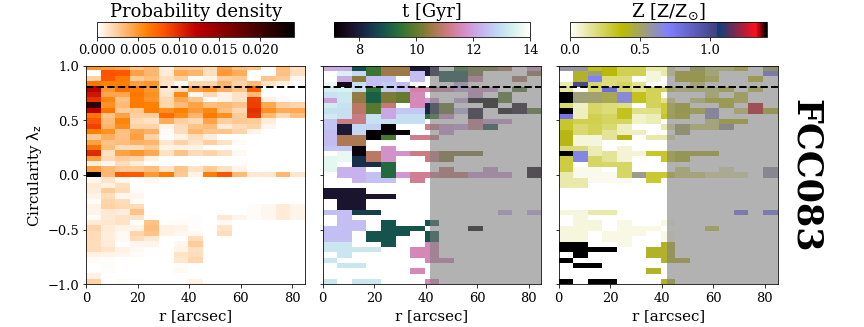}
    \caption{
    Best-fit population-orbit superposition model (top panels) and orbital decomposition (bottom panels) of FCC\,083. The content of the top and bottom panels is the same as in Figs.~\ref{img:fitting177} and \ref{phasespace177}, respectively.}
    \label{img:fitting083}
\end{figure*}

% \begin{figure*}
%     \centering{
%         \includegraphics[width=1.5\columnwidth, clip=true, trim=0 0 0 0]{{Figures/FCC113_SN40new6_fitting}.png}}
%         \includegraphics[width=0.5\columnwidth, clip=true, trim=0 0 0 0]{{Figures/FCC113_SN40new6_phase-space}.png}
%     \caption{\DD{LTG FCC~113. Similar to Fig.~???\ref{img:fitting153} and Fig.~???\ref{phasespace153} as we show for FCC 153, but we do not show the age and metallicity.}}
%     \label{img:fitting113}
% \end{figure*}

\begin{figure*}
    \centering{
        \includegraphics[width=1.5\columnwidth, clip=true, trim=80 0 60 0]{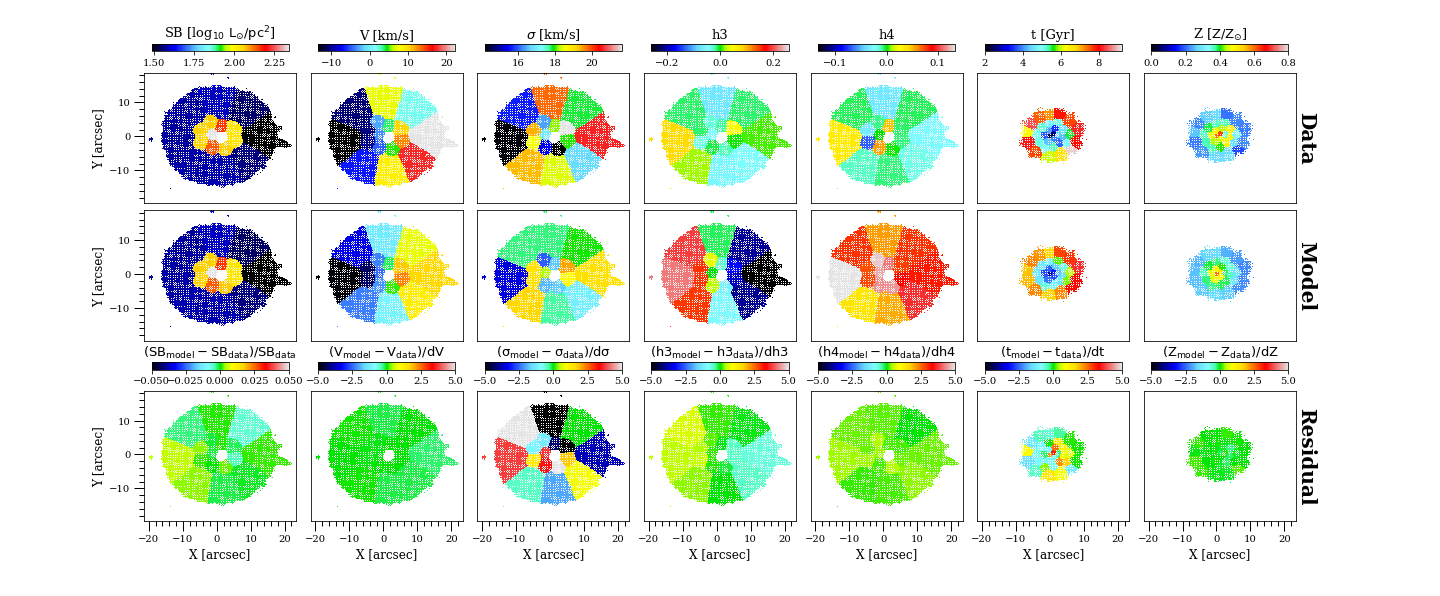}}
        \includegraphics[width=1.05\columnwidth, clip=true, trim=20 0 20 0]{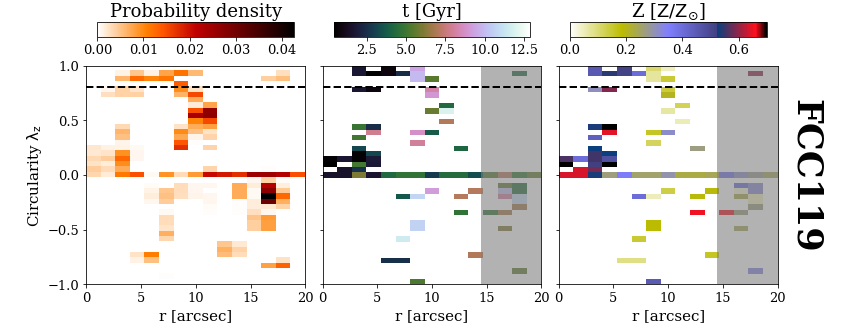}
    \caption{Best-fit population-orbit superposition model (top panels) and orbital decomposition (bottom panels) of FCC\,119. The content of the top and bottom panels is the same as in Figs.~\ref{img:fitting177} and \ref{phasespace177}, respectively.}
    \label{img:fitting119}
\end{figure*}

\begin{figure*}
    \centering{
        \includegraphics[width=1.5\columnwidth, clip=true, trim=80 0 60 0]{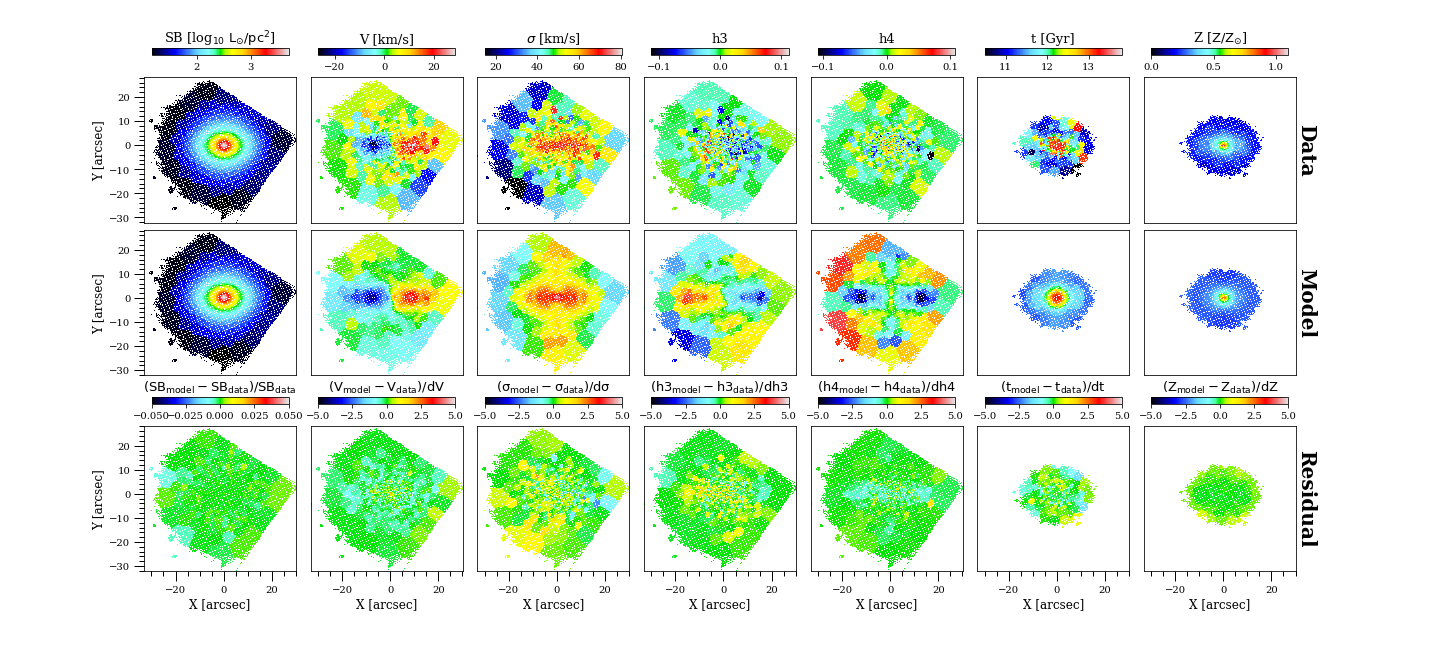}}
        \includegraphics[width=1.05\columnwidth, clip=true, trim=20 0 20 0]{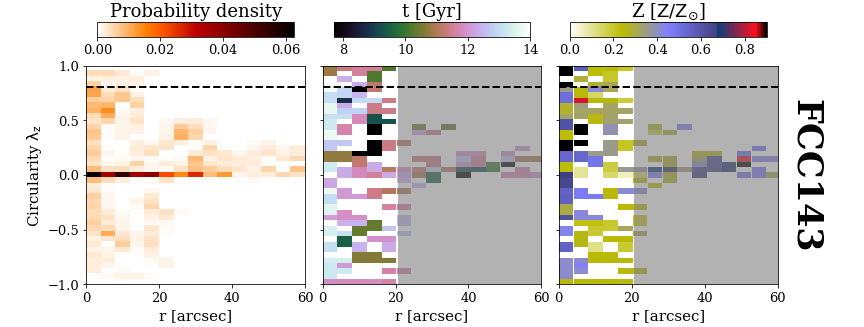}
    \caption{Best-fit population-orbit superposition model (top panels) and orbital decomposition (bottom panels) of FCC\,143. The content of the top and bottom panels is the same as in Figs.~\ref{img:fitting177} and \ref{phasespace177}, respectively.}
    \label{img:fitting143}
\end{figure*}

\begin{figure*}
    \centering{
        \includegraphics[width=1.5\columnwidth, clip=true, trim=80 0 60 0]{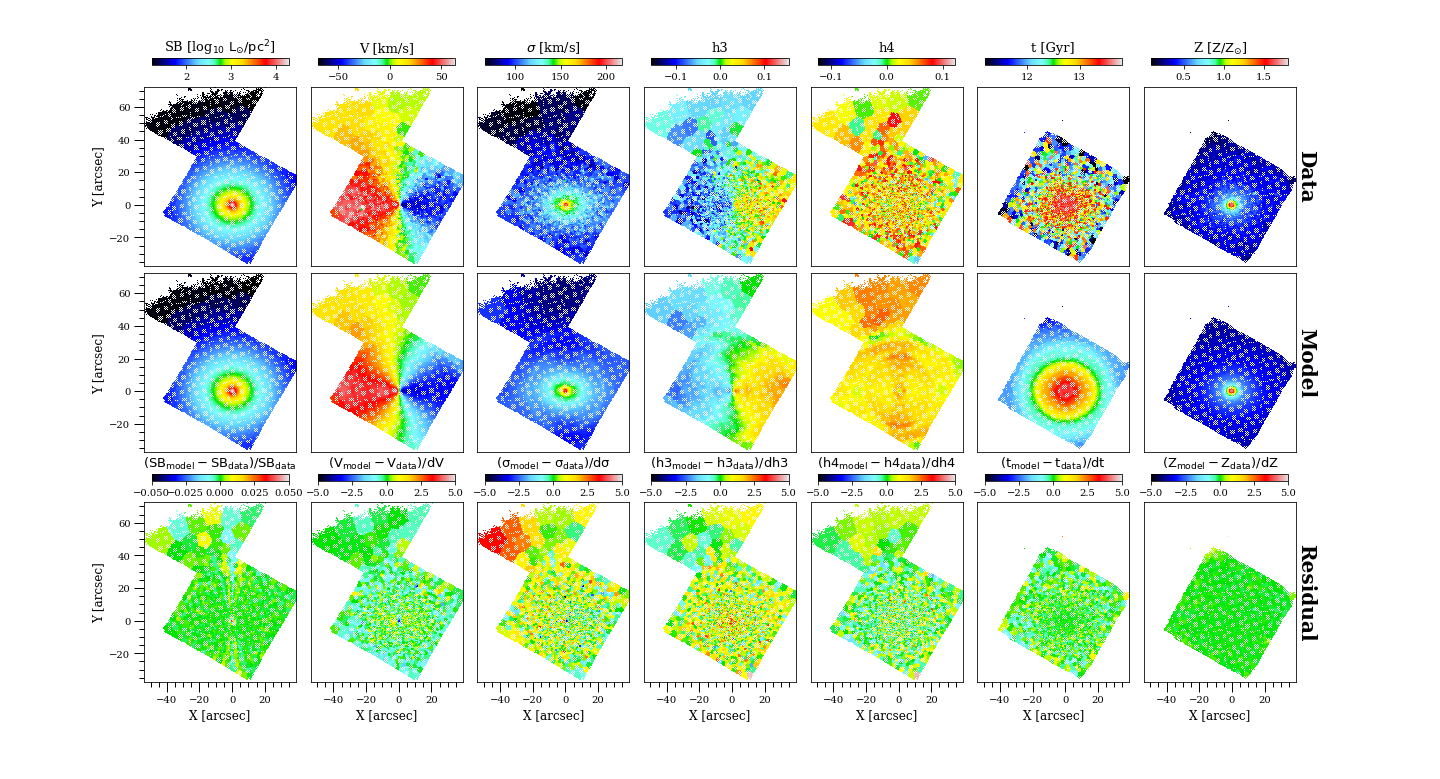}}
        \includegraphics[width=1.05\columnwidth, clip=true, trim=20 0 20 0]{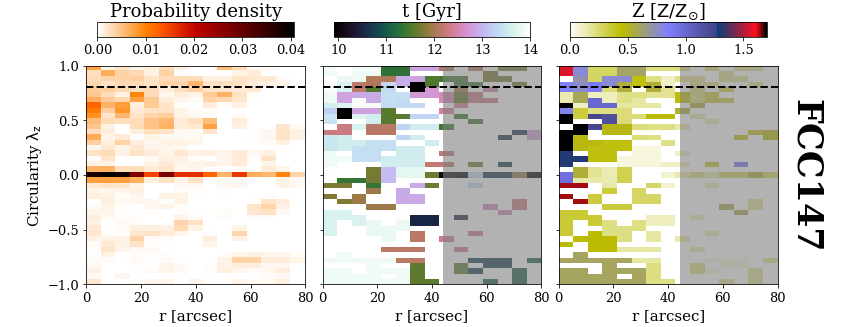}
    \caption{Best-fit population-orbit superposition model (top panels) and orbital decomposition (bottom panels) of FCC\,147. The content of the top and bottom panels is the same as in Figs.~\ref{img:fitting177} and \ref{phasespace177}, respectively.}
    \label{img:fitting147}
\end{figure*}

\begin{figure*}
    \centering{
        \includegraphics[width=1.5\columnwidth, clip=true, trim=80 0 60 0]{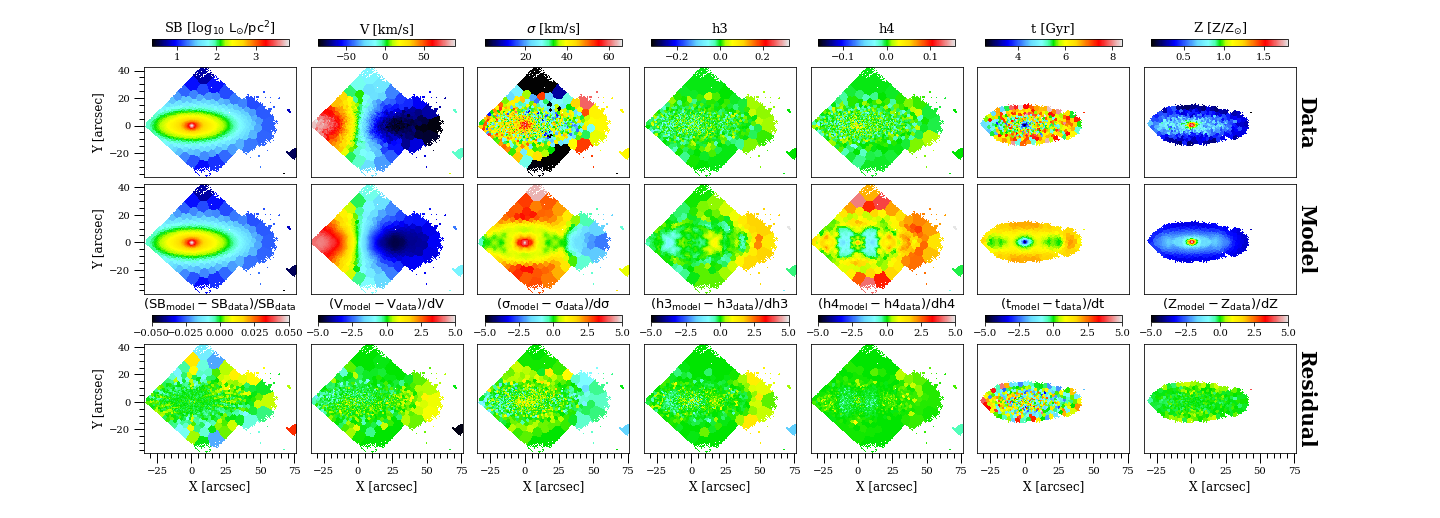}}
        \includegraphics[width=1.05\columnwidth, clip=true, trim=20 0 20 0]{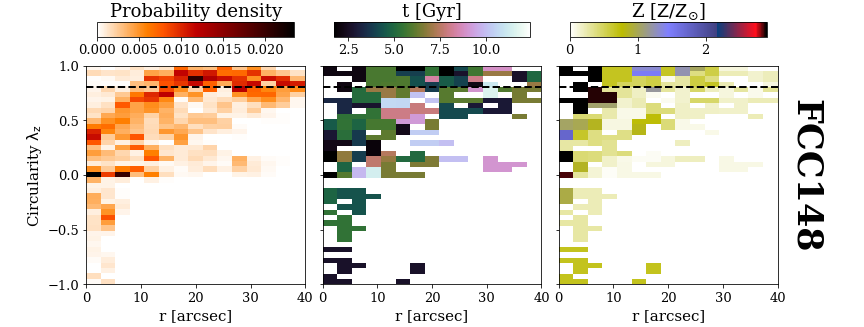}
    \caption{Best-fit population-orbit superposition model (top panels) and orbital decomposition (bottom panels) of FCC\,148. The content of the top and bottom panels is the same as in Figs.~\ref{img:fitting177} and \ref{phasespace177}, respectively.}
    \label{img:fitting148}
\end{figure*}

\begin{figure*}
    \centering{
        \includegraphics[width=1.5\columnwidth, clip=true, trim=80 0 60 0]{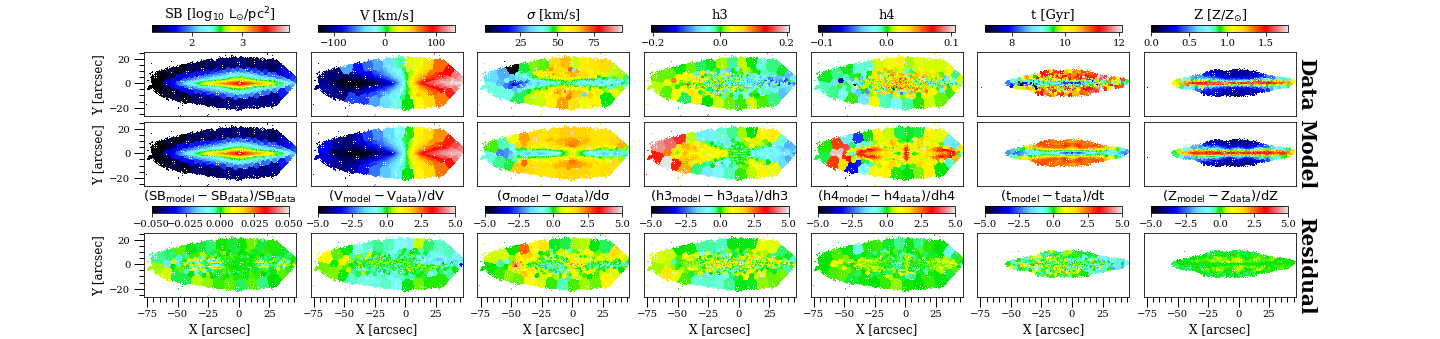}}
        \includegraphics[width=1.05\columnwidth, clip=true, trim=20 0 20 0]{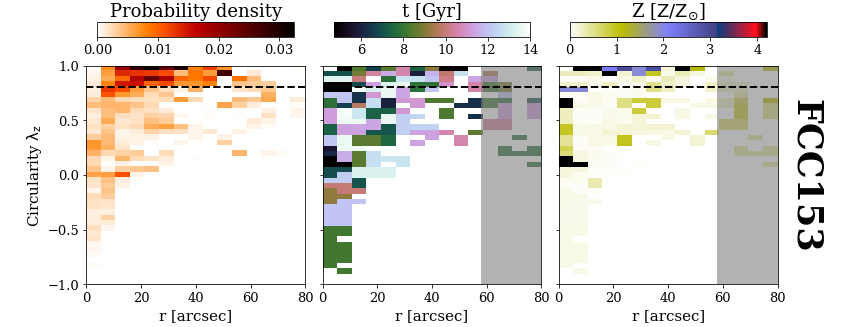}
    \caption{Best-fit population-orbit superposition model (top panels) and orbital decomposition (bottom panels) of FCC\,153. The content of the top and bottom panels is the same as in Figs.~\ref{img:fitting177} and \ref{phasespace177}, respectively.}
    \label{img:fitting153}
\end{figure*}

\begin{figure*}
    \centering{
        \includegraphics[width=1.5\columnwidth, clip=true, trim=80 0 60 0]{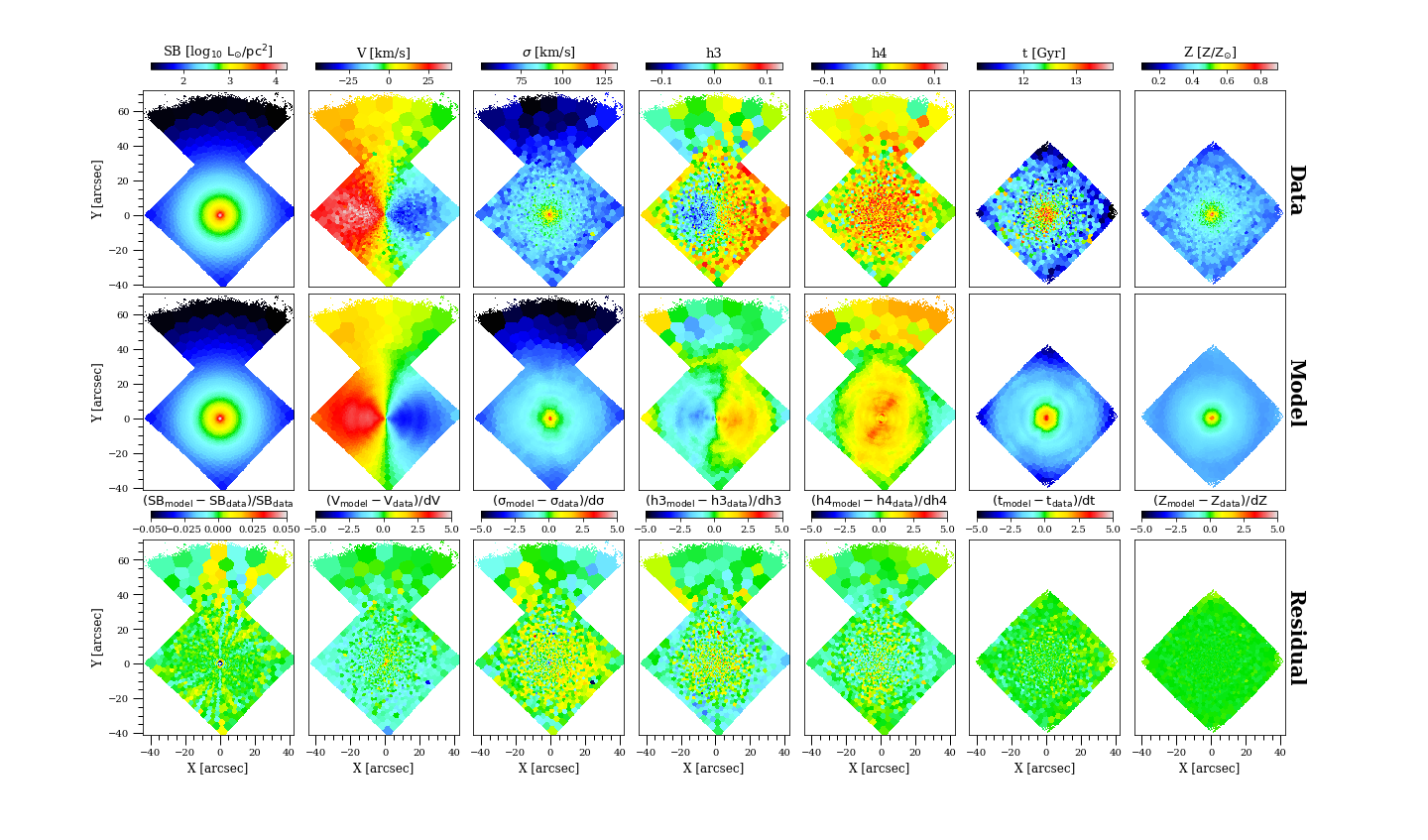}}
        \includegraphics[width=1.05\columnwidth, clip=true, trim=20 0 20 0]{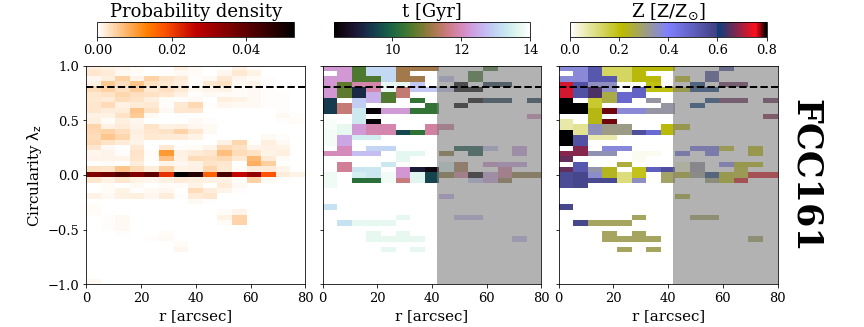}
    \caption{Best-fit population-orbit superposition model (top panels) and orbital decomposition (bottom panels) of FCC\,161. The content of the top and bottom panels is the same as in Figs.~\ref{img:fitting177} and \ref{phasespace177}, respectively.}
    \label{img:fitting161}
\end{figure*}

\begin{figure*}
    \centering{
        \includegraphics[width=1.5\columnwidth, clip=true, trim=80 0 60 0]{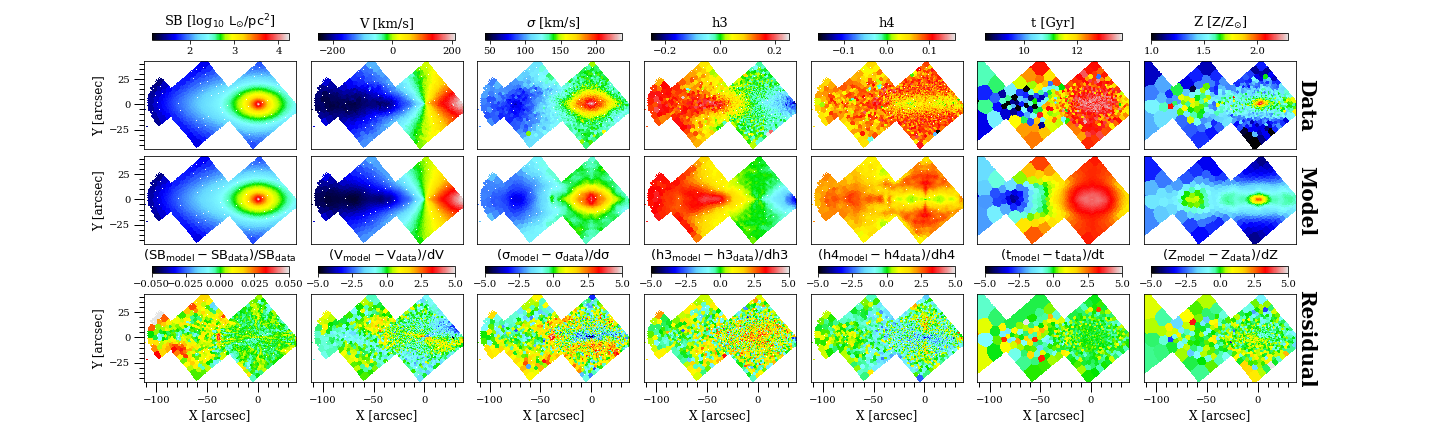}}
        \includegraphics[width=1.05\columnwidth, clip=true, trim=20 0 20 0]{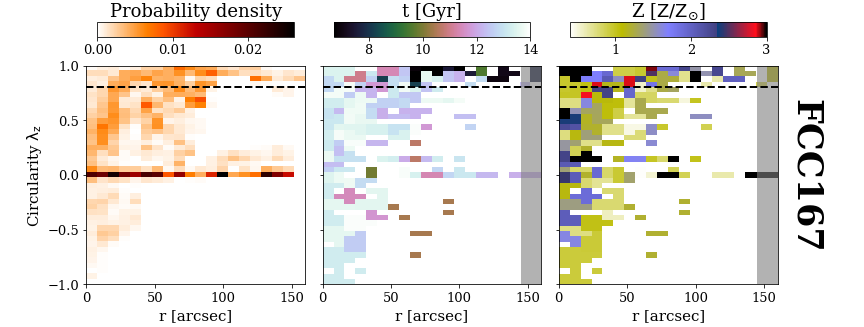}
    \caption{Best-fit population-orbit superposition model (top panels) and orbital decomposition (bottom panels) of FCC\,167. The content of the top and bottom panels is the same as in Figs.~\ref{img:fitting177} and \ref{phasespace177}, respectively.}
    \label{img:fitting167}
\end{figure*}

\begin{figure*}
    \centering{
        \includegraphics[width=1.5\columnwidth, clip=true, trim=80 0 60 0]{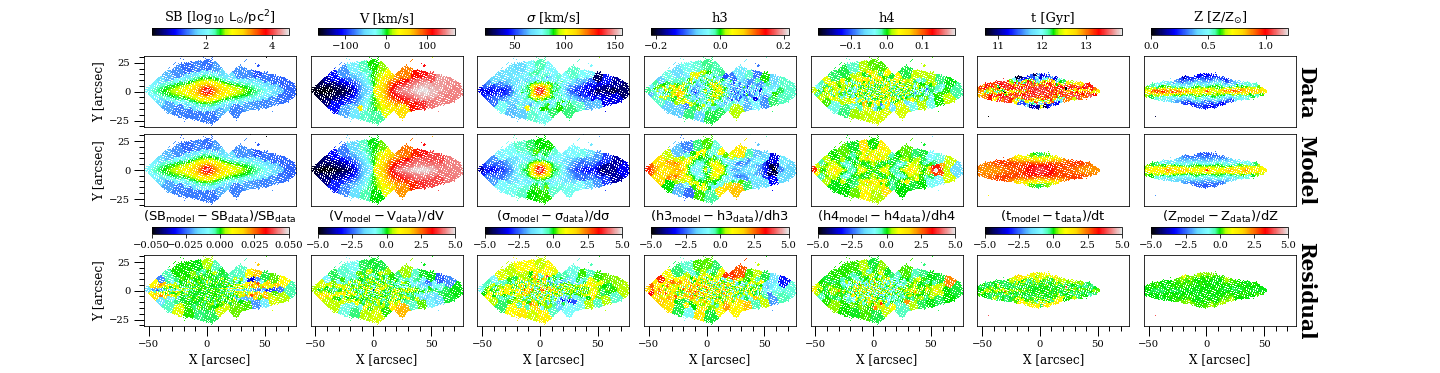}}
        \includegraphics[width=1.05\columnwidth, clip=true, trim=20 0 20 0]{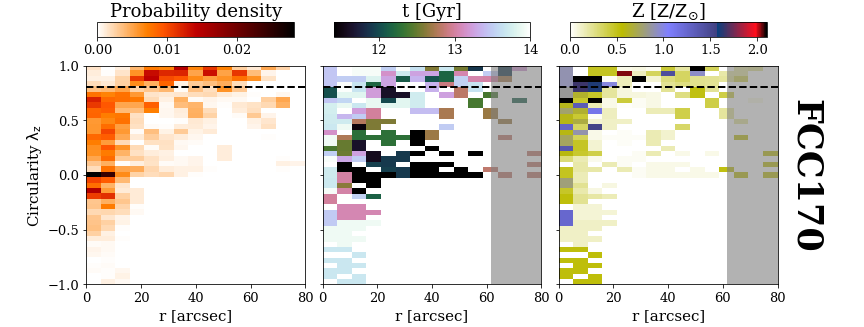}
    \caption{Best-fit population-orbit superposition model (top panels) and orbital decomposition (bottom panels) of FCC\,170. The content of the top and bottom panels is the same as in Figs.~\ref{img:fitting177} and \ref{phasespace177}, respectively.}
    \label{img:fitting170}
\end{figure*}

\begin{figure*}
    \centering{
        \includegraphics[width=1.5\columnwidth, clip=true, trim=80 0 60 0]{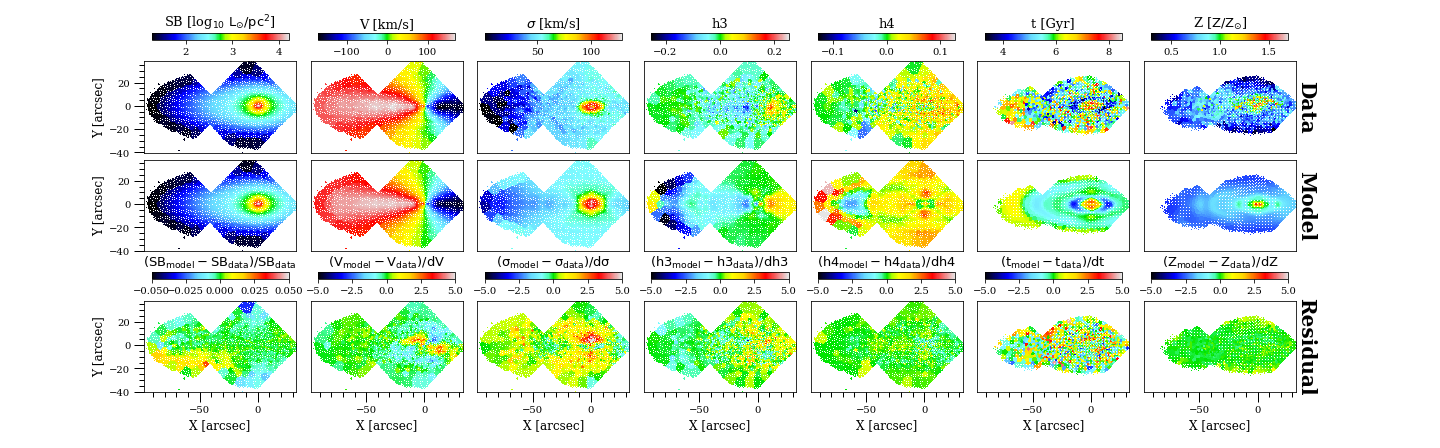}}
        \includegraphics[width=1.05\columnwidth, clip=true, trim=20 0 20 0]{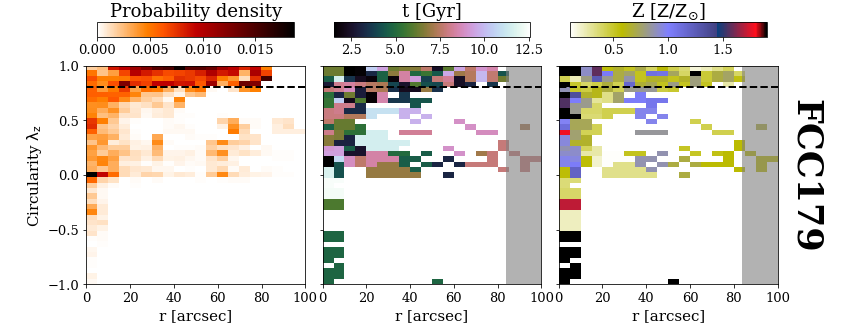}
    \caption{Best-fit population-orbit superposition model (top panels) and orbital decomposition (bottom panels) of FCC\,179. The content of the top and bottom panels is the same as in Figs.~\ref{img:fitting177} and \ref{phasespace177}, respectively. We find a possible weak bar structure at |X| $\lesssim20$ arcsec and |Y| $\lesssim 10$ arcsec.}
    \label{img:fitting179}
\end{figure*}

\begin{figure*}
    \centering{
        \includegraphics[width=1.5\columnwidth, clip=true, trim=80 0 60 0]{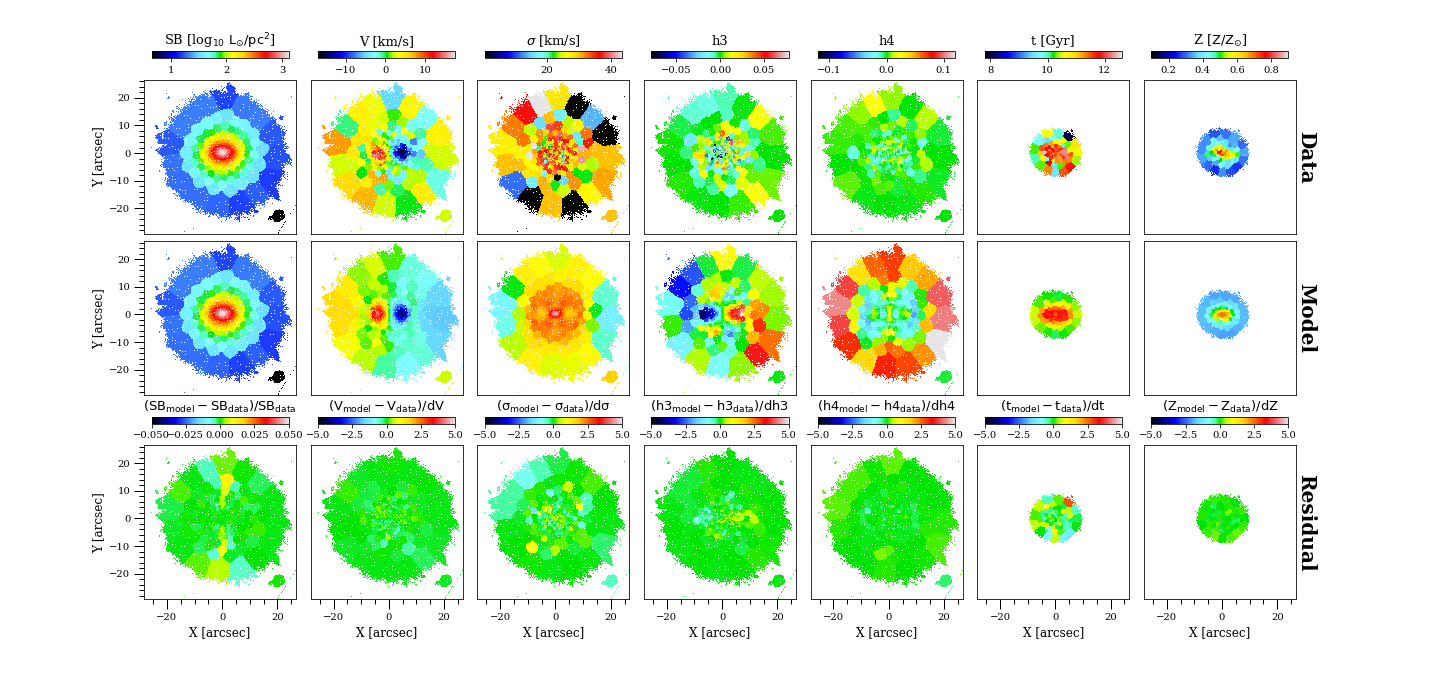}}
        \includegraphics[width=1.05\columnwidth, clip=true, trim=20 0 20 0]{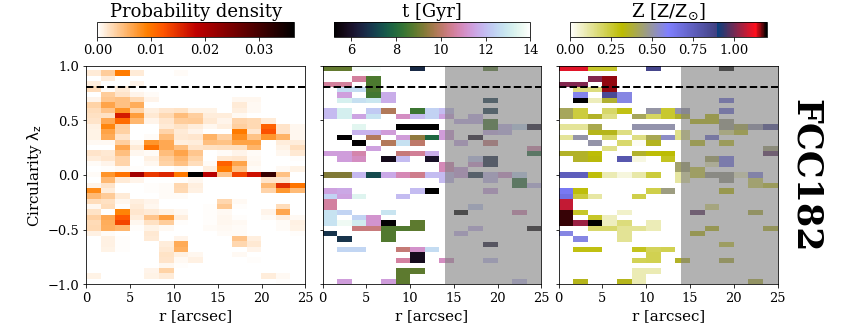}
    \caption{Best-fit population-orbit superposition model (top panels) and orbital decomposition (bottom panels) of FCC\,182. The content of the top and bottom panels is the same as in Figs.~\ref{img:fitting177} and \ref{phasespace177}, respectively.}
    \label{img:fitting182}
\end{figure*}

% \begin{figure*}
%     \centering{
%         \includegraphics[width=1.5\columnwidth, clip=true, trim=80 0 60 0]{{Figures/FCC184_SN200_fitting}.png}}
%         \includegraphics[width=1.05\columnwidth, clip=true, trim=20 0 20 0]{{Figures/FCC184_SN200_phase-space}.png}
%     \caption{Same as in Fig.~?????????\ref{img:fitting083}, but for FCC~184. A strongly bared galaxy.}
%     \label{img:fitting184}
% \end{figure*}

\begin{figure*}
    \centering{
        \includegraphics[width=1.5\columnwidth, clip=true, trim=80 0 60 0]{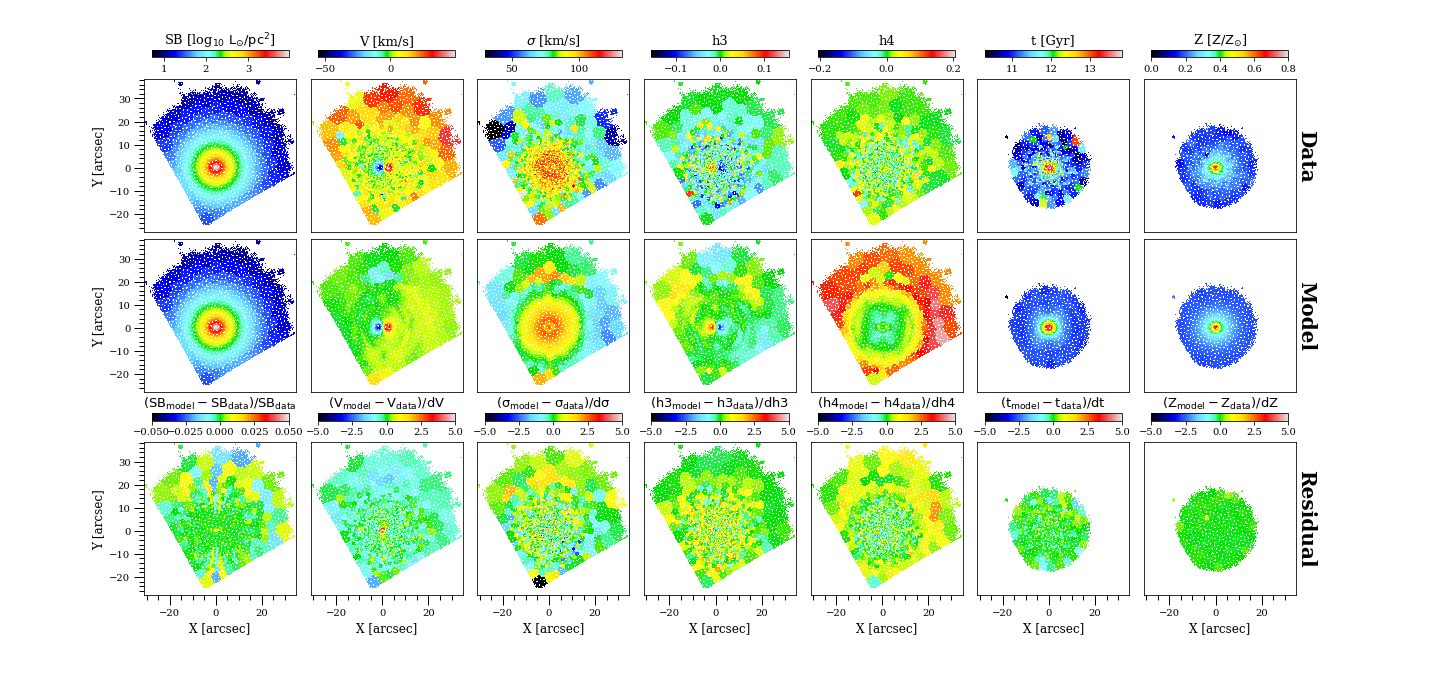}}
        \includegraphics[width=1.05\columnwidth, clip=true, trim=20 0 20 0]{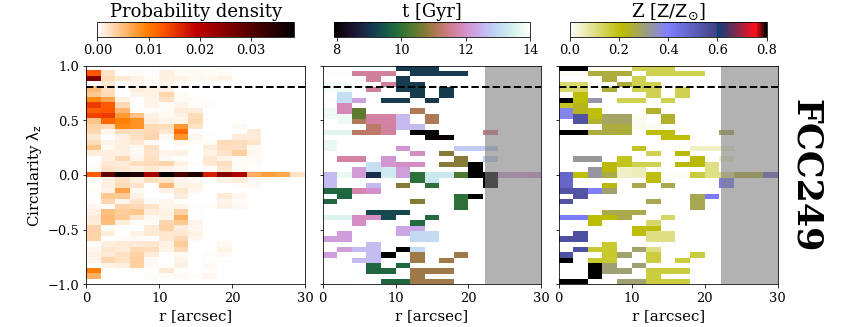}
    \caption{Best-fit population-orbit superposition model (top panels) and orbital decomposition (bottom panels) of FCC\,249. The content of the top and bottom panels is the same as in Figs.~\ref{img:fitting177} and \ref{phasespace177}, respectively.}
    \label{img:fitting249}
\end{figure*}

\begin{figure*}
    \centering{
        \includegraphics[width=1.5\columnwidth, clip=true, trim=80 0 60 0]{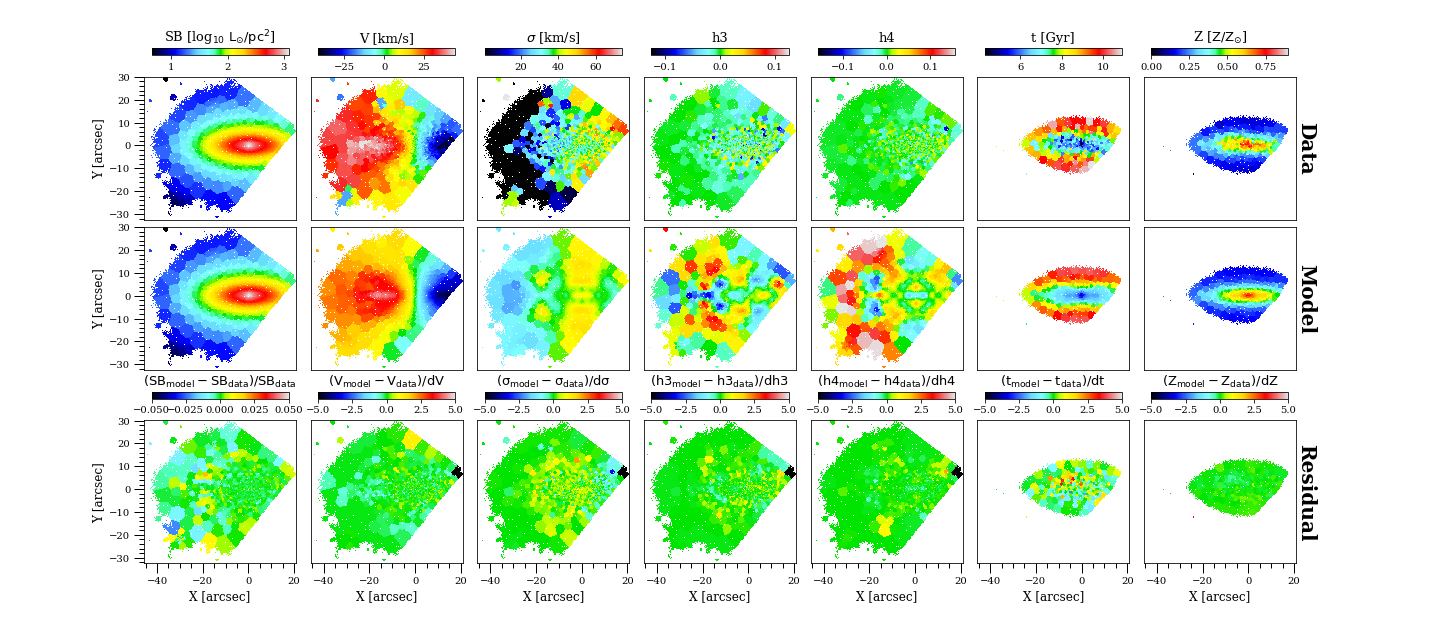}}
        \includegraphics[width=1.05\columnwidth, clip=true, trim=20 0 20 0]{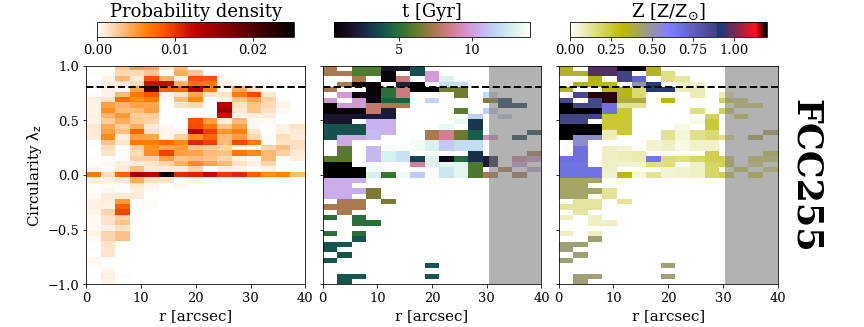}
    \caption{Best-fit population-orbit superposition model (top panels) and orbital decomposition (bottom panels) of FCC\,255. The content of the top and bottom panels is the same as in Figs.~\ref{img:fitting177} and \ref{phasespace177}, respectively.}
    \label{img:fitting255}
\end{figure*}

\begin{figure*}
    \centering{
        \includegraphics[width=1.5\columnwidth, clip=true, trim=0 0 0 0]{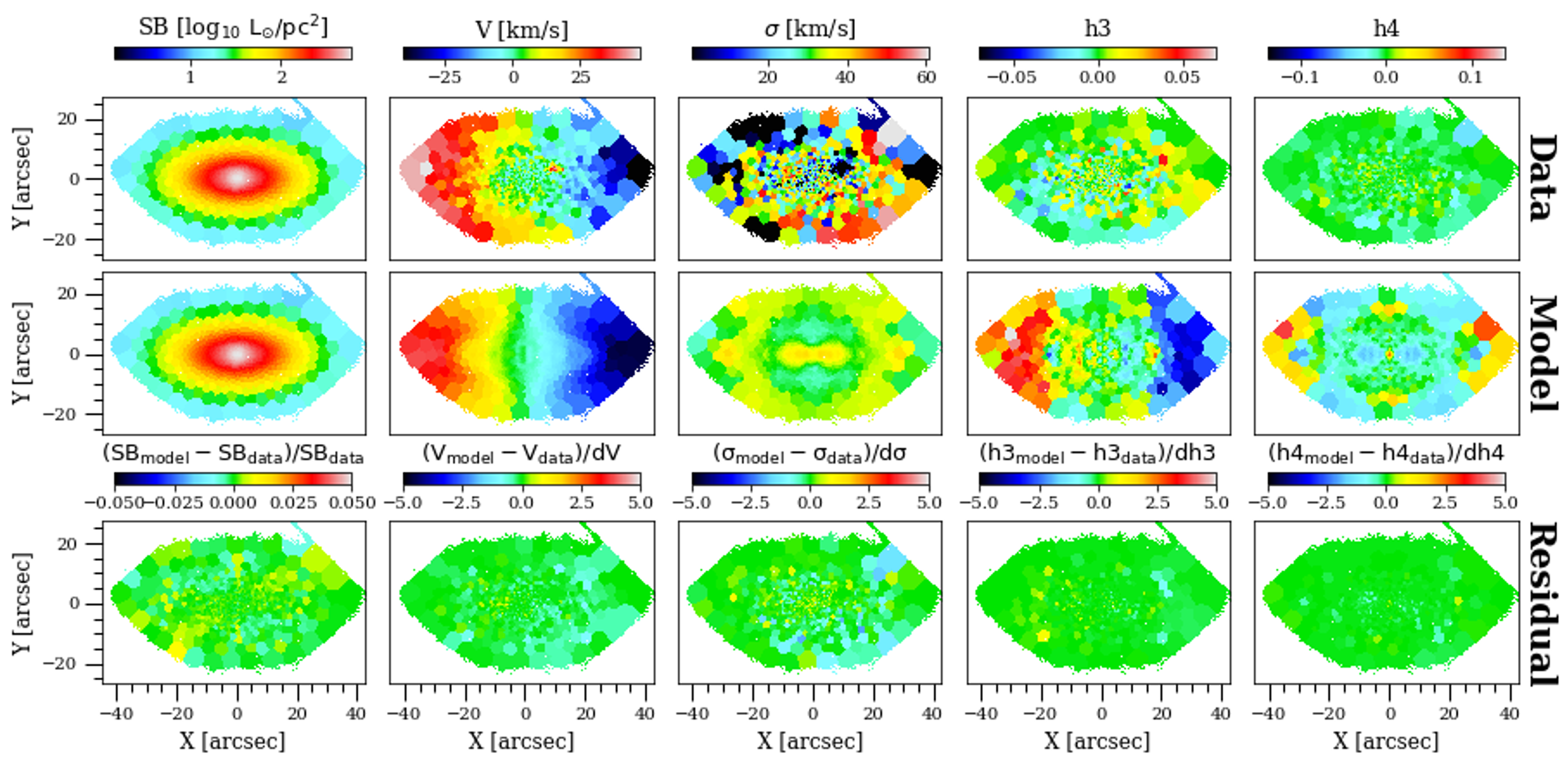}}
        \includegraphics[width=0.5\columnwidth, clip=true, trim=0 0 0 0]{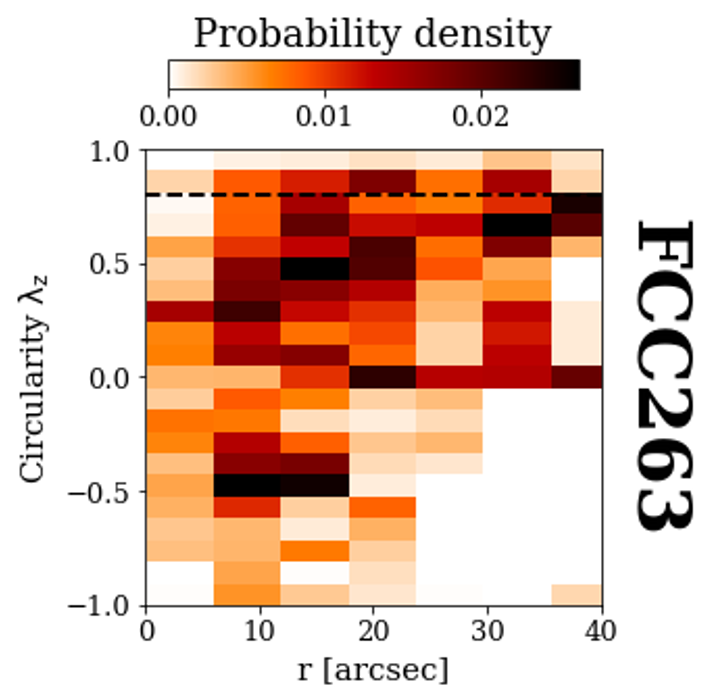}
    \caption{Best-fit orbit superposition model (left panels) and orbital decomposition (right panels) of FCC\,263. Details are similar to Figs.~\ref{img:fitting177} and \ref{phasespace177} as we show for FCC 177, but we do not show the age and metallicity.}
    \label{img:fitting263}
\end{figure*}

\begin{figure*}
    \centering{
        \includegraphics[width=1.5\columnwidth, clip=true, trim=80 0 60 0]{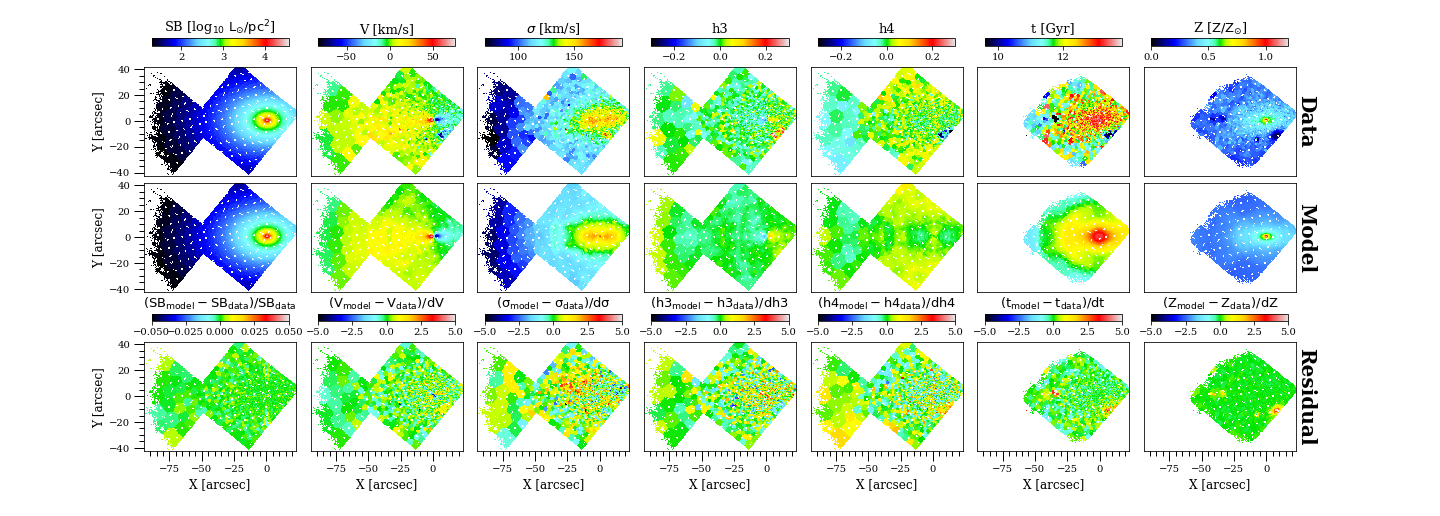}}
        \includegraphics[width=1.05\columnwidth, clip=true, trim=20 0 20 0]{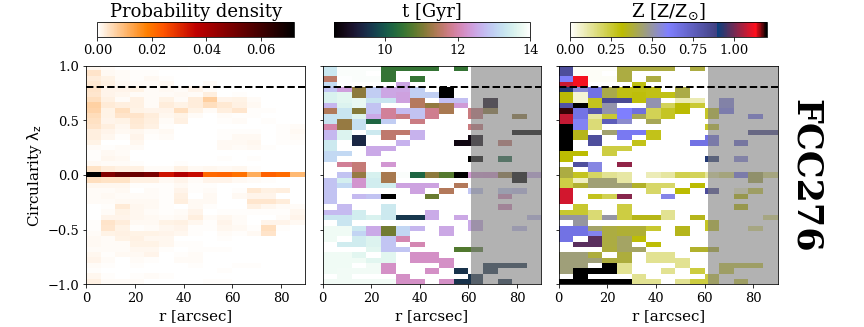}
    \caption{Best-fit population-orbit superposition model (top panels) and orbital decomposition (bottom panels) of FCC\,276. The content of the top and bottom panels is the same as in Figs.~\ref{img:fitting177} and \ref{phasespace177}, respectively.}
    \label{img:fitting276}
\end{figure*}

\begin{figure*}
    \centering{
        \includegraphics[width=1.5\columnwidth, clip=true, trim=0 0 0 0]{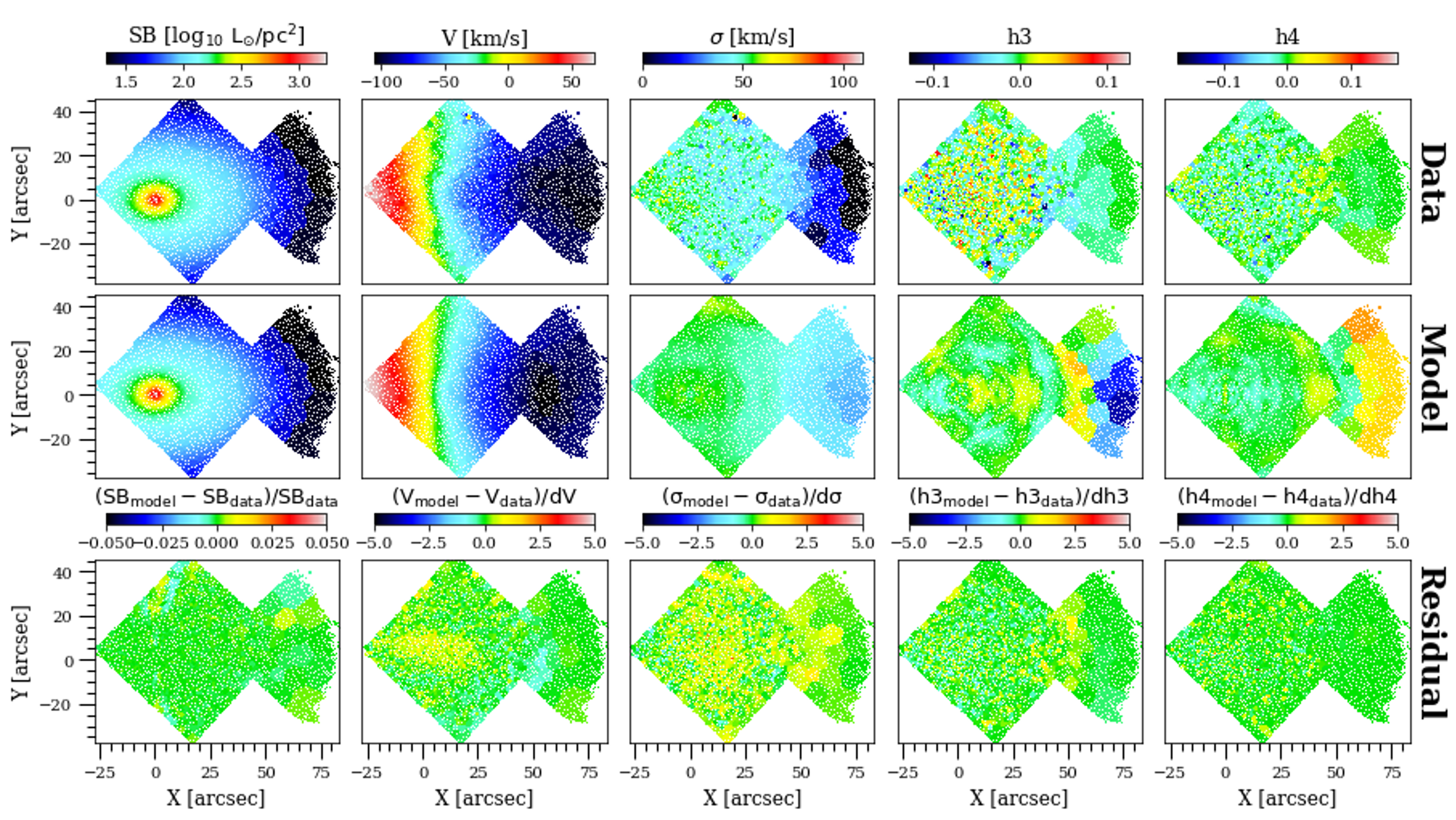}}
        \includegraphics[width=0.5\columnwidth, clip=true, trim=0 0 0 0]{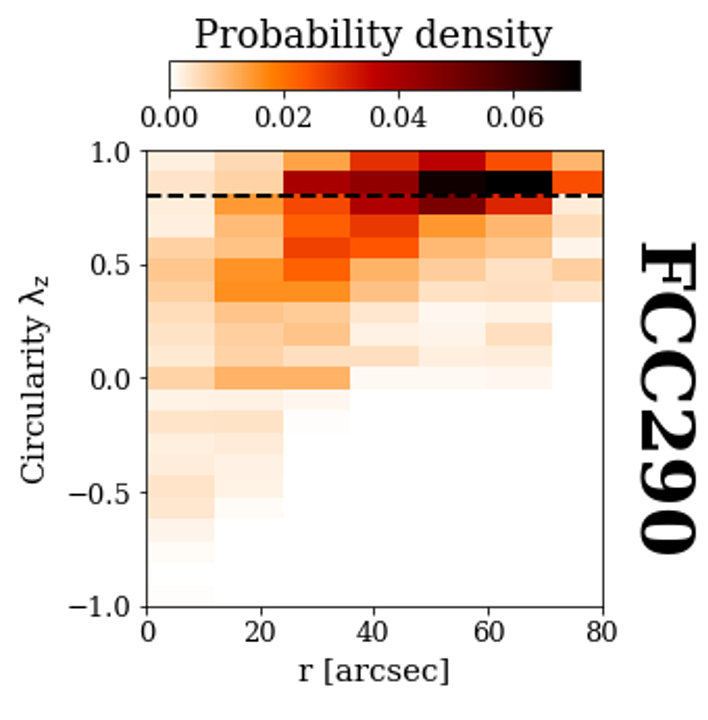}
    \caption{Best-fit orbit superposition model (left panels) and orbital decomposition (right panels) of FCC\,290. Details are similar to Figs.~\ref{img:fitting177} and \ref{phasespace177} as we show for FCC 177, but we do not show the age and metallicity.}
    \label{img:fitting290}
\end{figure*}

\begin{figure*}
    \centering{
        \includegraphics[width=1.5\columnwidth, clip=true, trim=80 0 60 0]{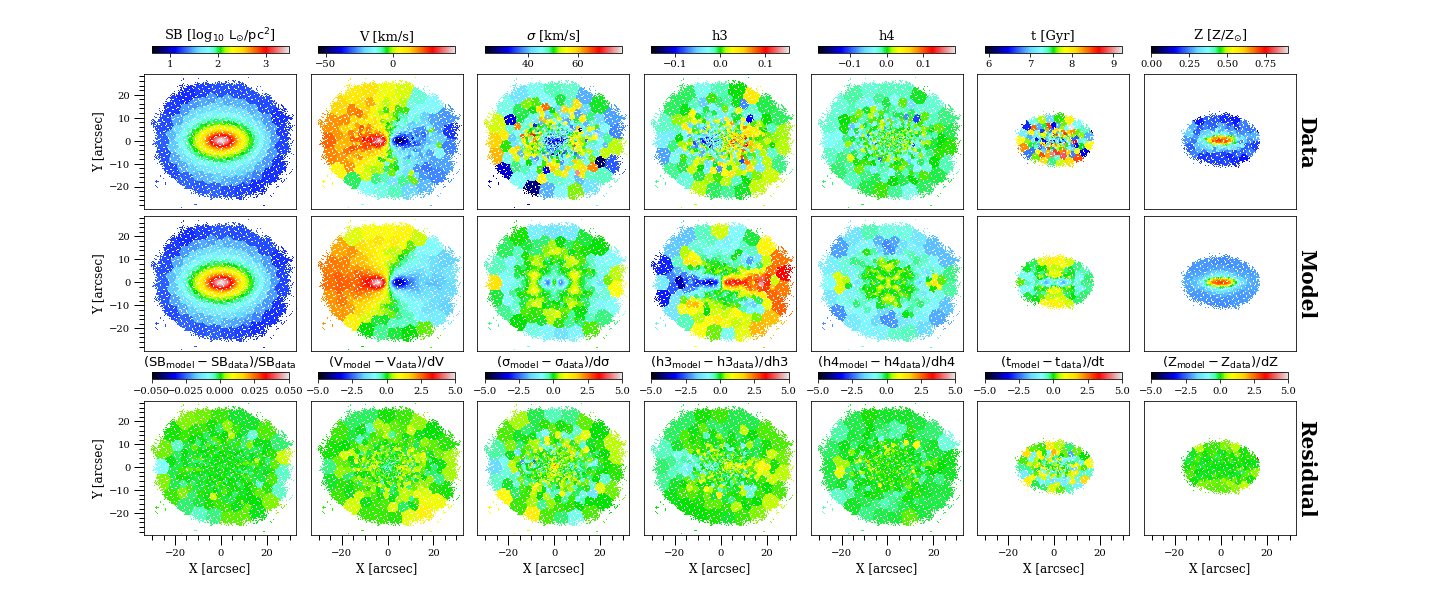}}
        \includegraphics[width=1.05\columnwidth, clip=true, trim=20 0 20 0]{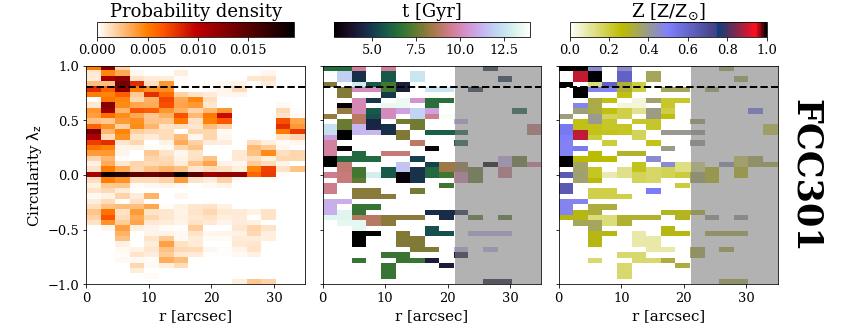}
    \caption{Best-fit population-orbit superposition model (top panels) and orbital decomposition (bottom panels) of FCC\,301. The content of the top and bottom panels is the same as in Figs.~\ref{img:fitting177} and \ref{phasespace177}, respectively.}
    \label{img:fitting301}
\end{figure*}

\begin{figure*}
    \centering{
        \includegraphics[width=1.5\columnwidth, clip=true, trim=0 0 0 0]{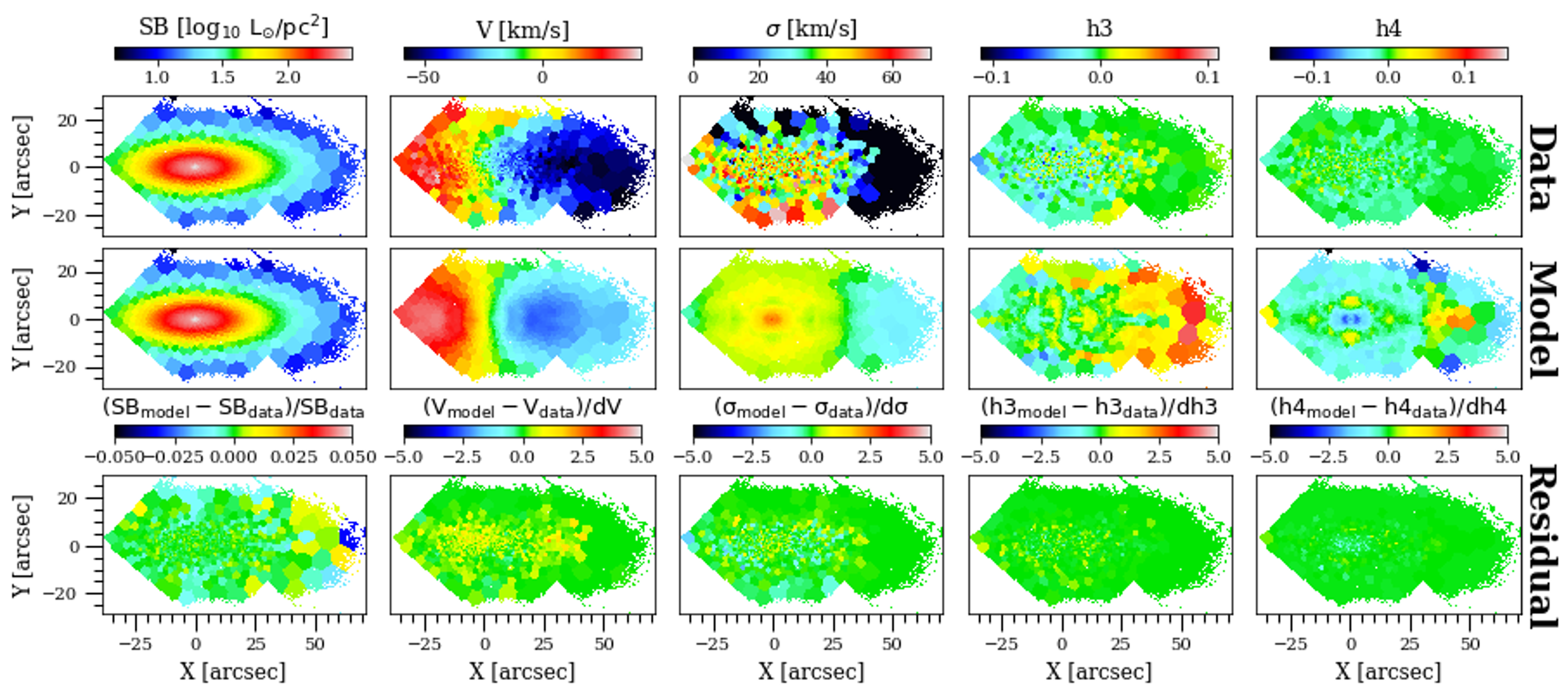}}
        \includegraphics[width=0.5\columnwidth, clip=true, trim=0 0 0 0]{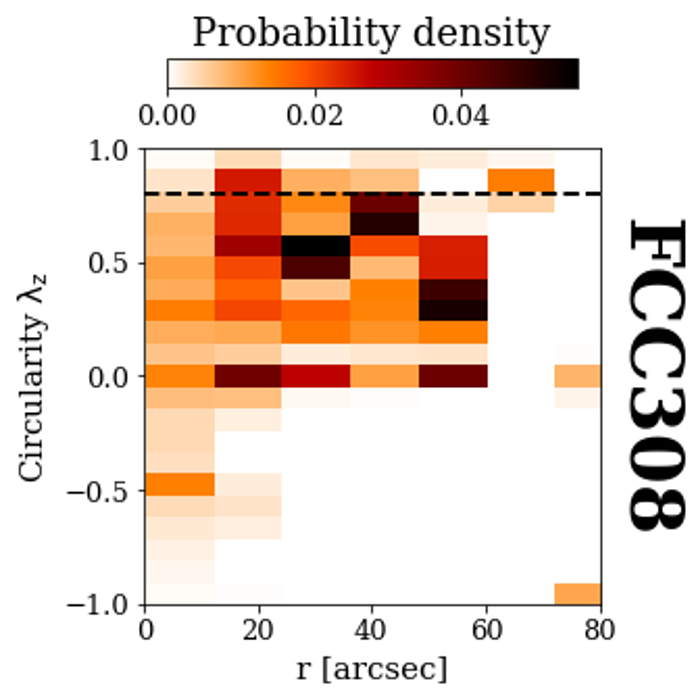}
    \caption{Best-fit orbit superposition model (left panels) and orbital decomposition (right panels) of FCC\,308. Details are similar to Figs.~\ref{img:fitting177} and \ref{phasespace177} as we show for FCC 177, but we do not show the age and metallicity.}
    \label{img:fitting308}
\end{figure*}

% \begin{figure*}
%     \centering{
%         \includegraphics[width=1.5\columnwidth, clip=true, trim=80 0 60 0]{{Figures/FCC310_SN60_fitting}.png}}
%         \includegraphics[width=1.05\columnwidth, clip=true, trim=20 0 20 0]{{Figures/FCC310_SN60_phase-space}.png}
%     \caption{Same as in Fig.~?????????????????????????????????????????????????????????\ref{img:fitting083}, but for FCC~310. A strongly bared galaxy.}
%     \label{img:fitting310}
% \end{figure*}

\begin{figure*}
    \centering{
        \includegraphics[width=1.5\columnwidth, clip=true, trim=0 0 0 0]{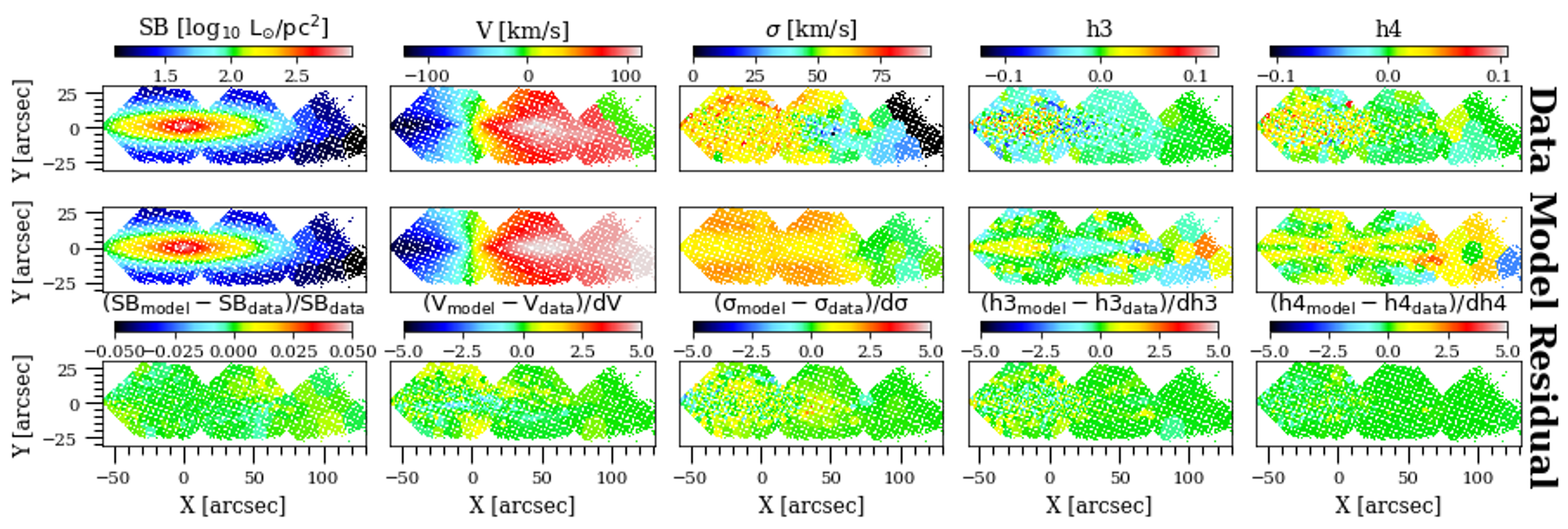}}
        \includegraphics[width=0.5\columnwidth, clip=true, trim=0 0 0 0]{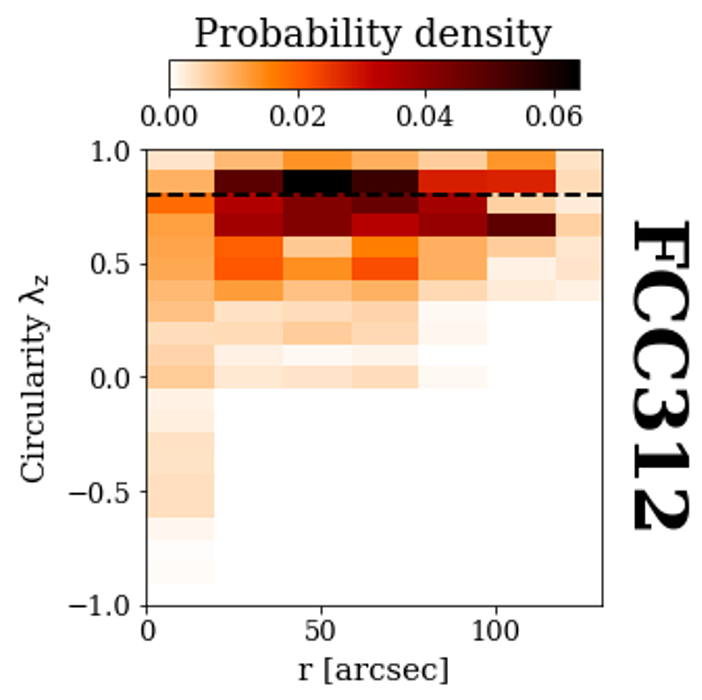}
    \caption{Best-fit orbit superposition model (left panels) and orbital decomposition (right panels) of FCC\,312. Details are similar to Figs.~\ref{img:fitting177} and \ref{phasespace177} as we show for FCC 177, but we do not show the age and metallicity.}
    \label{img:fitting312}
\end{figure*}

\section{Grid parameters of FCC~177}

\begin{figure*}
    \centering{
        \includegraphics[width=2.\columnwidth, clip=true, trim=0 0 0 0]{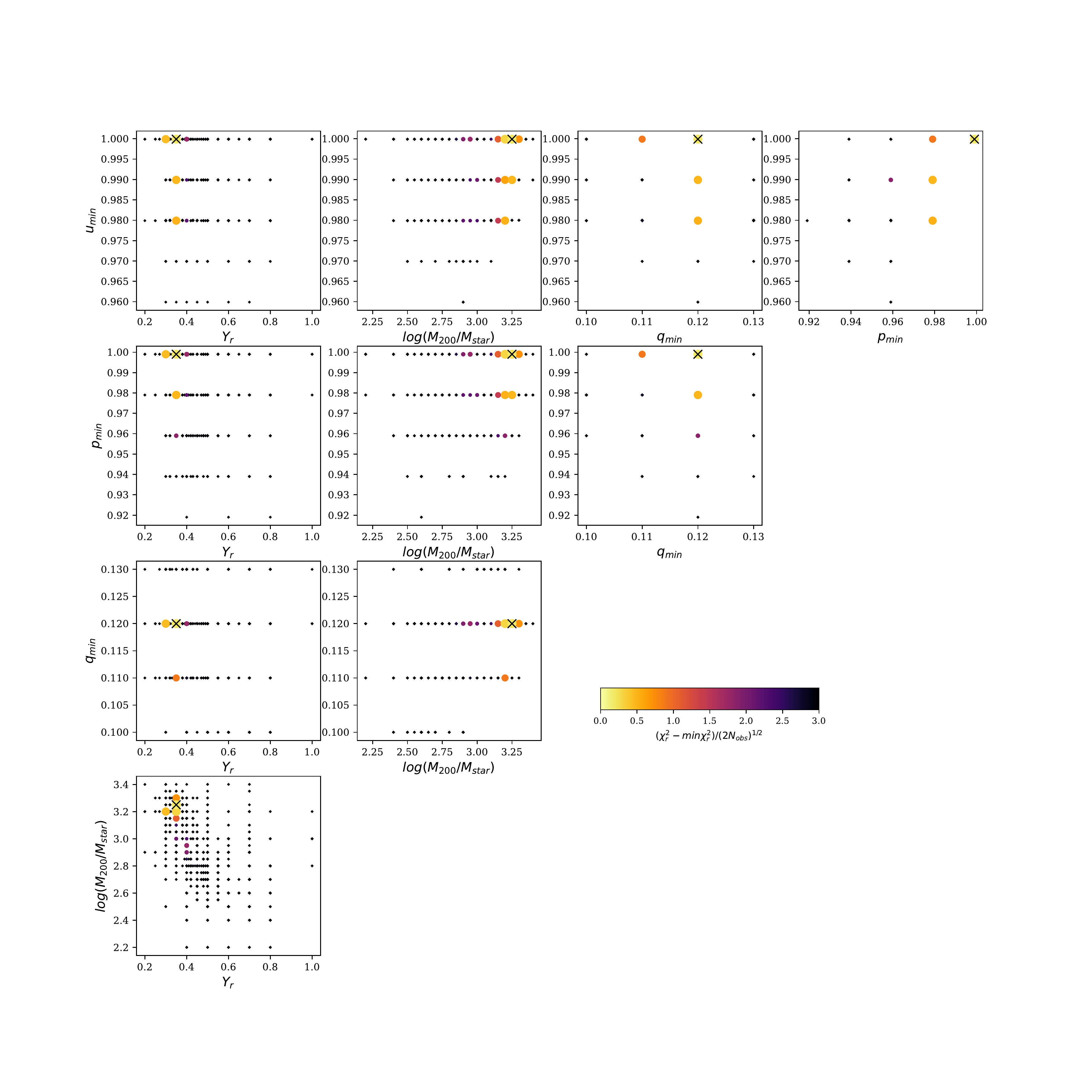}}
    \caption{Optimization over the 5D parameter-space for FCC~177. Each points indicates an exploration of the parameter-space with the smaller orbit sampling. Color represents the $\chi^2$ of each parameter set. The best-fit parameter is indicated by the yellow most point with a cross symbol.}
    \label{img:grid177}
\end{figure*}

\end{appendix}

\end{document}